\documentclass[fleqn,usenatbib]{mnras}

\usepackage{newtxtext,newtxmath}

\usepackage[T1]{fontenc}

\DeclareRobustCommand{\VAN}[3]{#2}
\let\VANthebibliography\thebibliography
\def\thebibliography{\DeclareRobustCommand{\VAN}[3]{##3}\VANthebibliography}

\usepackage{graphicx}	
\usepackage{amsmath}	
\usepackage{hyperref}

\title[Precise warm-gas outflow conditions in IC\;5063]{Precise physical conditions for the warm gas outflows in the nearby active galaxy IC\;5063}

\author[L.R. Holden et al.]{
Luke R. Holden,$^{1}$\thanks{E-mail: lholden2@sheffield.ac.uk}
Clive N. Tadhunter,$^{1}$
Raffaella Morganti,$^{2,3}$
Tom Oosterloo$^{2,3}$
\\
$^{1}$Department of Physics $\&$ Astronomy, University of Sheffield, S6 3TG Sheffield, UK. \\
$^{2}$ASTRON, the Netherlands Institute for Radio Astronomy, Oude Hoogeveensedijk 4, 7991 PD, Dwingeloo, The Netherlands.\\
$^{3}$Kapteyn Astronomical Institute, University of Groningen, Postbus 800, 9700 AV Groningen, The Netherlands.\\
}

\date{Accepted XXX. Received YYY; in original form ZZZ}

\pubyear{2021}

\begin{document}
\label{firstpage}
\pagerange{\pageref{firstpage}--\pageref{lastpage}}
\maketitle

\begin{abstract}
AGN-driven outflows are now routinely used in models of galaxy evolution as a feedback mechanism, however many of their properties remain highly uncertain. Perhaps the greatest source of uncertainty is the electron density of the outflowing gas, which directly affects derived kinetic powers and mass outflow rates. Here we present spatially-resolved, wide spectral-coverage Xshooter observations of the nearby active galaxy IC\;5063 ($z = 0.001131$), which shows clear signatures of outflows being driven by shocks induced by a radio jet interacting with the ISM. For the first time, we use the higher critical-density transauroral [SII] and [OII] lines to derive electron densities in spatially-resolved observations of an active galaxy, and present evidence that the lines are emitted in the same spatial regions as other key diagnostic lines. In addition, we find that the post-shock gas is denser than the pre-shock gas, possibly due to shock compression effects. We derive kinetic powers for the warm ionised outflow phase and find them to be below those required by galaxy evolution models; however, other studies of different gas phases in IC\;5063 allow us to place our results in a wider context in which the cooler gas phases constitute most of the outflowing mass. We investigate the dominant ionisation and excitation mechanisms and find that the warm ionised outflow phase is dominated by AGN-photoionisation, while the warm molecular phase has composite AGN-shock excitation. Overall, our results highlight the importance of robust outflow diagnostics and reinforce the utility of the transauroral lines for future studies of outflows in active galaxies.
\end{abstract}




\section{Introduction}

The extreme amounts of energy released by an active galactic nucleus (AGN) may couple to the interstellar medium (ISM) of the host galaxy in various different ways. This process, known as AGN-feedback, has been invoked as a way to regulate observed scaling relations between SMBH and host galaxy properties \citep{Silk1998, Fabian1999}. Theoretical models of galaxy evolution now routinely require that a relatively large fraction ($\sim$0.5--10 per cent; \citealt{DiMatteo2005, Springel2005, Hopkins2010}) of the AGN bolometric luminosity ($L_\mathrm{Bol}$) powers galaxy-wide outflows as a feedback mechanism.

Testing these models with direct observations is crucial. Indeed, AGN-driven outflows have been observed in active galaxies in different gas phases including cold molecular (e.g. CO; \citealt{Alatalo2011, Cicone2014, Morganti2015, Oosterloo2017}), warm molecular (e.g. H$_2$; \citealt{Tadhunter2014}), neutral (e.g. HI, NaID; \citealt{Morganti1998, Oosterloo2000, Rupke2005, Morganti2005}) and warm ionised (e.g. \citealt{VillarMartin1999, Holt2003, Nesvadba2006, Zaurin2013, Harrison2014, Concas2017, Rose2018, Tadhunter2019}).  It has been proposed that the different phases may represent stages of a cooling sequence after the initial shock-acceleration of an outflow heats the gas and destroys the molecular and neutral components \citep{VillarMartin1999, Zubovas2014, Tadhunter2014}. Following the heating, the gas would cool and be observable as the warm ionised phase at T$\sim10^4$\;K, then the neutral phase at \mbox{T\textless10$^4$\;K} followed by the cool molecular phase at \mbox{T\textless100\;K}. However, in reality the physical relation between these phases remains unclear, and conclusions drawn from observations of a single phase should be made with care (see discussion in \citealt{Cicone2018}).

Many studies have focused on estimates of the so-called `coupling efficiency' ($\epsilon_\mathrm{f} = \dot{E}_\mathrm{kin}/L_\mathrm{bol}$; the ratio of outflow kinetic power to AGN bolometric luminosity) by using observations of the warm ionised phase (e.g. \citealt{Liu2013, Harrison2014, Rose2018, Tadhunter2019}). This is typically done to compare values of $\epsilon_\mathrm{f}$ derived from observation to those used in galaxy evolution models in an attempt to verify the models. However, many key properties of the warm ionised phase that are required to derive coupling efficiencies are uncertain, and a large range of values have been found (see Figure 2 in \citealt{Harrison2018}).

Perhaps the greatest source of uncertainty in deriving outflow kinetic powers and coupling efficiencies is the electron density ($n_\mathrm{e}$) of the outflowing gas. Commonly used electron density diagnostic techniques that make use of the `traditional' [SII]$\lambda\lambda$6717,6731 and  [OII]$\lambda\lambda$3726,3729 doublet ratios are only sensitive up to $n_\mathrm{e}\sim10^{3.5}$\;cm$^{-3}$, leading to values of $n_\mathrm{e}\sim$10$^{2-3}$\;cm$^{-3}$ being often estimated or assumed (e.g. \citealt{Liu2013, Harrison2014, Fiore2017, Mingozzi2019}). Furthermore, the doublets suffer from blending issues when the line profiles are broad and complex, as is generally the case with outflows. However, using alternative density diagnostics, such as the `transauroral' (TR) [OII](3726 + 3729)/(7319 + 7331) and [SII](4068 + 4076)/(6717 + 6731) diagnostic ratios, higher electron densities in the range of \mbox{10$^3$\;\textless\;$n_\mathrm{e}$\;\textless\;$10^{5.5}$\;cm$^{-3}$} have been found \citep{Holt2011, Rose2018, Santoro2018, Spence2018, Santoro2020, Davies2020}. Studies which make use of other techniques of electron density estimation, such as using the ionisation parameter with infrared-based estimates of the outflow radius \citep{Baron2019b}, similarly find much higher electron densities than typically assumed. 

Higher values for electron density, such as those derived from the transauroral lines, lead to substantially lower mass outflow rates and coupling efficiencies that are potentially below those required by models of galaxy evolution. However, to date, it has not yet been demonstrated that the transauroral lines are emitted on the same radial scales as other key outflow diagnostic lines (e.g. [OIII] and H$\mathrm{\beta}$) that are used to determine kinematics and outflow masses within a given galaxy: the clouds emitting the key diagnostic lines may be distinct and lie at different spatial positions to those emitting the transauroral lines. Previous transauroral line studies of AGN-driven outflows have been spatially unresolved, concentrating on single aperture spectra of near-nuclear regions, and have thus been unable to test this (e.g. \citealt{Holt2011, Rose2018, Santoro2018, Santoro2020}).

In addition to the uncertain properties of the warm ionised phase, the acceleration mechanisms for outflows across all phases are also unclear. Several different mechanisms have been proposed, including the launching of a fast wind from the accretion disk which interacts with the larger-scale ISM and drives an outflow (e.g. \citealt{Hopkins2010}; the `radiative' mode), and relativistic radio-plasma jets launched directly from the central supermassive black hole interacting with the host ISM and inducing shocks, which in turn accelerate outflows (e.g. \citealt{Wagner2011, Gaibler2012, Mukherjee2018}). The latter mechanism is different from the commonly considered `maintenance' mode role of radio jets, which is thought to play an important role in heating hot gas on large scales, because here the jet is instead directly affecting the cooler phases of the ISM on smaller scales (a few kpc) and launching outflows in manner that is more akin to the radiative mode. In this context it is notable that, for a large sample of SDSS-selected AGN, \citet{Mullaney2013} find the highest degree of kinematic disturbance in [OIII] emission line profiles for objects of intermediate radio power (L$_\mathrm{1.4\;GHz} = 10^{23-25}$\;WHz$^{-1}$). This suggests that the feedback effect of relativistic jets is not confined to the highest power jets, but is also important at lower jet powers.

Investigating the ionisation of the outflowing gas may provide insights into which acceleration mechanisms are dominant in different object types. For example, definitive evidence of shock-ionised outflows would conclusively show that the outflows are being accelerated by shocks. Several diagnostic diagrams making use of emission line ratios have been used to discriminate between different ionisation mechanisms. Perhaps the most well-known of these are the `BPT' diagrams \citep{Baldwin1981}, which make use of ratios of the most prominent optical emission lines to define empirical regions in the ratio-ratio planes corresponding to different dominant warm ionised gas ionisation mechanisms. Similarly, diagnostic diagrams have been proposed for the near-infrared (NIR), making use of the \mbox{[FeII]$\mathrm{\lambda}$12570/Pa$\mathrm{\beta}$} and \mbox{H$_2$1--0S(1)$\mathrm{\lambda}$21218/Br$\mathrm{\gamma}$} ratios \citep{Larkin1998, Rodriguez-Ardila2005, Riffel2013a, Colina2015}. These may be used to discriminate between AGN photoionisation and shock ionisation/excitation for both the warm ionised and warm molecular gas phases.

Detailed single-object studies are ideal for investigating the physical conditions and ionisation mechanisms of outflowing gas. Therefore, here we present a spatially-resolved, detailed spectroscopic study of the local Seyfert galaxy IC\;5063, which has an intermediate radio power, and outflow regions that are clearly resolved in ground-based observations. Our principal goal is to investigate the extent to which the transauroral lines can be used as a density diagnostic in spatially-resolved studies of AGN-driven outflows, and subsequently, their use in non spatially-resolved studies. We also investigate the dominant ionisation mechanisms for the gas, which is thought to be accelerated by jet-ISM induced shocks.

The structure of the report is as follows. In Section \ref{section: ic_5063}, we describe IC\;5063, and give an overview of the past studies of the object. We describe our Xshooter observations, data reduction methods and emission line fitting procedure in Section \ref{section: observations_and_reduction}. The methodology and results for the UVB+VIS Xshooter data is presented in Section \ref{section: uvb_vis_results}; for the NIR data, this is given in Section \ref{section: nir_results}. Finally, we discuss the interpretation and implications of our results in Section \ref{section: discussion}, and give our conclusions in Section \ref{section: conclusions}.

Throughout this work, we assume a cosmology of $H_\mathrm{0}$ = 70\;km\,s$^{-1}$Mpc$^{-1}$, $\Omega_\mathrm{m}$ = 0.3 and ${\Omega}_\mathrm{\lambda}$ = 0.7. At the redshift of IC\;5063 ($z=0.01131$; \citealt{Tadhunter2014}), this corresponds to a luminosity distance of 48.9\;Mpc and an angular scale of 0.231\;kpc/arcsec.

\section{IC\;5063}
\label{section: ic_5063}

\begin{figure}
    \includegraphics[width=\linewidth]{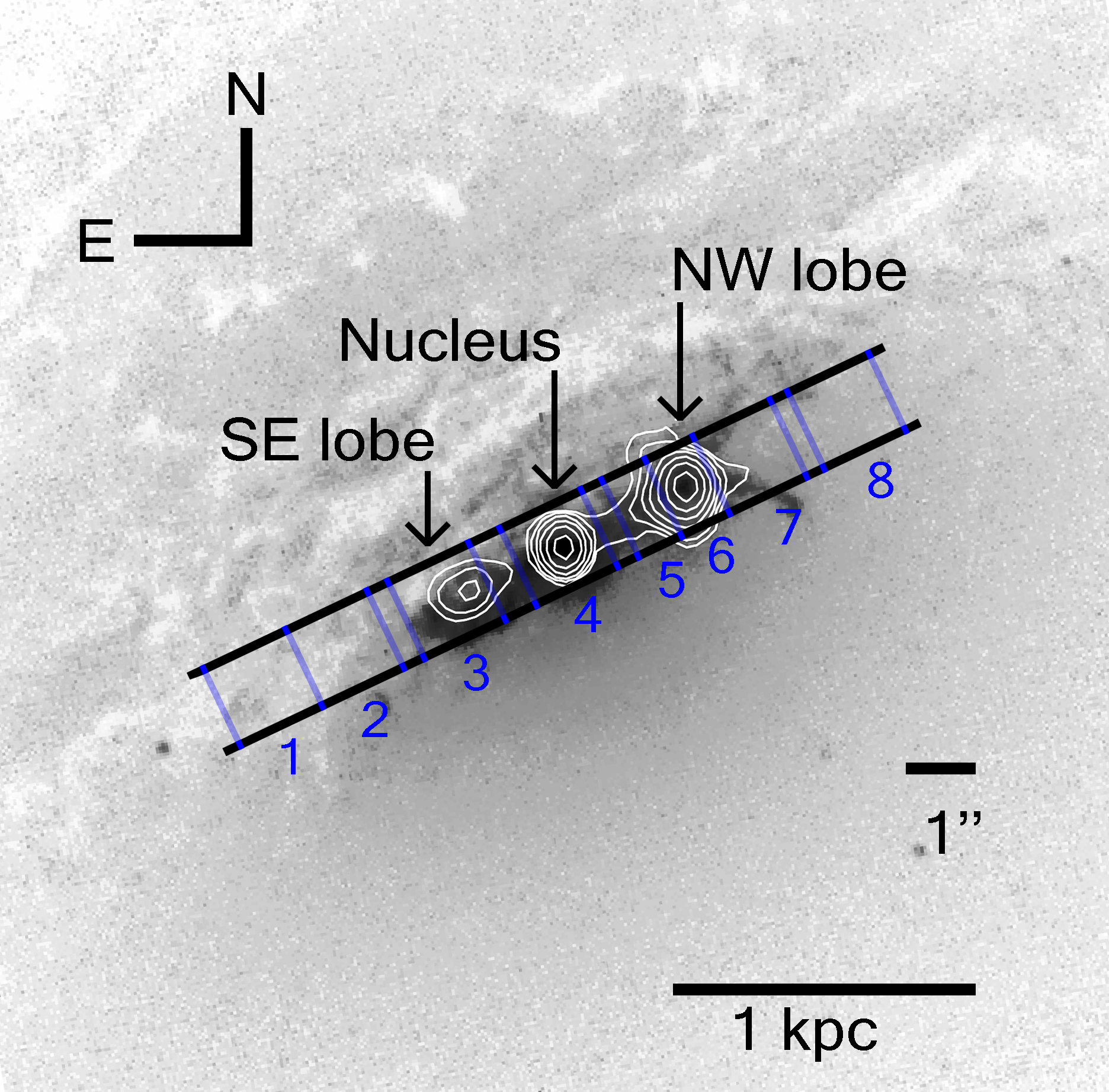}
    \caption{Archival HST WFPC2 optical image of the central $\sim$3\;kpc of IC\;5063 (Cycle 4; SNAP:5479; PI Malkan), taken through the F606W filter, with 17\;GHz radio emission from ATCA \citep{Morganti2007} shown as white contours. The slit is overlaid in black, with the borders of each aperture along the slit (Section \ref{section: aperture_selection}; Figure \ref{fig: apertures}) shown and labelled in blue. The NW and SE radio lobes are labelled, along with the nucleus. Note that the F606W filter admits strong [OIII] and H$\mathrm{\alpha}$+[NII] emission lines, as well as continuum emission. In this negative image, the inner part of the large-scale dust lane can be seen as the lighter-toned filamentary structures to the north of the nucleus, whereas the inner emission-line structures appear as the darker grey or black features close to the nucleus.}
    \label{fig: ic5063_hst_17ghz}
\end{figure}

IC\;5063 is a nearby ($z$=0.01131) early-type galaxy with a Type\;2 Seyfert nucleus \citep{Danziger1981, Inglis1993}. The galaxy hosts a gaseous disk that is detected out to a radius of $\sim$28kpc in HI 21cm emission \citep{Morganti1998}, is associated with prominent dust lanes, and may be the result of a merger \citep{Morganti1998}. The inner parts of this disk (within a few kpc) are also detected in cold CO \citep{Morganti2015} and warm H$_2$ molecular gas \citep{Tadhunter2014}, and [OIII] emission lines from warm, ionised gas are detected in HST images along the edges of the dust lane to a radius of $\sim$10\;kpc \citep{Morganti1998}. IC\;5063 has a radio power at the upper end of the range for Seyfert galaxies (P$_\mathrm{1.4\;GHz}=3\times10^{23}$\;WHz$^{-1}$), and the radio emission is concentrated in a triple-structure parallel to the dust lanes, which consists of a nucleus and two radio lobes $\sim2$\;arcsec ($\sim$0.5\;kpc) on either side of the nucleus to the south east (SE) and north west (NW) (Figure \ref{fig: ic5063_hst_17ghz}).

Observations have revealed fast (\mbox{$\Delta V$\textgreater700\;km\,s$^{-1}$}) outflows spatially associated with the NW radio lobe across a range of gas phases: warm ionised \citep{Morganti2007, Sharp2010, Congiu2017, Venturi2021}; neutral \citep{Morganti1998, Oosterloo2000}; warm molecular \citep{Tadhunter2014} and cold molecular \citep{Morganti2013, Morganti2015, Dasyra2016, Oosterloo2017}. In addition, IC\;5063 has been the target of X-ray observations \citep{Vignali1997, Tazaki2011, Travascio2021}.

IC\;5063 is unusual in the sense that its radio jets are propagating in the plane of its disk \citep{Morganti1998, Oosterloo2000, Morganti2015, Mukherjee2018}. Considering that the outflows in the galaxy are spatially associated with the bright radio features, particularly at the NW radio lobe, it is thought that jet-ISM interactions are the main outflow acceleration mechanism in the galaxy, with these interactions being particularly strong due to the jet propagating in the denser ISM in the disk \citep{Tadhunter2014, Morganti2015}. Hydrodynamic simulations by \citet{Mukherjee2018} explain the gas kinematics in IC\;5063 by assuming that a relatively low-power jet (P$_\mathrm{jet}$=10$^{37-38}$\;W) percolates through the path of least resistance in the disk. As it does so, jet-ISM interactions accelerate gas perpendicular to the jet direction. This is supported by observations of a giant low-ionisation loop \citep{Maksym2020}. Furthermore, Deep Chandra observations of IC\;5063 presented by \citet{Travascio2021} have revealed extended X-ray emission in a bi-conical structure in the direction of the radio jets, with the soft X-ray emission increasing along the jet towards the nucleus, as well as evidence for dense molecular clouds in the region of the NW radio lobe being responsible for stopping the jet.

IC\;5063's relatively close proximity, the strength of its jet-cloud interactions, and the wealth of previous studies of all gas phases make it ideal for understanding the acceleration mechanisms of outflows in active galaxies, as well as the physical relationship between the different phases. Furthermore, its outflowing regions can be spatially resolved, allowing density diagnostics to be thoroughly tested, in particular the extent to which a lack of spatial resolution may be a problem for the transauroral line technique. For these reasons, IC\;5063 was chosen to be the target of deep observations with the Very Large Telescope's (VLT) Xshooter spectrograph in 2018.

\section{Observations and Data Reduction}
\label{section: observations_and_reduction}

\subsection{Xshooter observations of IC\;5063}

IC\;5063 was observed using VLT/Xshooter in service mode on the nights of the 13th and 16th June 2018 in four observing blocks during photometric and dark sky conditions. The observations were done in long-slit mode (slit length = 11\;arcsec), with slit widths of 1.3\;arcsec for the UVB arm (wavelength range: 3200--5600\;\AA) and 1.2\;arcsec for the VIS (5500--10200\;\AA) and NIR (10200--24750\;\AA) arms. The slit was aligned along the galaxy's radio axis (PA=115$^{\circ}$), meaning that the radio hotspots ($\sim$4\;arcsec separation) were contained in the 11\;arcsec slit length (see Figure \ref{fig: ic5063_hst_17ghz}). The data has a pixel scale of 0.16\;arcsec/pixel for the UVB and VIS arms and 0.21\;arcsec/pixel for the NIR arm. Separate sky exposures were taken by nodding off the target object with a 30 arcsecond spatial offset and using an ABBA observing pattern. This facilitated accurate sky subtraction during data reduction. Each observing block thus consisted of 6 exposures (3 on-object and 3 on-sky), with a total integrated object exposure time of 5400\;s. In addition, standard calibration files (e.g. bias frames, dark frames and standard star observations) were provided along with the science data for use in data reduction.

In order to determine the instrumental broadening of our Xshooter observations, Gaussian profiles were fitted to the Galactic NaD absorption lines at 5890\;{\AA} and 5896\;{\AA}. Since the gas responsible for this absorption is quiescent (i.e non-outflowing) within the Milky Way, any broadening of these lines will be due entirely to the instrumental setup. From the galactic NaD lines, we measure an instrumental width of FWHM$_\mathrm{inst}$ = 0.825$\pm$0.016\;{\AA}, corresponding to a highest velocity resolution of 42.0$\pm$0.8\;km\,s$^{-1}$ at 5900\;\AA.

\subsection{Seeing estimates}

We made estimates of the seeing during the four observing blocks by fitting Gaussian profiles to the spatial slices of stars in the acquisition images taken with the Sloan r' filter for each observing block. The width of each spatial slice was 1.2\;arcsec, taken to match the Xshooter slit widths. We also took seeing estimates from the Paranal DIMM (Differential Image Motion Monitor) during the four observing blocks, as measured near zenith with a filter centred on $\sim$5000\;\AA. Seeing estimates measured from the acquisition images and the DIMM are given in Table \ref{tab: seeing}.

\begin{table}
\centering
\begin{tabular}{l|l|l}
\hline
Observing Block & \parbox{1cm}{DIMM \\ (arcsec)} & \parbox{1.5cm}{Acq. Image \\ (arcsec)} \\ \hline
1               & 0.83                                                         & 2.08$\pm$0.03                                                  \\
2               & 1.06                                                         & 0.93$\pm$0.01                                           \\
3               & 0.69                                                         & 0.93$\pm$0.01                                               \\
4               & 0.74                                                         & 0.72$\pm$0.01                                              
\end{tabular}
\caption{Seeing estimates for each observing block of the VLT/Xshooter observations of IC\;5063, taken on the nights of the 13th (Block 1) and 16th (blocks 2, 3 and 4) of June 2018. `Max DIMM' is the maximum seeing recorded by the VLT observatory DIMM during the time-frame of a given observing block whilst `Acq. Image' is the seeing measured by fitting 1D Gaussian profiles to 1.2\;arcsec wide spatial slices centred on stars in the r$^\prime$-band acquisition images. All values are given in arcseconds.}
\label{tab: seeing}
\end{table}

\subsection{Reduction of the Xshooter data}
\label{section: data_reduction}

The European Southern Observatory's (ESO) pipeline for Xshooter data reduction, \textsc{ESOReflex} (version 2.11.0; \citealt{Freudling2013}), was used for the the initial stage of data reduction. Used with the calibration data provided by ESO, this produced a single two-dimensional bias-subtracted, flat-fielded, order-merged, wavelength-calibrated and flux-calibrated long-slit spectrum for each Xshooter arm (UVB, VIS, NIR) in each observing block, giving a total of 12 spectra (one for each arm per block).

Residual bad pixels, cosmic rays and other cosmetic defects in each spectrum were then cleaned via interpolation using the \textsc{CLEAN} command from the \textsc{STARLINK FIGARO} package \citep{Currie2014}, and a 2nd-order correction was applied to the NIR data to improve the night-sky line subtraction.

Telluric correction of the VIS and NIR arm data to remove atmospheric absorption features was performed using ESO's \textsc{molecfit} software (version 1.5.9: \citealt{Smette2015, Kausch2015}). Atmospheric models were created using fits to key telluric absorption features which were then used to create synthetic transmission spectra for the VIS and NIR data. The 2D spectra were subsequently telluric corrected by dividing by the corresponding transmission spectrum. After verifying that they were spatially aligned, the cleaned, 2nd-order corrected (for the NIR arm) and telluric corrected 2D spectra for the four observing blocks were then median-combined using \textsc{IRAF} \textsc{imcombine} \citep{Tody1986, Tody1993} into a single 2D spectrum for each arm.

Extinction due to dust in the Milky Way galaxy was corrected using the Galactic extinction maps presented by \citet{Schlegel1998} and re-calibrated by \citet{Schlafly2011}. The mean color excess value in the direction of IC\;5063 \mbox{($E(B-V)_\mathrm{mean}$ = 0.0526$\pm$0.0012)} from these maps was found using the NASA/IPAC Infrared Science Archive reddening lookup tool. This color excess value was used with the $R_\mathrm{v}$=3.1 extinction law presented by \citet{Cardelli1989} (hereafter CCM89) to correct for Galactic extinction. The spectra were then shifted into the rest-frame of IC\;5063 using the \textsc{IRAF} \textsc{dopcor} command with the redshift measured by \citet{Tadhunter2014} ($z=0.01131$).

\subsubsection{Aperture selection and extraction}
\label{section: aperture_selection}

\begin{figure*}
	\includegraphics[width=\linewidth]{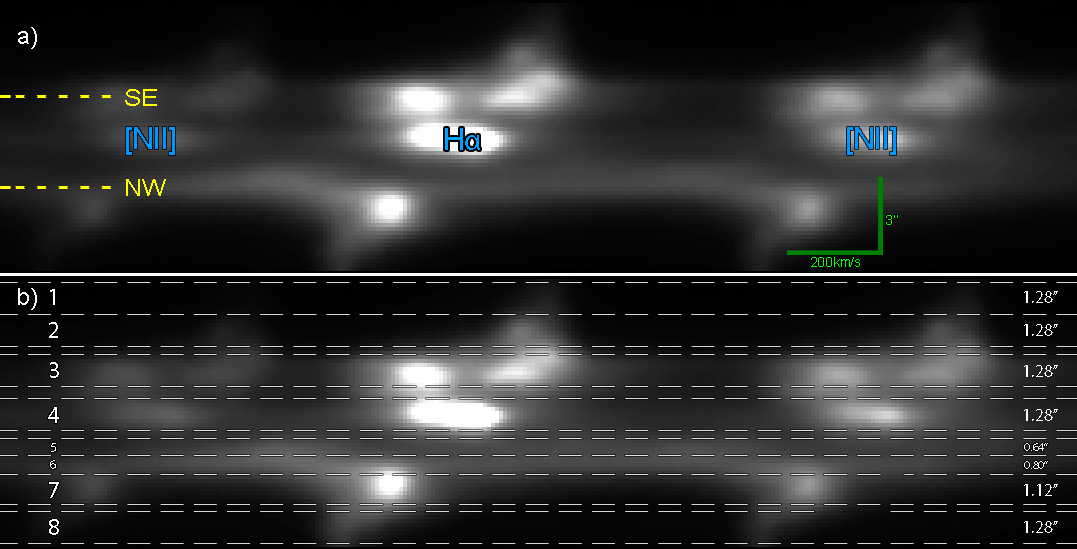}
	\caption{\textbf{a}: Two-dimensional spectrum of the H$\mathrm{\alpha}$+[NII] line features at $\sim$6560\;{\AA} in the VIS arm of the reduced Xshooter data taken of IC\;5063, showing the line profiles in the inner regions of the galaxy. The spectral axis is horizontal and the spatial axis is vertical; the velocity and spatial resolution is shown as a labelled green axis. The centroids of the radio lobes --- measured from 24.8\;GHz radio continuum imaging presented by \citet{Morganti2007} --- are shown as yellow dashed lines, and each lobe is labelled. \textbf{b}: The same as in a), but with apertures selected for the UVB and VIS arms marked with longer white dashed lines and labelled as `1-8' on the left; the spatial width of each aperture in arcseconds is given on the right.}
	\label{fig: apertures}
\end{figure*}

Highly disturbed kinematics and complex line profiles can be seen in the regions coincident with the radio source (Figure \ref{fig: apertures}a), consistent with \citet{Morganti2007}. The reduced and merged two-dimensional spectra for the UVB and VIS Xshooter arms were divided into apertures (groupings of pixel rows) covering these important features of the inner regions of IC\;5063. The apertures chosen for the UVB and VIS arms are shown in Figure \ref{fig: apertures}b, with Aperture 4 covering the nucleus, and apertures 3, 5, 6 and 7 covering distinct kinematic regions along the radio jets. Aperture 3 approximately covers the SW radio lobe, Aperture 5 covers a region of disturbed emission-line kinematics between the nucleus and the maximum extent of the NW lobe, and apertures 6 and 7 cover the NW radio lobe. Once the apertures were selected, the same pixel rows in both the UVB and VIS data were extracted and added to produce two one-dimensional spectra for each aperture (one UVB and one VIS). Before extraction, Gaussian fitting of spatial regions free of line features in both the UVB and VIS arms was performed to ensure that the spectra in the two arms were closely spatially aligned. Due to slight differences in flux calibration, a small ($\sim$5\;per cent) correction was applied to the fluxes of the VIS data for each extracted aperture in order to bring them into line with those of the corresponding UVB aperture. This was done by measuring the average flux between 5440--5500\;{\AA} for the UVB arm and 5600--5660\;{\AA} for the VIS arm --- the ratio of these average fluxes was then used as a correction factor to match the flux scales across both arms.

Finally, the extracted and corrected UVB and VIS 1D spectra were combined to produce a single 1D spectrum for each aperture, covering the full wavelength range of both arms. In the course of this, the spectra for both arms were resampled to a common wavelength grid with steps of $\Delta\lambda$ = 1{\;\AA} using the \textsc{SpectRes} \citep{Carnall2017}, \textsc{specutils} \citep{Earl2021} and \textsc{Astropy} \citep{AstropyCollaboration2013, AstropyCollaboration2018} \textsc{Python} packages. $\Delta\lambda$ = 1{\;\AA} was chosen for the resampling because this was the smallest wavelength step allowed by the base stellar templates used in our stellar continuum modelling (Section \ref{section: starlight}). 

Because of the different spatial pixel scale of Xshooter's NIR arm relative to its UVB and VIS arms, we consider the NIR data separately. Apertures for the NIR data were derived from the UVB+VIS apertures by first fitting two Gaussians to a spatial slice of the continuum near the [OIII]$\lambda\lambda$5007,4959 doublet and taking the narrowest to be the peak of the continuum spatially along the slit. We then calculated the distances from the centre of each UVB+VIS aperture to this continuum peak. The same procedure was used to identify the continuum peak in NIR arm using a continuum slice adjacent to the near-infrared HeI$\lambda$10830 line. The ratio of the VIS+UVB pixel scale (0.16\;arcsec/pixel) to the NIR pixel scale (0.21\;arcsec/pixel) was then used to convert the aperture distances and widths from the UVB+VIS arm to the NIR arm. We note that the NIR apertures are not exactly the same as the UVB+VIS apertures due to the different pixel scales of the arms and the fact that the apertures are only a few pixels wide, along with uncertainties in the Gaussian fits to the continuum spatial slices and the fact that we only measure aperture positions to 0.1\;pixels.

\subsubsection{Stellar continua modelling and subtraction}
\label{section: starlight}

Following the selection, extraction and correction of the apertures, the underlying stellar continua in each UVB+VIS aperture were modelled and subtracted using the \textsc{STARLIGHT} stellar spectral synthesis code (version 4: \citealt{CidFernandes2005, Mateus2006}), making use of the \textsc{STELIB} empirical stellar templates \citep{LeBorgne2003} and the stellar population synthesis model presented by \citet{Bruzual2003}. This was done to ensure measured emission line fluxes used in data analysis were as accurate as possible and did not suffer from the effects of underlying stellar absorption.

In order to ensure good fits to the stellar continua, any spectral features not associated with a stellar component (such as AGN emission lines and any residual telluric absorption) were excluded from the fitting process. The normalising region for the fits was chosen to be 4740--4780{\;\AA} and the wavelength range covered was 3220--8000{\;\AA}.

The adequacy of the resulting fits was checked visually by closely inspecting key absorption features that do not suffer emission-line contamination, such as the MgI absorption feature at 5167\;{\AA} and the CaII K absorption feature at 3934\;{\AA} (see \mbox{Appendix \ref{section: starlight_appendix}}), as well as the fit to the overall continuum shape. After being deemed acceptable, the modelled stellar spectra were subtracted from the data in each aperture. 

\subsubsection{Emission line fitting}
\label{section: emission_line_fitting}

The profiles of key emission lines in each aperture were fit with Gaussian profiles and low-order polynomial continuum fits using \textsc{Python} scripts written using the \textsc{NumPy} \citep{Harris2020}, \textsc{Pandas} \citep{reback2020pandas}, \textsc{AstroPy} and \textsc{specutils} packages. In all cases, several Gaussians were required to properly fit the emission line profiles. Our intention was to use as few Gaussians as possible while still producing an adequate fit to the line profiles. Therefore, further Gaussian components were only added if they significantly reduced the mean-square residuals and $\chi^2$ of the fit. Furthermore, we used a sum-of-squares f-test \citep{Montgomery2012} to ensure that the relative decrease in residuals was statistically significant (at the $\alpha$=0.01 significance level), taking into account the degrees of freedom of the total model when adding additional Gaussians.

\begin{figure*}
	\includegraphics[width=\linewidth]{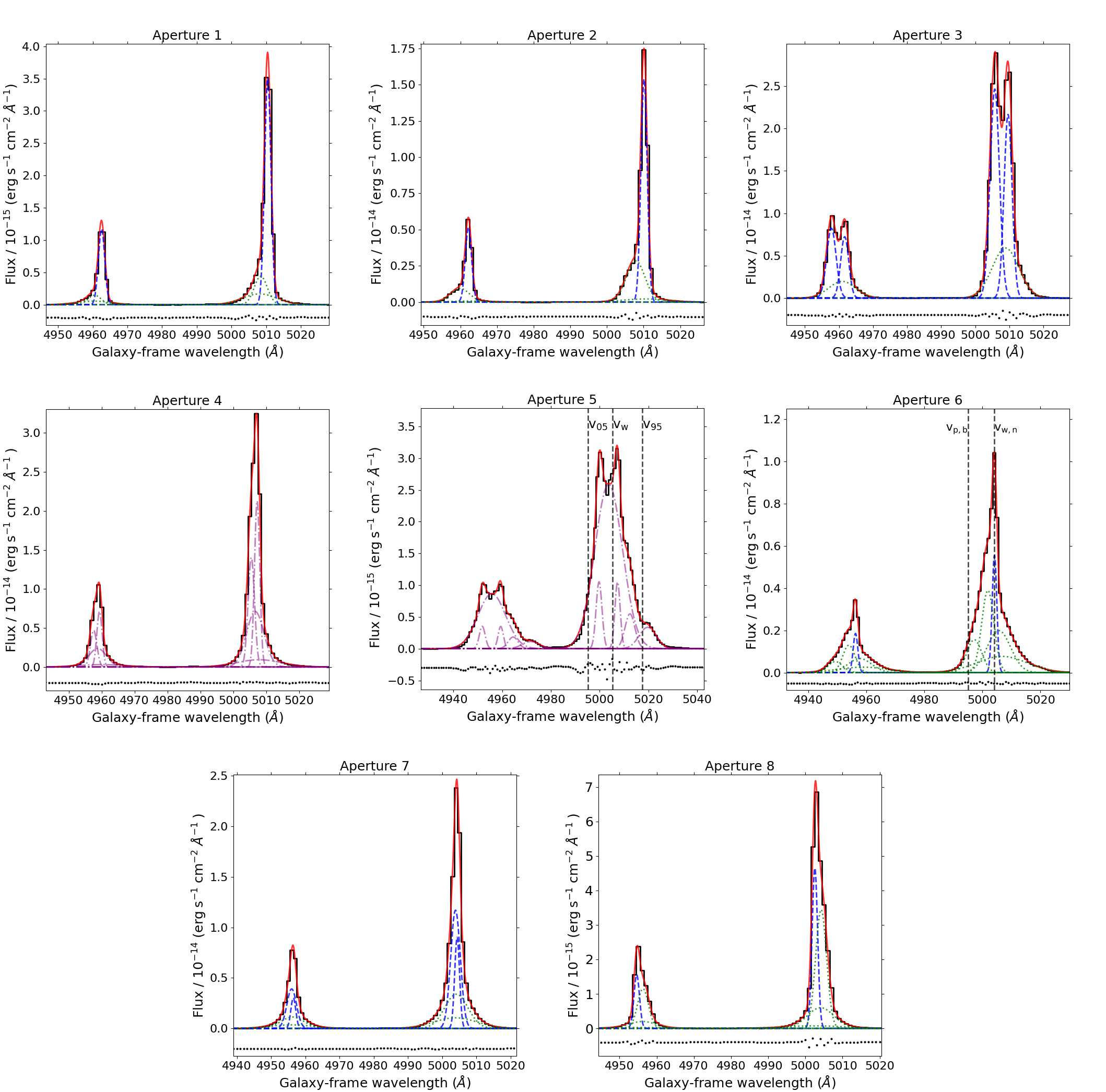}
	\caption{[OIII]$\lambda\lambda$4959,5007 rest-frame line profiles and models for the apertures covering the inner regions of IC\;5063. The observed line profiles are shown in black, the overall fits to the profiles are shown as red solid lines, the Gaussian components comprising the total narrow component are shown as blue dashed lines and the Gaussian components comprising the total broad component are shown as dotted green lines. Fitting residuals are shown as black dots below each line profile. In the cases of apertures 4 and 5, each Gaussian component is shown as a purple dash-dotted line, as there is no distinction is made between broad and narrow. The wavelengths corresponding to the percentile ($v_\mathrm{p}$; $v_\mathrm{05}$ and $v_\mathrm{95}$) and the flux weighted velocities ($v_\mathrm{w}$) for Aperture 5 are shown as dashed grey lines. For Aperture 6, dashed grey lines mark the wavelengths of the percentile velocity of the broad component ($v_\mathrm{p,b}$) and flux-weighted velocity of the narrow component ($v_\mathrm{w,n}$) --- these are used to calculate the outflow velocity ($v_\mathrm{out}$ = $v_\mathrm{p,b}$ - $v_\mathrm{w,n}$). Aperture 3 presents two clearly split narrow components, which are labelled `Narrow 1' (red-most) and `Narrow 2' (blue-most).}
	\label{fig: o3_models}
\end{figure*}

Emission-line profile models for the combined UVB+VIS data were produced by first fitting the [OIII]$\lambda\lambda$4959,5007 doublet. Gaussian components were simultaneously fitted to both lines in the doublet, with each pair of Gaussians having the same FWHM, separation (49.9\AA) and intensity ratio (1:2.99) defined by atomic physics \citep{Osterbrock2006}. The resulting multiple-Gaussian fits to the [OIII]$\lambda\lambda$4959,5007 doublet are hereafter referred to as the `[OIII] models' and are shown for each aperture in Figure \ref{fig: o3_models}. For each aperture, we identify narrow (FWHM\;\textless\;200\;kms$^{-1}$) features that have kinematics consistent with gravitational (rotational) motions in the host galaxy (as deduced from large-scale HI\;21\;cm and CO kinematics: \citealt{Morganti1998, Morganti2015}), and broad components (FWHM$_\mathrm{w}$\;\textgreater500\;kms$^{-1}$: see Section \ref{section: kinematics}) that we identify as an outflow. The broad components have complex, often multi-peaked profiles that required fitting with a combination of Gaussians of different widths. Note that we do not consider the \textit{individual} Gaussians used to fit the broad components to have a ready interpretation as physically distinct kinematic components. Rather, they are required to account for the \textit{total} flux in the broad part of the line profiles.


Whereas in most apertures the [OIII] profiles can be fitted by one total broad and one total narrow component (often modelled with more than one Gaussian), Aperture 3 is an exception to this because there are two clearly separated `narrow' components and a single broad component in the [OIII]$\lambda\lambda$4959,5007 line profiles. In this case, the two narrow components are considered separately, labelled `Narrow 1' (redmost) and `Narrow 2' (bluemost). In apertures 4 and 5, it was found that the profiles are dominated by broad components, and therefore the total line fluxes (narrow + broad) are considered instead.

\begin{figure*}
	\includegraphics[width=\linewidth]{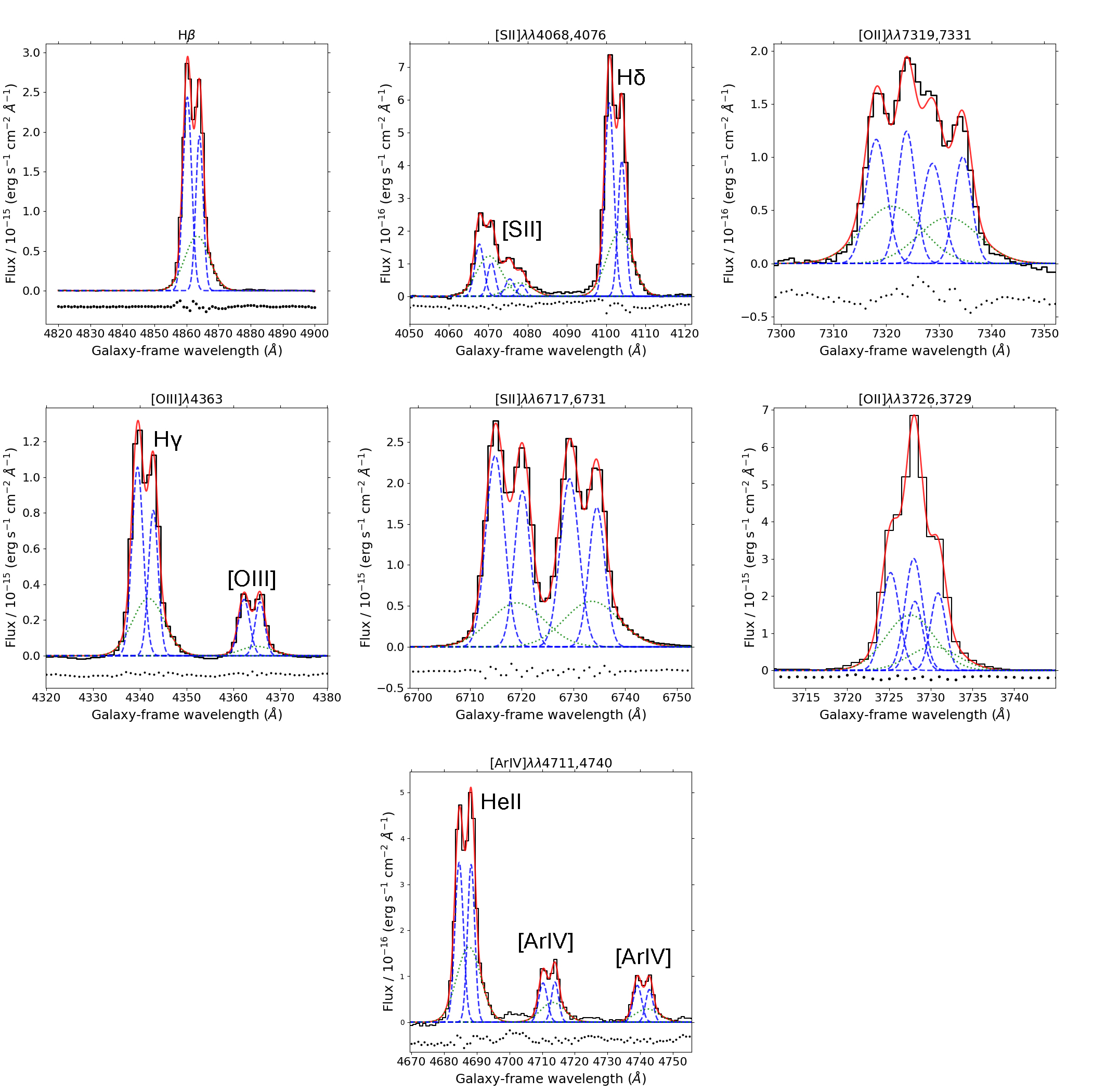}
	\caption{Fits to key diagnostic emission lines in Aperture 3 of our UVB+VIS spectra using the [OIII] model for Aperture 3 shown in Figure \ref{fig: o3_models}. Where lines from multiple atoms are present, we label them by atom and species. The diagnostic lines shown here are used throughout this work to derive key outflow parameters.}
	\label{fig: o3_models_all_lines}
\end{figure*}

The [OIII] models were used to constrain the fits to the other UVB+VIS emission lines used in our analysis, namely H$\mathrm{\beta}$, H$\mathrm{\gamma}$, [OIII]$\lambda$4363, [OII]$\lambda$3726,3729, [OII]$\lambda\lambda$7319,7331, [SII]$\lambda\lambda$4068,4076,  [SII]$\lambda\lambda$6717,6731, [ArIV]$\lambda\lambda$4711,4740 and HeII$\lambda$4686 --- these fits for Aperture 3 are shown in Figure \ref{fig: o3_models_all_lines}. We note that we did not produce fits for H$\mathrm{\alpha}$ and the [NII]$\lambda\lambda$6548,6583 doublet in apertures with complex, broad emission-line profiles, as the lines were significantly blended and we were thus unable to precisely measure the fluxes of the lines due to degeneracy issues. For the lines that we did fit, including the transauroral [SII]$\lambda\lambda$4068,4076 and [OII]$\lambda\lambda$7319,7331 doublets, we find that the [OIII] models fit them well in all apertures. The centroid wavelength of the brightest narrow component was allowed to vary by $\pm$1{\;\AA} in order to provide a better fit within the limit of the $\Delta\lambda$ = 1{\;\AA} resampling. For some of the weaker emission lines with much lower fluxes than the [OIII]$\lambda\lambda$4959,5007 doublet, the fainter Gaussian components were sometimes dropped, since the line profile features they accounted for were extremely faint relative to the continuum.

In contrast, we find that the [OIII] models do not describe emission lines in the NIR arm well, such as [FeII]$\lambda$12570, [FeII]$\lambda$16400, H$_2 \lambda$21218, Pa$\beta$, Br$\gamma$ and HeI$\lambda$10830. This is likely a result of the lack of stellar continua subtraction for the NIR data, which is due to the limited wavelength range of our \textsc{STARLIGHT} fits (3220--8000\;\AA), the lower cosmetic quality of the near-IR spectra, and the difficulty in exactly matching the NIR apertures to the UVB+VIS apertures. Therefore, we fit each line in the NIR arm independently. \\

\section{Properties of the outflowing gas}
\label{section: results}

\subsection{Analysis of the UVB+VIS apertures}
\label{section: uvb_vis_results}

\subsubsection{Velocity shifts and widths}
\label{section: kinematics}

We used the Doppler shifts and widths of broad and narrow components of the [OIII] models, relative to the lab wavelength of [OIII]$\lambda$5007, to determine outflow and quiescent (non-outflowing) gas kinematics. We first removed the instrumental broadening (as measured from the Galactic NaD lines) from the width of each component in quadrature.

When deriving outflow kinematics, we note that projection effects must be carefully considered. Therefore, two methods for estimating the velocity of the outflows were used, giving `minimal' and `maximal' velocities, following the methodology presented by \citet{Rose2018}. Our first method was to calculate flux-weighted mean velocity shifts and widths --- this was done for the total broad and narrow components by weighting the centroid velocities and Full Width Half Maxima (FWHM) of each constituent component by its total flux:

\begin{equation}
v_w = \frac{\Sigma_i(F_i \times \Delta V_i)}{\Sigma_i{F_i}}\text{, and}
\label{eq: flux_velocity}
\end{equation}

\begin{equation}
\mathrm{FWHM_w} = \frac{\Sigma_i(F_i{\times}\mathrm{FWHM_i})}{\Sigma_i{F_i}},
\label{eq: flux_width}
\end{equation}
\noindent
where $F_\mathrm{i}$, FWHM$_\mathrm{i}$ and $\Delta V_\mathrm{i}$ are the fluxes, FWHM and velocity shifts respectively. Velocity shifts were calculated using the wavelength shifts of a given centroid to the lab wavelength of [OIII]$\lambda$5007, and for the broad components are likely to under-estimate the true outflow velocities because projection effects have not been taken into account.

We also derived percentile velocities by using the far wings of the broad component profiles, as in \citet{Rose2018}. Here, it is assumed that all line broadening is due to different projections of the velocity vectors, along our line of sight, rather than intrinsic velocity dispersion in the gas in each volume element. In this case, the extended velocity wings represent gas moving directly along our line of sight, and potentially give a better estimate of the true outflow velocity. Therefore, we do not report velocity widths in this case. Percentile velocity shifts were determined for the broad components by taking the shift of the wavelength which contains either 5 or 95 per cent of the total flux of the line profile relative to the lab wavelength of [OIII]$\lambda$5007 (whichever was greater) --- we label these velocities as v$_\mathrm{p}$. As an example, in Figure \ref{fig: o3_models} we show both percentile ($v_\mathrm{p}$) velocities and the flux-weighted ($v_\mathrm{w}$) velocity for the [OIII] profile of Aperture 5.

\begin{table*}
\begin{tabular}{lllllll}
\hline
Component        & Distance (arcsec) & Distance (kpc) & v$_\mathrm{p}$ (km\,s$^{-1}$) & v$_\mathrm{w}$ (km\,s$^{-1}$) & FWHM$_\mathrm{w}$ (km\,s$^{-1}$)    & v$_\mathrm{out}$ (km\,s$^{-1}$) \\ \hline
AP1 Narrow       & $-$4.72             & $-$1.09          & ---    & 218$\pm$5    & 120$\pm$3  &                  \\ \hline
AP2 Narrow       & $-$3.44             & $-$0.79          & ---    & 197$\pm$3    & 110$\pm$2  &                  \\ \hline
AP3 Narrow 1     & $-$1.84             & $-$0.43          & ---    & 166$\pm$3    & 158$\pm$4  &                  \\
AP3 Narrow 2     & $-$1.84             & $-$0.43          & $-$202$\pm$2   & $-$68$\pm$1    & 187$\pm$3  & $-$233$\pm$4$^a$   \\
AP3 Broad        & $-$1..84            & $-$0.43          & 507$\pm$23   & 121$\pm$8    & 549$\pm$36 & 341$\pm$23   \\ \hline
AP4 Total        & 0                 & 0              & 264$\pm$41   & $-$7$\pm$2     & 296$\pm$16 &                  \\ \hline
AP5 Total        & 1.36              & 0.31           & $-$699$\pm$173 & $-$102$\pm$8   & 670$\pm$27 & $-$699$\pm$173$^b$ \\ \hline
AP6 Narrow       & 2.08              & 0.48           & ---   & $-$193$\pm$4   & 102$\pm$4  &                  \\
AP6 Broad        & 2.08              & 0.48           & $-$714$\pm$156 & $-$198$\pm$15  & 614$\pm$58 & $-$521$\pm$156 \\ \hline
AP7 Narrow       & 3.04              & 0.70           & ---  & $-$166$\pm$4   & 163$\pm$4  &                  \\
AP7 Broad        & 3.04              & 0.70           & $-$624$\pm$85  & $-$166$\pm$13  & 627$\pm$57 & $-$457$\pm$85   \\ \hline
AP8 Narrow       & 4.48              & 1.03           & ---  & $-$197$\pm$17 & 157$\pm$18 &                  \\ \hline
\end{tabular} \\
$^a$ calculated using the flux-weighted velocities (v$_\mathrm{w}$) for AP3N1 (quiescent gas) and AP3N2 (outflowing gas). \\
$^b$ the same as the percentile velocity (v$_\mathrm{p}$) since there is no corresponding narrow component.
\caption{The kinematics for each component including the percentile velocities (v$_\mathrm{p}$; the velocity that contains 5 or 95\;per\;cent of the flux of the emission line profile), flux-weighted velocities (v$_\mathrm{w}$; determining used flux-weighted velocity averages) and corresponding flux-weighted velocity widths (FWHM$_\mathrm{w}$). The outflow velocity (v$_\mathrm{out}$) is taken to be the difference between the percentile velocity of the outflowing broad components and the flux-weighted velocity of the quiescent narrow components at each position. The distances from the centre of Aperture 4 to the centre of the aperture for each component are also shown in both arcseconds and kpc. For Aperture 3, `Narrow 1' and 'Narrow 2' denote the two narrow components of the split line profile (see Figure \ref{fig: o3_models}), with `Narrow 2' being the blue-most narrow component}
\label{tab: kinematics}
\end{table*}

\begin{figure*}
	\includegraphics[width=\linewidth]{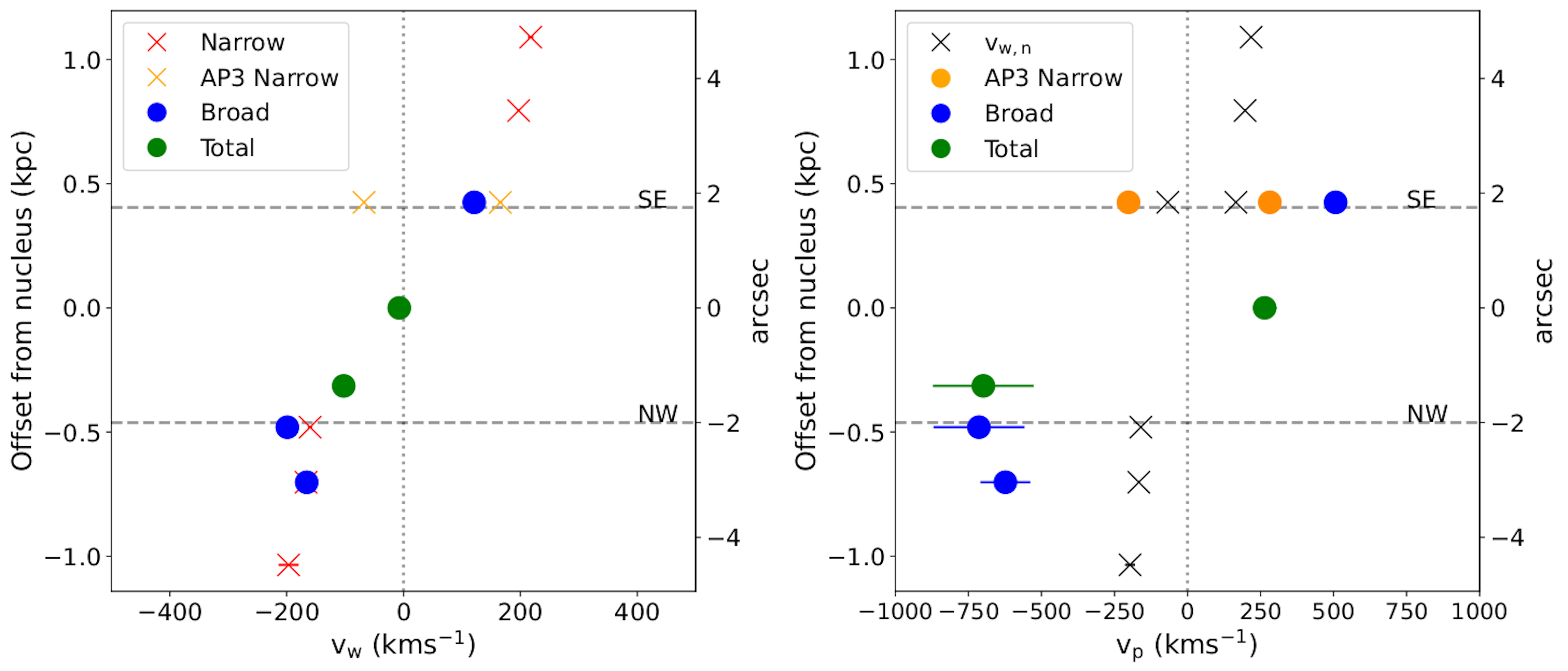}
	\caption{Velocity shifts for narrow (red crosses), broad (blue circles) and total (green circles) kinematic components in each aperture, shown spatially across IC\;5063. The split narrow components in Aperture 3 are shown in orange. The position of each component spatially corresponds to the centre of the aperture in which it was measured. Left: flux-weighted velocities of the narrow and broad components --- the narrow components are taken to represent the galaxy's quiescent rotating gas disk. Right: percentile velocity shifts (v$_\mathrm{p}$) for the broad components, taken to represent outflowing gas, compared with the weighted velocities of the narrow components ($v_\mathrm{w,n}$, shown for reference as black crosses). The dashed grey lines mark the centroids of the NW and SE radio lobes, measured from 24.8\;GHz continuum images presented by \citet{Morganti2007}. A velocity shift of zero relative to the galaxy's rest-frame is marked with a grey dotted line.}
	\label{fig: velocities}
\end{figure*}

\begin{figure}
	\includegraphics[width=\linewidth]{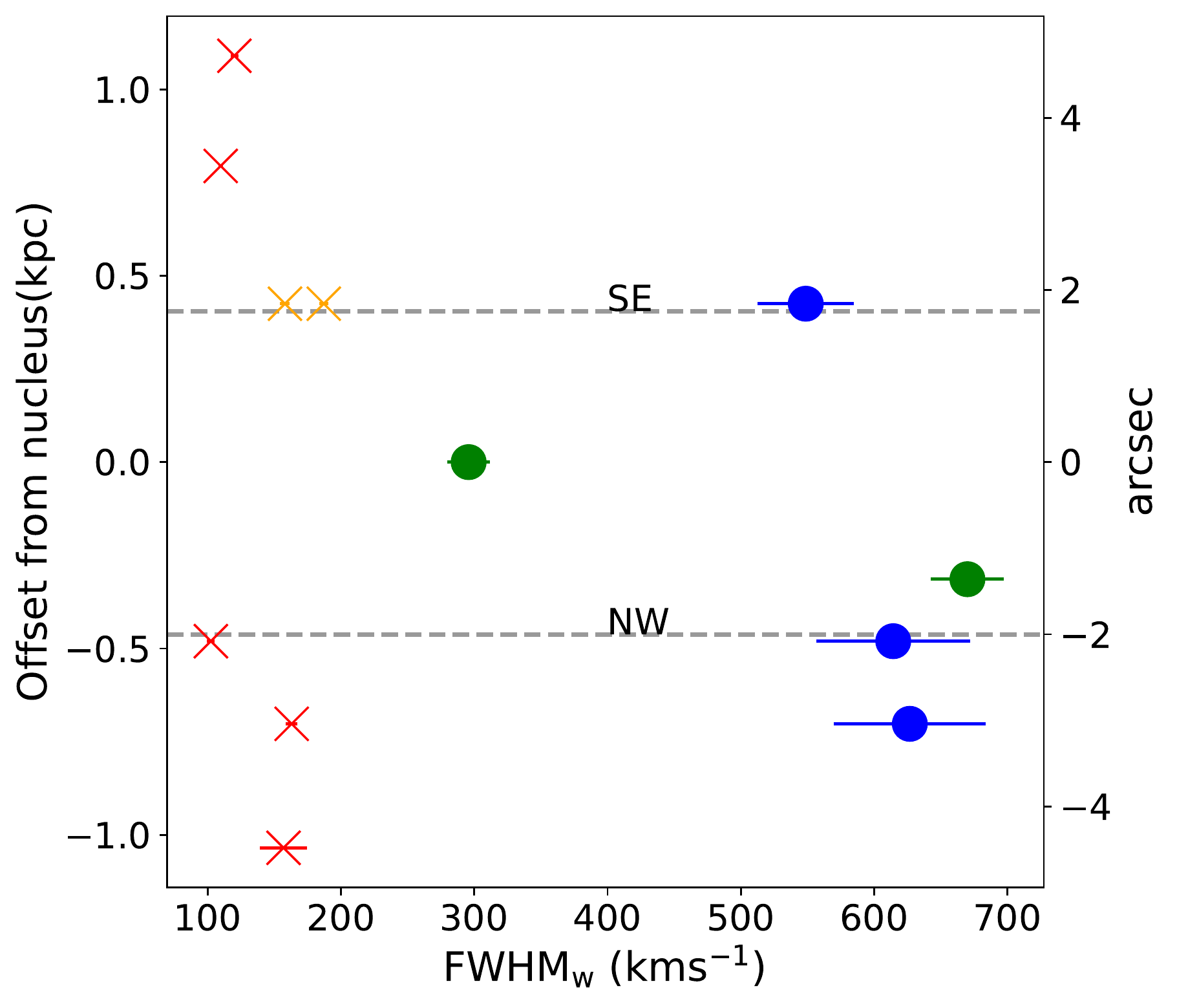}
	\caption{Flux-weighted velocity widths (FWHM$_\mathrm{w}$) as determined in the minimal velocity case, in which they are assumed to be entirely due to turbulence within the outflow. The colour and marker scheme and dashed lines are the same as in Figure \ref{fig: velocities}.}
	\label{fig: fwhm}
\end{figure}

Figure \ref{fig: velocities} shows a position-velocity (PV) diagram of the percentile (v$_\mathrm{p}$) and flux-weighted (v$_\mathrm{w}$) velocity shifts, as measured in each aperture, and Figure \ref{fig: fwhm} shows the flux-weighted velocity widths. In the flux-weighted case (v$_\mathrm{w}$; right panel of Figure \ref{fig: velocities}), the velocity shifts of the narrow components (with the exception of the blue-most narrow component in Aperture 3) appear to follow a rotation curve with amplitude $\pm$218\;km\,s$^{-1}$, corresponding to the rotational motion of the galaxy's disk (as seen in HI\;21cm and CO observations: \citealt{Morganti1998, Morganti2015, Oosterloo2017}). Deviations from this rotation curve can be seen in the percentile velocity shifts of the broad components and the total line profile in Aperture 5 (which is dominated by broad components), indicating that the broad components represent outflowing gas. Interestingly, this deviation is not seen in the flux-weighted case, indicating that the outflows are kinematically symmetric relative to the disk rotation. 

One of the narrow components observed in the line profile of Aperture 3 (`AP3 Narrow 2') shows significant deviation ($\sim$350\;km\,s$^{-1}$) from the rotation curve that the other narrow components follow, indicating a distinct kinematic component at this position which may represent outflowing or inflowing gas. 

From these kinematics, the broad components in apertures 3, 6 and 7, along with the `AP3 Narrow 2' component and the total line profile of Aperture 5, are taken to represent outflowing gas. We note that the broad components in apertures 1, 2 and 8 are most likely due spill-over from the broad profiles in the central apertures due to seeing effects, not locally outflowing gas, and as such are not considered to be due to intrinsic outflow broadening at these locations. Since the kinematics of the narrow components are consistent with quiescent gas in a rotating disk, we thus take the outflow velocity in each aperture to be the difference between the percentile velocity (v$_\mathrm{p}$) of the broad component (outflow) and the flux-weighted velocity (v$_\mathrm{w}$) of the narrow component (quiescent) --- in Figure \ref{fig: o3_models}, we show an example of this for Aperture 6. For apertures 4 and 5, where we only detect broad components, we take the outflow velocity to be the percentile velocity. The derived flux-weighted, percentile and outflow velocities (along with the velocity widths) are given in Table \ref{tab: kinematics}.

\subsubsection{Spatial distributions and kinematics of optical diagnostic lines}
\label{section: spatial_optical}

\begin{table}
\centering
\begin{tabular}{l|l|l}
\hline
Emission Line           & \parbox{1cm}{Centroid \\ (pixels)} & \parbox{1cm}{Centroid \\ (arcsec)} \\ \hline
[OIII]$\lambda$4959 & 12.3$\pm$0.1                                                                                                       & 1.97$\pm$0.02                                                                                                             \\
H$\beta$                & 12.5$\pm$0.1                                                                                                       & 2.00$\pm$0.02                                                                                                             \\
{[}OII{]}$\lambda$7319  & 11.9$\pm$0.1                                                                                                       & 1.90$\pm$0.02                                                                                                             \\
{[}SII{]}$\lambda$4069  & 12.0$\pm$0.1                                                                                                       & 1.92$\pm$0.02                                                                                                            
\end{tabular}
\caption{Centroids of Gaussian fits to spatial slices between \mbox{$-$600\;km\,s$^{-1}$\;\textless\;$v$\;\textless\;$-$400\;km\,s$^{-1}$} of the transauroral [OII] and [SII] emission lines, along with H$\mathrm{\beta}$ and [OIII]$\lambda$4959, relative to the spatial position of the continuum centre.}
\label{tab: spatial_vis}
\end{table}

\begin{figure}
	\includegraphics[width=\linewidth]{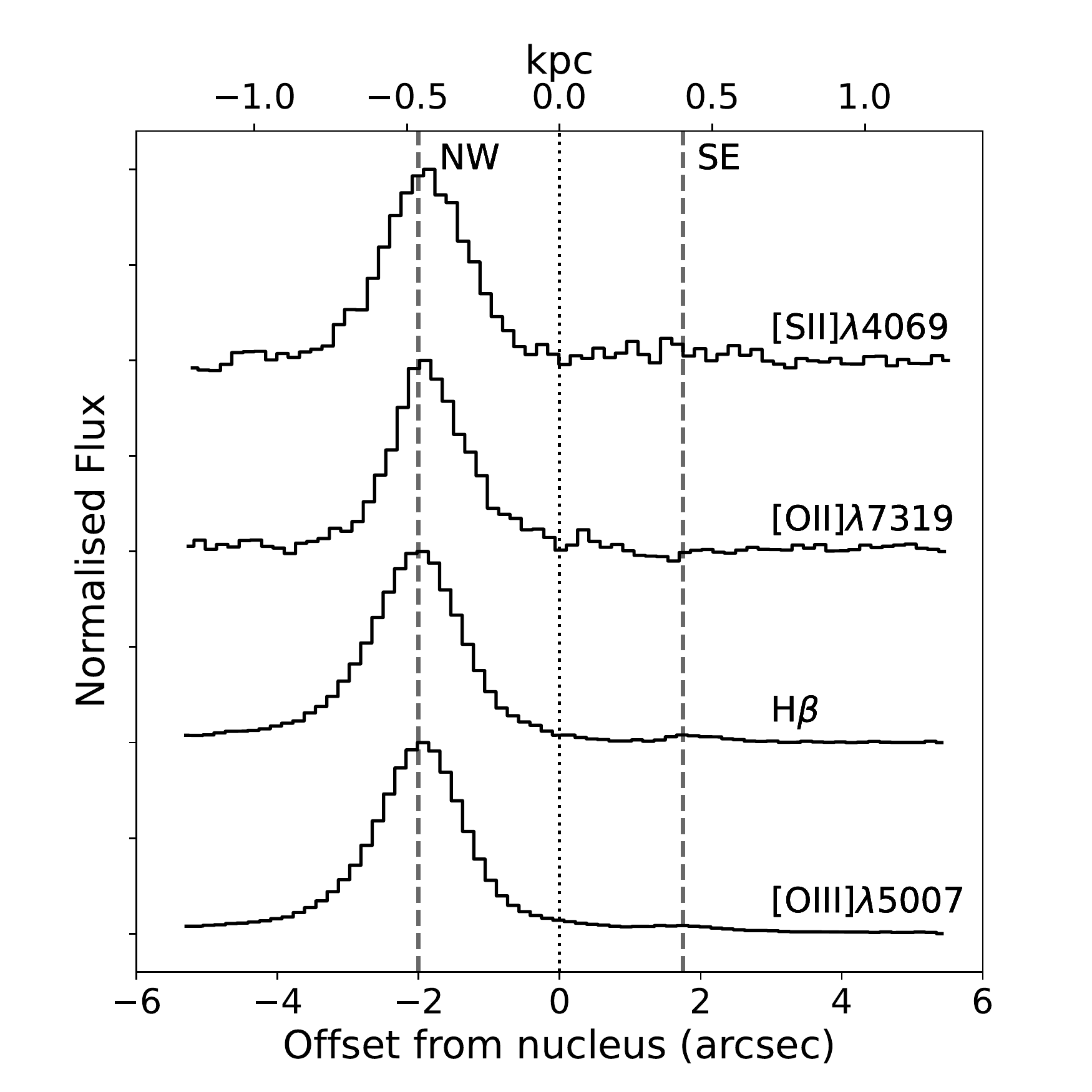}
	\caption{Spatial distributions of the blue wings (\mbox{$-$600\;km\,s$^{-1}$\;\textless\;$v$\;\textless\;$-$400\;km\,s$^{-1}$}) of the transauroral [SII] and [OII] lines, H$\mathrm{\beta}$ and [OIII]$\lambda$4959 emission lines. The spatial flux distribution of the transauroral lines can be seen to closely follow those of [OIII] and H$\mathrm{\beta}$. The dotted line shows the position of the nucleus and the dashed lines show the positions of the centroids of the NW and SE radio lobes, as measured from 24.8\;GHz imaging presented by \citet{Morganti2007}.}
	\label{fig: spatial_optical}
\end{figure}

Previous studies that make use of the transauroral lines to derive electron densities have been spatially unresolved (e.g. \citealt{Holt2011, Rose2018, Santoro2018, Spence2018, Santoro2020, Davies2020}), and it has not yet been verified that the transauroral lines are emitted by the same clouds as other key diagnostic lines such as [OIII]$\lambda$5007 and H$\mathrm{\beta}$. However, we found that fitting our [OIII] models to these lines works well, as does fitting them to the transauroral [SII] and [OII] doublets (Figure \ref{fig: o3_models_all_lines}). This shows that the lines have similar profiles and thus kinematics, potentially indicating that they are emitted by the same cloud systems. To verify this further, we extracted spatial slices of the blue wings of several key emission lines in the range \mbox{$-$600\;km\,s$^{-1}$\;\textless\;$v$ \;\textless\;$-$400\;km\,s$^{-1}$}, avoiding any narrow components of the line profiles. We also extracted slices of continuum, free from any emission lines, both blueward and redward of each emission line with slice widths of 20\;\AA. The two continuum slices were added and scaled to the same number of pixel columns extracted from the blue wing of each line --- the average continuum slice was then subtracted from that of the blue wing. The spatial position of the nucleus at the wavelength of each line was determined using Gaussian fits to the continuum spatial slices, and the peak positions of the extended emission line structures were then measured relative to this estimated nucleus position using Gaussian fits. We present the spatial flux distributions in Figure \ref{fig: spatial_optical} and give the peak centroid positions of the extended line emission from the Gaussian fits in Table \ref{tab: spatial_vis}.

We find that the spatial flux profiles for the transauroral [OII] and [SII] lines are similar to the [OIII]$\lambda$5007 and H$\mathrm{\beta}$ lines. Furthermore, values of the fitted Gaussian centroids for the peak each line are within 0.6\;pixels, a remarkable result given the known uncertainties from the Gaussian fitting process and the unknown systematic uncertainties from the continuum subtraction process, which will affect the lower-flux transauroral lines more than the brighter [OIII] and H$\mathrm{\beta}$ lines. Alongside the [OIII] models fitting the transauroral line profiles well, we take this as evidence that the transauroral lines are emitted in the same spatial locations as other key diagnostic lines. However, we cannot entirely rule out the lines being emitted by different clouds within the same spatial apertures.

\subsubsection{Transauroral line diagnostics}
\label{section: tr_lines}

The high spectral resolution and wide wavelength coverage of the Xshooter observations allow the use of a technique first presented by \citet{Holt2011}, which employs the traditional and transauroral [OII] and [SII] lines to simultaneously derive values for electron density and reddening. This is done by comparing the following line ratios to those expected from photoionisation modelling:
\begin{align*}
TR([OII]) = F(3726 + 3729) / F(7319 + 7331),
\end{align*}
\begin{align*}
TR([SII]) = F(4068 + 4076) / F(6717 + 6731).
\end{align*} 

A major advantage of this technique is that the total line fluxes of the doublets are used for the ratios, instead of the flux ratios of lines \textit{within} the doublets (as is the case for traditional techniques of electron density estimation) which are sensitive to blending effects at larger line widths. We use the [OIII] models as a basis for the fits to the TR([OII]) and TR([SII]) ratios. In order to avoid additional issues due to degeneracy and blending owing to the complex kinematics, we further constrain these fits for a given kinematic component in several ways. First, the widths of doublet lines were forced to be equal, and the wavelength separations between them were set equal to those from atomic physics. Second, we constrained the intensity ratios of doublet lines in the following ways in order to ensure that we were not modelling unphysical values.
\begin{itemize}
	\item We ensured that the ratios of the [OII]$\lambda\lambda$3726,3729, [SII]$\lambda\lambda$4049,4076 and [SII]$\lambda\lambda$6717,6731 doublets fell within the range of their theoretical values \citep{Osterbrock2006, Rose2018}. If a measured intensity ratio was above or below this permitted range, it was forced to be the maximum or minimum allowed theoretical value, respectively.
	\item The intensity ratio of [OII](7319/7330) was set to be 1.24, because this ratio does not vary with density \citep{Rose2018}. Note that this doublet is actually two separate doublets ([OII]$\lambda\lambda$7319,7320 and [OII]$\lambda\lambda$7330,7331); however, we model them as single lines as their separation ($\sim$1\;{\AA}) is much lower than the widths of our narrowest kinematic components.
\end{itemize}

\begin{figure}
	\includegraphics[width=\linewidth]{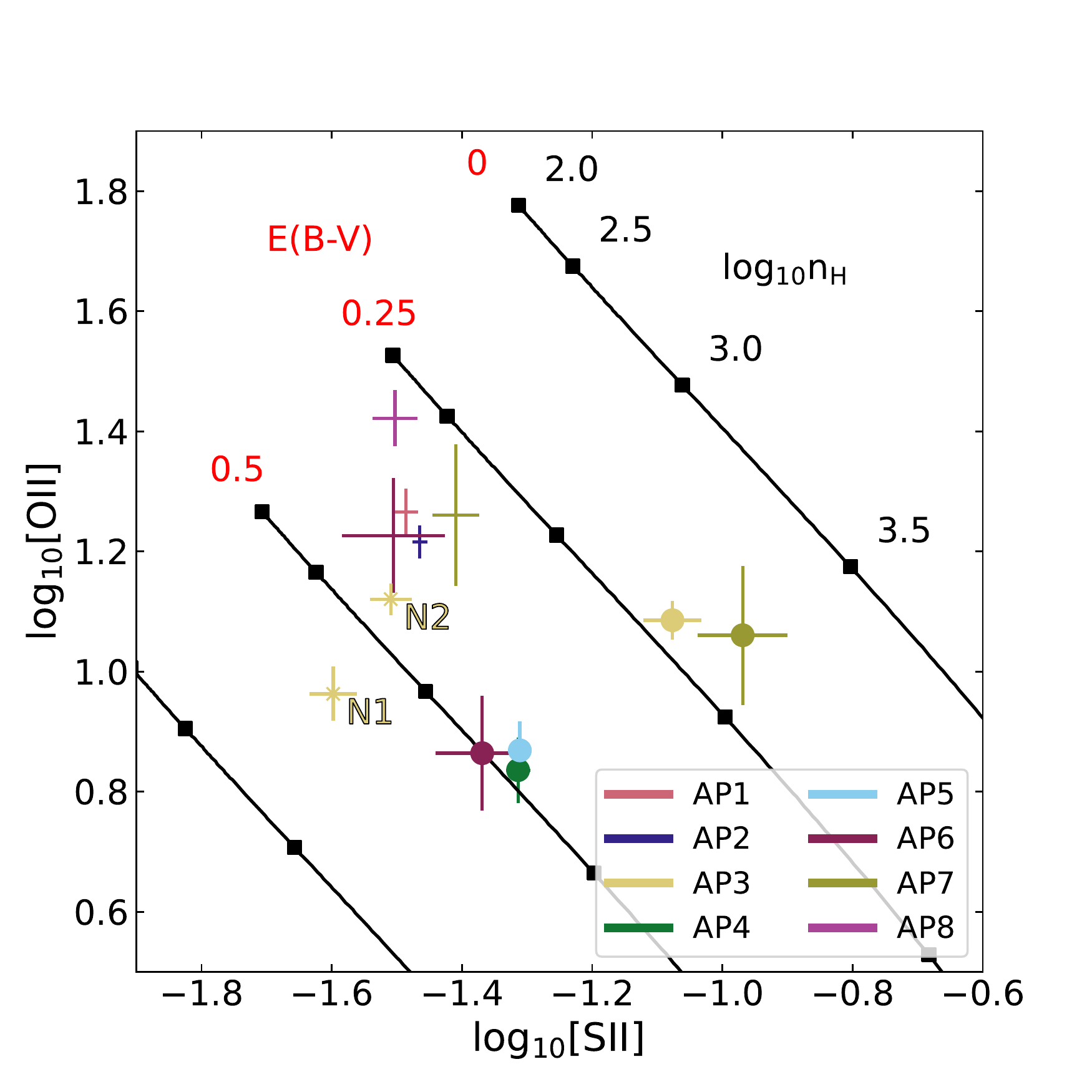}
	\caption{log$_{10}$ values of TR line ratios measured for the total narrow (crosses) and broad (circles) components in each aperture. Values for the total line profiles, as we measure in apertures 4 and 5, are also shown as circles. The two narrow components in Aperture 3 are marked as `N1' and `N2'. Overlaid is a grid of simulated TR line ratios created using the \textsc{CLOUDY} photoionisation code (version C17.02, \citealt{Ferland2017}) for different values of electron density and reddened using the CCM89 reddening law, indicated by the joined series of black squares.}
	\label{fig: tr_grid}
\end{figure}

The \textsc{CLOUDY} photoionisation code (version C17.02: \citealt{Ferland2017}) was used to create single-slab, plane-parallel, radiation-bounded and solar-composition models of photo-ionised gas with no dust depletion. We assumed that the central photoionising continuum followed a power law of shape $F_v$ $\propto$ $v^{-1.5}$. Ionisation parameters ($U$) were estimated for each aperture using estimated [OIII]/H$\mathrm{\beta}$ and [NII]/H$\mathrm{\alpha}$ line ratios with the relation presented by \citet{Baron2019b}, from which we found ionisation parameters in the range \mbox{$-$2.90\;\textless\;log$U$\;\textless\;$-$2.45}. Therefore, log$U=-$2.75 was chosen for use in the \textsc{CLOUDY} models. We varied the electron densities in 0.1\;dex steps in the range \mbox{2.0\;\textless\;log$_{10}n_\mathrm{e}$\;\textless\;5.0}, and used the CCM89 reddening law with $R_\mathrm{v}$=3.1 to redden these simulated TR ratios in order to create a grid with varying electron density and reddening. The resulting grid, along with the measured TR ratios in each aperture, is shown in Figure \ref{fig: tr_grid}. The densities and reddenings derived from this method are presented in Table \ref{tab: densities} --- note that these values were determined using a finer grid than is shown in Figure \ref{fig: tr_grid}.

Using the transauroral line ratios, we find that the broad components have significantly higher (\mbox{\textgreater3$\sigma$}) electron densities (\mbox{3.17\;\textless\;log$_{10}n_\mathrm{e}$\;\textless\;3.43}) than the narrow components (\mbox{2.12\;\textless\;log$_{10}n_\mathrm{e}$\;\textless\;2.59}) in all apertures where we measure both. In apertures 4 and 5 (where we used total line fluxes), we find electron densities similar to those of the broad components in other apertures. Furthermore, in apertures where we only measure a narrow component (1, 2 and 8), we find densities similar to narrow components in other apertures. The reddenings that we derive from the transauroral lines are moderately high (\mbox{0.17\;\textless\;E(B-V)$_{TR}$\;\textless\;0.51}), with no clear distinction between the values for narrow and broad components.

\begin{table*}
\def\arraystretch{1.5}
\begin{tabular}{lllllll}
\hline
Component        & log$_{10}n_\mathrm{e}$ (TR)     & log$_{10}n_\mathrm{e}$ ([OII]) & log$_{10}n_\mathrm{e}$ ([SII]) & log$_{10}n_\mathrm{e}$ ([ArIV]) & E(B-V)$_{TR}$          & E(B-V)$_{H\mathrm{\beta}}$ \\ \hline
AP1 Narrow       & 2.53$^{+0.06}_{-0.07}$ & \textless2.13                                     & 1.97$^{+0.08}_{-0.10}$                    & 3.70$^{-0.27}_{+0.24}$                     & 0.37$^{+0.02}_{-0.03}$ & 0.331$\pm$0.010   \\ \hline
AP2 Narrow       & 2.65$^{+0.03}_{-0.04}$ & 2.02$^{+0.11}_{-0.14}$                    & \textless2.15                                     & \textless3.71                                      & 0.38$^{+0.02}_{-0.02}$ & 0.311$\pm$0.007   \\ \hline
AP3 Narrow 1     & 2.59$^{+0.05}_{-0.05}$ & --- & --- & --- & --- & ---                   \\
AP3 Narrow 2     & 2.55$^{+0.03}_{-0.03}$ & --- & --- & --- & --- & ---                  \\
AP3 Broad        & 3.30$^{+0.05}_{-0.05}$ & --- & 2.84$^{+0.09}_{-0.09}$                    & --- & 0.22$^{+0.03}_{-0.03}$ & 0.215$\pm$0.017   \\ \hline
AP4 Total        & 3.25$^{+0.05}_{-0.05}$ & --- & ---  & 3.79$^{+0.18}_{-0.19}$                     & 0.48$^{+0.03}_{-0.03}$ & 0.507$\pm$0.026   \\ \hline
AP5 Total        & 3.23$^{+0.04}_{-0.05}$ & --- & --- & \textless5.97                                      & 0.46$^{+0.02}_{-0.02}$ & 0.342$\pm$0.019   \\ \hline
AP6 Narrow       & 2.55$^{+0.16}_{-0.20}$ & --- & --- & --- & 0.40$^{+0.05}_{-0.05}$ & 0.172$\pm$0.029   \\
AP6 Broad        & 3.17$^{+0.09}_{-0.09}$ & --- & --- & --- & 0.50$^{+0.06}_{-0.05}$ & 0.399$\pm$0.062   \\ \hline
AP7 Narrow       & 2.69$^{+0.15}_{-0.19}$ & --- & --- & --- & 0.33$^{+0.06}_{-0.07}$ & 0.265$\pm$0.016   \\
AP7 Broad        & 3.43$^{+0.09}_{-0.09}$ & --- & --- & --- & 0.17$^{+0.22}_{-0.07}$ & 0.270$\pm$0.037   \\ \hline
AP8 Narrow       & 2.12$^{+0.14}_{-0.12}$ & 2.07$^{+0.11}_{-0.12}$                    & 2.02$^{+0.14}_{-0.17}$                    & \textless4.03                                      & 0.30$^{+0.03}_{-0.03}$ & 0.270$\pm$0.029   \\ \hline
\end{tabular}
\caption{log$_{10}$ electron density values determined using the TR, traditional and [ArIV] line ratios for the gas in IC\;5063. E(B-V) values are also shown, determined using the transauroral technique (TR) and the H$\mathrm{\gamma}$/H$\mathrm{\beta}$ ratios with Case B recombination theory and the CCM89 reddening law. 3$\sigma$ upper limits are shown where the measured line ratio was not 3$\sigma$ from the lower or upper limit of the traditional line ratios. We are not able to determine values for electron density using the traditional and [ArIV] line ratios in every aperture due to line blending and low signal relative to the continuum --- these cases are shown here with a dash. Distances for each aperture are given in the same convention as Table \ref{tab: kinematics}.}
\label{tab: densities}
\end{table*}

Note that the position of the TR line-ratio grid generated from photoionisation modelling potentially depends on the ionisation parameter, ionising continuum spectral index and metallicity used in the model. \citet{Santoro2020} show that varying the ionisation parameter in the range \mbox{$-$3.8\;\textless\;log$U$\;\textless\;$-$2} and gas metallicities in the range \mbox{0.5\;Z$_\odot$\;\textless\;$Z$\;\textless\;2\;$Z_\odot$} with different SED shapes can change the derived electron density values by 0.1--0.7\;dex and reddening values by 0.1--0.2 mag. However, for lower density clouds (\mbox{\textless\;10$^4$\;cm$^{-3}$}) such as those we measure here, the effect of varying these parameters on the electron density is reduced to 0.1--0.3\;dex \footnote{We do not include this variation in the uncertainties presented in Table \ref{tab: densities} and Figure \ref{fig: tr_balmer_reddening}}. This corresponds to a maximum factor of two in derived electron density, which is much less than the potential order-of-magnitude inaccuracy incurred by using lower-critical density diagnostics for higher density clouds, as well as uncertainties associated with line blending within the traditional doublets.

\subsubsection{Traditional line ratio electron densities}

In addition to the electron densities determined using the transauroral technique, the traditional [OII]3726/3729 and [SII]6716/6731 emission line ratios and the higher critical density, higher ionisation [ArIV]4711/4740 ratio were used to provide independent estimates of electron density, allowing for the different techniques to be compared. The fluxes of the lines in each doublet were constrained using the [OIII] models, as shown in Figure \ref{fig: o3_models_all_lines}. We were unable to fit certain doublets in some apertures due to line blending, and in the case of the [ArIV] doublet, the low fluxes of the lines relative to the underlying stellar continua. We therefore do not report derived densities for these cases.

In order to ensure that the electron densities derived from the traditional ratios were accurate, it was required that the measured line ratios were 3$\sigma$ away from the theoretical lower and upper ratio limits (\citealt{Osterbrock2006}: \mbox{0.41\;\textless\;[OII]$\frac{3729}{3726}$\;\textless\;1.50}, \mbox{0.30\;\textless\;[SII]$\frac{6717}{6731}$\;\textless\;1.45}; \citealt{Wang2004}: \mbox{0.117\;\textless\;[ArIV]$\frac{4711}{4740}$\textless\;1.50}). We did not calculate densities using line ratios that did not meet this criterion --- instead, we calculated upper limits by taking the ratio value that was 3$\sigma$ from the measured value. 

Where it was possible to make measurements of these ratios, the \textsc{fivel} script \citep{Shaw1995} was used to calculate values for electron density. Doing so required an estimate of the electron temperature, which was determined for each component in each aperture using the [OIII](4959 + 5007)/4363 ratio (Section \ref{section: electron_temperatures}). The electron densities we determined using all techniques discussed are shown in Table \ref{tab: densities}. 

Unfortunately, we are only able to determine densities using all techniques for the narrow components in apertures 1, 2 and 8. In Aperture 8, all density estimates are consistent within the uncertainties, however this is not the case for apertures 1 and 2, in which we find clear evidence for the TR lines producing densities that are higher than derived from traditional [SII] and [OII] ratios, but lower than those from the [ArIV] ratio.

Considering the broad components, we are only able to use a traditional ratio ([SII]) to adequately measure an electron density in Aperture 3, where the value is significantly lower (\mbox{\textgreater3$\sigma$}) than the density found using the transauroral lines. The difference in the densities determined using the traditional ratio and the TR lines is approximately a factor of three in this case, and we therefore find that Aperture 3 gives the best evidence for the TR lines producing higher densities than a traditional ratio for outflowing gas. In Aperture 4, we find that the [ArIV] ratio gives a higher density than the TR lines, however this difference is not significant to $3\sigma$. Likewise, we find that the [ArIV] gives a higher density than the TR lines in Aperture 5, however the [ArIV] density here is an upper limit.

\subsubsection{Recombination line reddenings}

Determining precise values for reddening at different spatial positions is crucial for properly correcting the luminosities of emission lines which are used to determine outflow properties. In order to compare reddening values found using the TR line ratio technique to a more commonly used method, values for E(B-V) using the Balmer decrement were also determined. This was done by comparing the observed line flux ratio of the H$\mathrm{\gamma}$ and H$\mathrm{\beta}$ recombination lines to the intrinsic ratio expected from Case B recombination theory \citep{Osterbrock2006} and the CCM89 reddening law. We do not use the H$\mathrm{\alpha}$/H$\mathrm{\beta}$ ratios to derive reddening values due to degeneracy issues related to blending between H$\mathrm{\alpha}$ and the [NII]$\lambda\lambda$6548,6583 doublet.

The H$\mathrm{\gamma}$/H$\mathrm{\beta}$ ratio can be sensitive to small errors in the subtraction of the underlying stellar continuum: \citet{Rose2018} show that a 10 per cent decrease in this ratio corresponds to a 60 per cent increase in the derived E(B-V) value. Therefore, while we present reddening values using both the TR and Balmer line techniques for comparison, only values derived from the TR technique were used to deredden the UVB+VIS spectra for further analysis.

\begin{figure}
	\includegraphics[width=\linewidth]{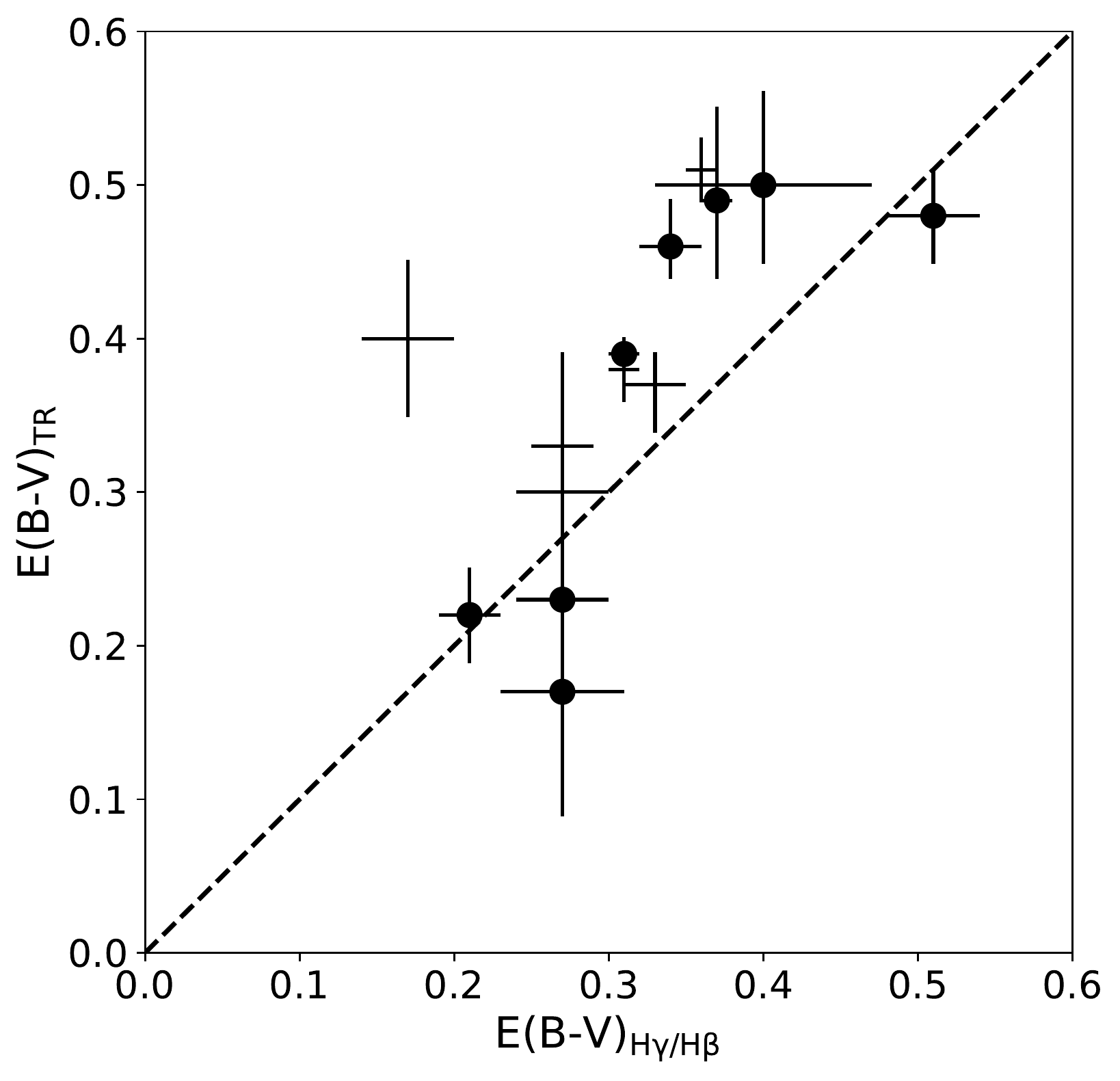}
	\caption{Reddening values derived from the Balmer decrement against values measured using the TR ratios. The black points represent the values measured in the various apertures and components, with broad components indicated with circles and the narrow components with crosses, while the black dashed line shows the one-to-one ratio. }
	\label{fig: tr_balmer_reddening}
\end{figure}

The results from the two methods of reddenings estimation are shown in Table \ref{tab: densities}, and we plot them against each other in \mbox{Figure \ref{fig: tr_balmer_reddening}}. It is found that transauroral method gives slightly higher values than using the H$\mathrm{\gamma}$/H$\mathrm{\beta}$ ratio, however the two methods are consistent within 3$\sigma$ for all apertures except 2 and 5 (Table \ref{tab: densities}). Both methods give a range of \mbox{0.17\;\textless\;E(B-V)\;\textless\;0.51}.

\subsubsection{Electron temperatures and gas ionisation mechanisms}
\label{section: electron_temperatures}

As an initial probe of the ionisation mechanism of the outflowing and quiescent warm ionised gas, we used emission lines resulting from forbidden transitions of the O$^{+2}$ ion to determine electron temperatures, since shock-ionised gas is expected to have a higher electron temperature than gas photoionised by an AGN \citep{Fosbury1978, VillarMartin1999}. Note that we do not use the standard BPT diagrams to investigate the ionisation mechanisms of the gas because the shock and photoionisation models overlap considerably in these diagrams (see \citealt{Zaurin2013} and \citealt{Santoro2018}), and H$\mathrm{\alpha}$ and the [NII]$\lambda\lambda$6548,6583 doublet are blended significantly in our apertures of interest. Electron temperatures were determined by measuring the extinction-corrected intensity ratio [OIII](5007+4959)/4363 (using the [OIII] models to fit the lines), and using the measured ratios and TR electron densities with the Lick Observatory \textsc{fivel} script. Values of electron temperature measured in this way for each kinematic component are presented in Table \ref{tab: temps}.

We find electron temperatures in the range \mbox{11500\;K\;\textless\;$T_e$\;\textless\;14000\;K}, with no clear distinction (\mbox{\textgreater3$\sigma$}) between broad and narrow components except in Aperture 3, in which the two narrow components present higher electron temperatures than the broad component.

\begin{table}
\def\arraystretch{1.5}
\centering
\begin{tabular}{llll}
\hline
Component & \parbox{1cm}{Distance\\ (arcsec)} & \parbox{1cm}{Distance\\ (kpc)} & \parbox{1.4cm}{Temperature\\ (K)} \\ \hline
AP1 Narrow   & $-$4.72                                                       & $-$1.21                                                    & 11934$^{+169}_{-207}$                                     \\ \hline
AP2 Narrow   & $-$3.44                                                       & $-$0.88                                                    & 13776$^{+145}_{-96}$                                      \\ \hline
AP3 Narrow 1 & $-$1.84                                                       & $-$0.47                                                    & 14117$^{+350}_{-293}$                                     \\
AP3 Narrow 2 & $-$1.84                                                       & $-$0.47                                                    & 13117$^{+185}_{-228}$                                     \\
AP3 Broad    & $-$1.84                                                       & $-$0.47                                                    & 11564$^{+204}_{-160}$                                     \\ \hline
AP4 Total    & 0                                                           & 0                                                        & 12578$^{+44}_{-44}$                                       \\ \hline
AP5 Total    & 1.24                                                        & 0.32                                                     & 11810$^{+208}_{-164}$                                     \\ \hline
AP6 Narrow   & 2.08                                                        & 0.53                                                     & 13921$^{+802}_{-666}$                                     \\
AP6 Broad    & 2.08                                                        & 0.53                                                     & 12980$^{+941}_{-664}$                                     \\ \hline
AP7 Narrow   & 3.04                                                        & 0.78                                                     & 11646$^{+457}_{-401}$                                     \\
AP7 Broad    & 3.04                                                        & 0.78                                                     & 13680$^{+586}_{-517}$                                     \\ \hline
AP8 Narrow   & 4.48                                                        & 1.15                                                     & 13537$^{+287}_{-282}$                                     \\ \hline
\end{tabular}
\caption{Electron temperatures of the gas in each component for each aperture derived using the [OIII](5007+4959)/4363 line ratio. The distances for each aperture are the same as given in Table \ref{tab: kinematics}.}
\label{tab: temps}
\end{table}

\begin{figure}
	\includegraphics[width=\linewidth]{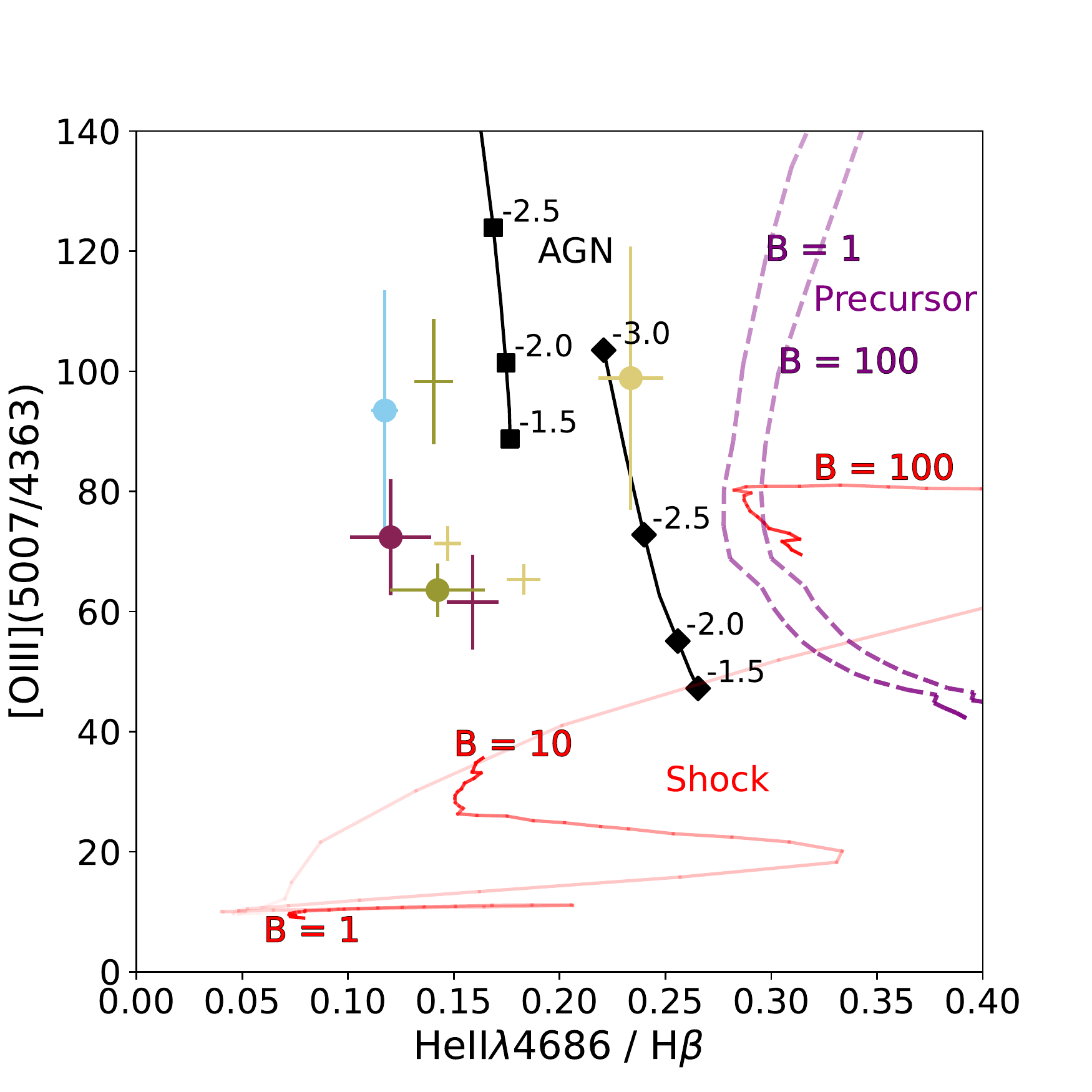}
	\caption{Measured HeII$\lambda$4686/H$\mathrm{\beta}$ and [OIII]5007/ [OIII]$\lambda$4363 line ratios for the outflowing gas in our apertures --- the colour and marker scheme for each aperture is the same as in Figure \ref{fig: tr_grid}. Also plotted are AGN photoionisation models (black markers and lines) for different ionisation parameters (labelled) and spectral indices (squares: $\mathrm{\alpha}$=1.5; diamonds: $\mathrm{\alpha}$=1.0) of solar composition (1\;Z$_\odot$) gas of density n$_\mathrm{e}$=10$^3$\;cm$^{-3}$. Line ratios produced by shock-ionisation models taken from the \textsc{Mappings III} library presented by \citet{Allen2008} for gas with a pre-shock electron density of 100\;cm$^{-3}$, magnetic fields of 1, 10 and 100\;$\mu G$ and shock velocities ranging from 100--1000\;km\,s$^{-1}$ are shown as purple dashed lines (pre-shock / precursor gas) and red solid (post-shock gas) lines. Lower shock velocities are shown with fainter lines and higher shock velocities are shown with darker lines.}
	\label{fig: heii_hbeta}
\end{figure}

To further compare with AGN photoionisation and shock models, we plotted [OIII](5007/4363) against HeII$\lambda$4686/H$\mathrm{\beta}$ in the diagnostic diagram shown in Figure \ref{fig: heii_hbeta}. The advantage of this diagram is that, not only are the shock and AGN photoionisation models more clearly separated than in BPT diagrams, but it allows us to determine the extent to which the presence of matter-bounded clouds might affect the results \citep{VillarMartin1999}. The AGN photoionisation models were those generated by \textsc{Cloudy}, as described in Section \ref{section: tr_lines}, for a radiation-bounded, solar-composition (1\;Z$_\odot$), dust-free gas with an electron density of $n_\mathrm{e}=10^3$\;cm$^{-3}$ (based on the densities we find using the TR ratios) and a central ionising continuum that follows a power law of shape $F_\mathrm{v}$ $\propto$ $v^{-1.5}$. We note that the [OIII](5007/4363) and HeII4686/H$\mathrm{\beta}$ ratios depend on these parameters, and we discuss the effects of varying them in \mbox{Appendix \ref{section: photoionisation_modelling}}. The solar-composition shock models were taken from the library presented by \citet{Allen2008}, which was created using the \textsc{Mappings III} code. In \mbox{Figure \ref{fig: heii_hbeta}}, we present the measured line ratios for each aperture in which we detect outflows, along with the modelled \textsc{Cloudy} ratios for different ionisation parameters (in the range \mbox{$-$3.0\;\textless\;log$U$\;\textless\;$-$1.5}), and the pre-shock (precursor) and post-shock (pure shock) models from \textsc{Mappings III} for a gas density of 100\;cm$^{-3}$, magnetic fields of 1, 10 and 100\;$\mathrm{{\mu}G}$ and different shock velocities (ranging from 100--1000\;km\,s$^{-1}$).

Comparing the measured line ratios to those from the AGN photoionisation and shock models, we find evidence for AGN photoionisation being dominant. We note that, while pure AGN photoionisation requires a sub-solar metallicity ($\sim$0.5\;Z$_\odot$) and higher ionisation parameters than we measure to explain some of the measured ratios, a different assumed spectral index and higher electron densities would also help improve consistency with the AGN photoionisation models (see Appendix \ref{section: photoionisation_modelling}). Therefore, a plausible combination of these parameters exists that produces line ratios consistent with our measurements. While we cannot rule out some contribution from shocks, we note that, unlike in \citet{VillarMartin1999}, we do not find that the [OIII](5007/4363) and HeII$\lambda$4686/H$\mathrm{\beta}$ ratios discriminate between broad and narrow components as would be expected if the gas was shock ionised. Therefore, we take this as evidence that both the quiescent and outflowing gas in the warm ionised phase are predominantly AGN-photoionised.

We find moderate ($\sim$0.10--0.25) HeII$\lambda$4686/H$\mathrm{\beta}$ ratios and low-to-moderate [OIII] temperatures (\mbox{60\;\textless\;[OIII](5007/4363)\;\textless\;120}; T$\sim$11500--14000\;K) for all components, both narrow and broad. These ratios rule out a major contribution from matter-bounded components \citep{Binette1996}.

\subsubsection{Mass outflow rates and kinetic powers}
\label{section: mout_fkin}

In order to facilitate comparisons between our observations, models of galaxy evolution and previous observations of the different gas phases in IC\;5063, we used the parameters for the outflows at different spatial positions to derive mass outflow rates, kinetic powers and coupling efficiencies.

The luminosities of the H$\mathrm{\beta}$ line for the broad component in each aperture were first calculated using
\begin{equation}
L(\mathrm{H\beta}) = F(\mathrm{H\beta})\times 4\pi D^2_L,
\end{equation}
where L(H$\mathrm{\beta}$) is the H$\mathrm{\beta}$ luminosity, F(H$\mathrm{\beta}$) is the TR-reddening corrected-flux of the H$\mathrm{\beta}$ line (measured from Gaussian fits using the [OIII] models) and $D_L$ is the luminosity distance of IC\;5063. These luminosities were then used to determine the mass of an outflow in a given aperture using

\begin{equation}
    M_\mathrm{out} = \frac{L(H\mathrm{\beta})m_\mathrm{p}}{\alpha^\mathrm{eff}_{\mathrm{H\beta}}hv_{\mathrm{H\beta}}n_\mathrm{e}},
    \label{eq: mout}
\end{equation}
where $M_\mathrm{out}$ is the total mass of the outflowing gas, $m_\mathrm{p}$ is the proton mass, $\alpha^{eff}_{H\beta}$ is the Case B recombination coefficient for H$\mathrm{\beta}$ (taken to be 3.03$\times10^{-14}$\;cm$^3$s$^{-1}$ for a gas with $n_\mathrm{e}=10^4$\;cm$^{-3}$ and $T_\mathrm{e}=10^4$\;K; \citealt{Osterbrock2006}) and $v_\mathrm{H\beta}$ is the frequency of the H$\mathrm{\beta}$ line. \\

The mass outflow rate was then calculated using the mass and the aperture crossing time:
\begin{equation}
    \dot{M}_\mathrm{out} = \frac{M_\mathrm{out}v_\mathrm{out}}{\Delta R},
    \label{eq: mout_rate}
\end{equation}
where $v_\mathrm{out}$ is the outflow velocity and $\Delta R$ is the width of the aperture. As discussed in Section \ref{section: kinematics}, we take the outflow velocity v$_\mathrm{out}$ to be the difference between the percentile ($v_\mathrm{p}$) velocity of the broad component and the flux-weighted ($v_\mathrm{w}$) velocity of the narrow component in a given aperture; we give outflow velocities for the relevant apertures in Table \ref{tab: kinematics}.

\begin{figure}
	\includegraphics[width=\linewidth]{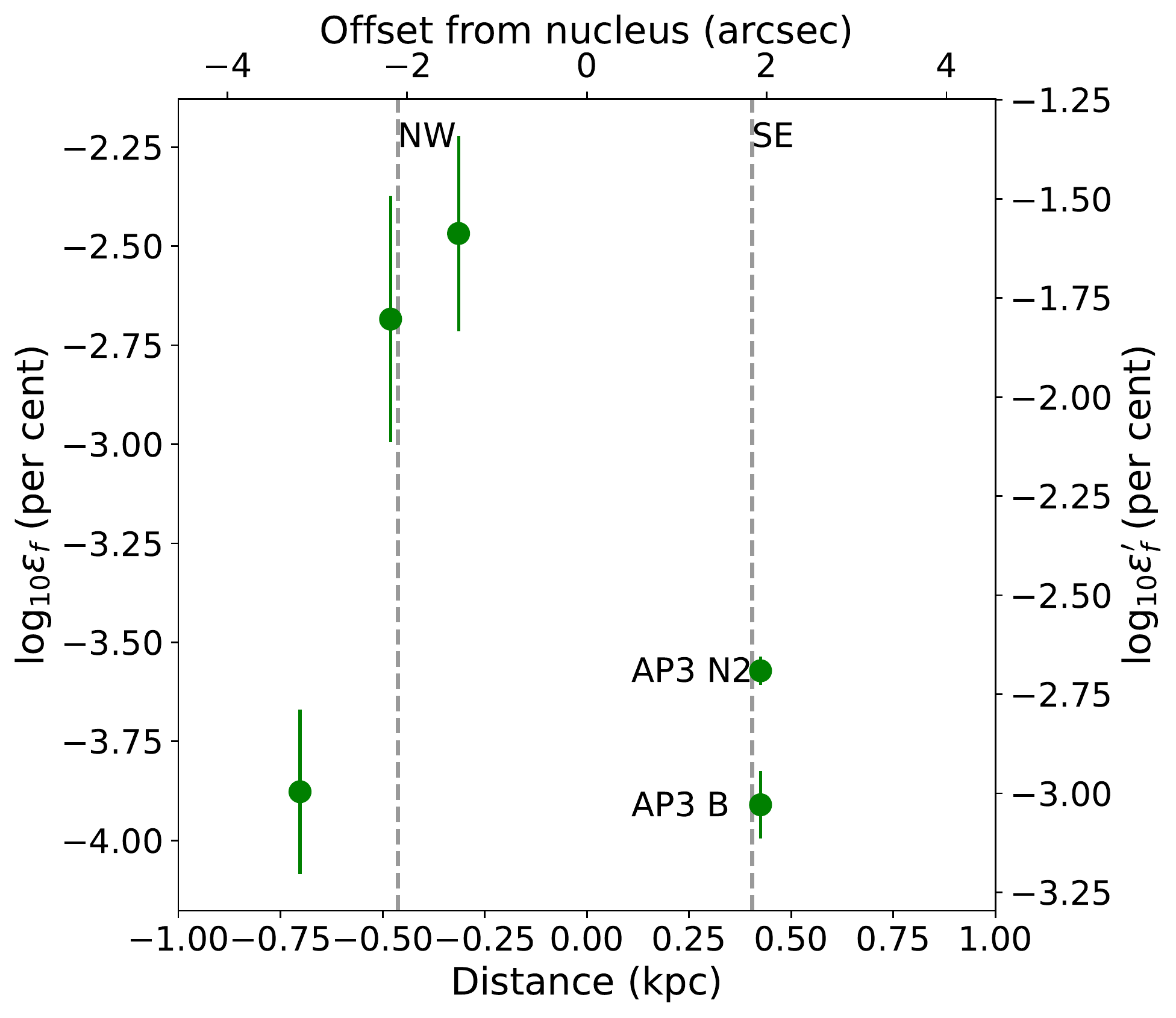}
	\caption{Percentage outflow coupling efficiencies in each aperture, calculated for two cases: using the nuclear bolometric luminosity of IC\;5063 given by \citet{Nicastro2003} and \citet{Morganti2007} ($\epsilon_\mathrm{f}$, left y-axis), and in the case that a jet of power P$_\mathrm{jet}$ = 10$^{37}$\;W \citep{Mukherjee2018} is the main driving mechanism ($\epsilon_\mathrm{f}^\prime$, right y-axis). The vertical dashed and dotted lines follow the same convention as Figure \ref{fig: velocities}, and the two kinematically blueshifted components found in Aperture 3 (AP3 Broad and AP3 Narrow 2) are labelled. The coupling efficiencies we find here in the bolometric luminosity case are below the $\epsilon_\mathrm{f}\sim 0.5$--$10$ per cent values used in AGN-feedback models of galaxy evolution (e.g. \citealt{DiMatteo2005, Hopkins2010}).}
	\label{fig: fkin}
\end{figure}

The outflow kinetic power was calculated using
\begin{equation}
    \dot{E}_\mathrm{kin} = \frac{M_\mathrm{out}v^2_\mathrm{out}}{2},
    \label{eq: ekin}
\end{equation}
The resulting outflow kinetic powers were then compared to AGN bolometric luminosity to give a coupling factor $\epsilon_\mathrm{f}$:
\begin{equation}
    \epsilon_\mathrm{f} = \frac{\dot{E}_\mathrm{kin}}{L_\mathrm{Bol}},
    \label{eq: fkin}
\end{equation}
\noindent
where we take $L_\mathrm{Bol}$ in Equation \ref{eq: fkin} to be the nuclear bolometric luminosity of IC\;5063: 7.6$\times 10^{37}$\;W \citep{Nicastro2003, Morganti2007}. The resulting coupling efficiencies are presented in Figure $\ref{fig: fkin}$ and Table $\ref{tab: mout_fkin}$. 

We find low mass outflow rates (\mbox{\textless\;0.4\;M$_\odot$\,yr$^{-1}$}) and coupling efficiencies (\mbox{$\epsilon_\mathrm{f}$\;$\ll$\;0.5 per cent}) in all apertures, much lower than those previously derived in some observational studies of samples of AGN (e.g. $\dot{M}_\mathrm{out}\sim$\;10--10,000\;M$_\odot$yr$^{-1}$: \citealt{Fiore2017}) and those required in galaxy evolution models ($\epsilon_\mathrm{f}\sim$0.5--10\;per\;cent: e.g. \citealt{DiMatteo2005, Hopkins2010}). Furthermore, we calculate another set of coupling efficiencies, which we label $\epsilon_\mathrm{f}^{\prime}$, by finding the ratio of outflow kinetic power to the minimal jet power used in modelling of the jet-ISM interactions in IC\;5063 by \citet{Mukherjee2018}: a value of P$_\mathrm{jet}$ = 10$^{37}$\;W. We find low values of $\epsilon_\mathrm{f}^{\prime}$ (\mbox{\textless\;0.001 per cent}), indicating that the kinetic power of the outflows accounts for an extremely small fraction of the total jet power, and demonstrating the feasibility of jet-ISM interactions being the outflow acceleration mechanism.

\begin{table*}
\begin{tabular}{lllll}
\hline
Component    & $\dot{M}_\mathrm{out}$ (M$_\odot$yr$^{-1}$) & $\dot{E}_\mathrm{kin}$ (W) & $\epsilon_\mathrm{f}$ (per cent) & $\epsilon_\mathrm{f}^\prime$ (per cent) \\ \hline
AP3 Narrow 2 & 1.2$\pm$0.1 $\times 10^{-1}$              & 2.0$\pm$0.2 $\times10^{32}$  & 2.7$\pm$0.2 $\times 10^{-4}$   & 2.0$\pm$0.2 $\times 10^{-3}$          \\
AP3 Broad    & 2.6$\pm$0.4 $\times10^{-2}$               & 9.4$\pm$1.8 $\times10^{31}$  & 1.2$\pm$0.2 $\times 10^{-4}$   & 9.4$\pm$1.8 $\times 10^{-4}$          \\ \hline
AP5 Total    & 1.7$\pm$0.5 $\times 10^{-1}$               & 2.6$\pm$4.3 $\times10^{33}$  & 3.4$\pm$1.9 $\times 10^{-3}$   & 2.6$\pm$1.5 $\times 10^{-2}$          \\ \hline
AP6 Broad    & 1.8$\pm$0.7 $\times 10^{-1}$              & 1.6$\pm$1.1 $\times10^{33}$  & 2.1$\pm$1.5 $\times 10^{-3}$   & 1.6$\pm$1.2 $\times 10^{-2}$          \\ \hline
AP7 Broad    & 1.5$\pm$0.5 $\times 10^{-2}$              & 1.0$\pm$0.5 $\times10^{32}$  & 1.3$\pm$0.6 $\times 10^{-4}$   & 1.0$\pm$0.5 $\times 10^{-3}$          \\ \hline
\end{tabular}
\caption{Mass outflow rates, kinetic powers and coupling efficiencies for the kinematic components associated with an outflow. The coupling efficiencies presented here are calculated using the nuclear bolometric luminosity of IC\;5063 presented by \citet{Nicastro2003} ($\epsilon_\mathrm{f}$), and for a jet power of P$_\mathrm{jet} = 10^{37}$\;W ($\epsilon^\prime_\mathrm{f}$) as used in modelling by \citet{Mukherjee2018}.}
\label{tab: mout_fkin}
\end{table*}

\subsection{Analysis of the NIR apertures}
\label{section: nir_results}

In order to obtain independent information on the gas ionisation/excitation and, potentially, acceleration mechanisms, we analysed the data from the NIR arm of our Xshooter observations of IC\;5063.

We found that the [OIII] models (Section \ref{section: emission_line_fitting}) did not fit the NIR line profiles well, and in many cases the line profiles found in the near-infrared were visibly different to those found in the UV and optical. Furthermore, we found that lines corresponding to different ionisation states (e.g. [FeII] and Pa$\mathrm{\beta}$) have differing line profiles within a given aperture. Therefore, unlike the UVB+VIS data, we did not first define a line profile model using a single prominent emission line. Instead, we fit each emission line independently and group Gaussian components into `broad' (\mbox{FWHM\;\textgreater\;200\;km\,s$^{-1}$}) or `narrow' (\mbox{FWHM\;\textless\;200\;km\,s$^{-1}$}). We therefore recognise that direct comparisons to the results found for the broad and narrow components of UVB+VIS emission lines should be made with care. Using these Gaussian fits, we measured several key prominent emission lines in the near-infrared, namely [FeII]$\lambda$12570, [FeII]$\lambda$16400, Pa$\beta$, Br$\gamma$ and HeI$\lambda$10830 --- which trace the warm ionised phase --- and H$_2$1--0S(1)$\lambda$21218, which traces the warm molecular phase.

\subsubsection{Warm molecular gas phase}

\begin{figure}
	\includegraphics[width=\linewidth]{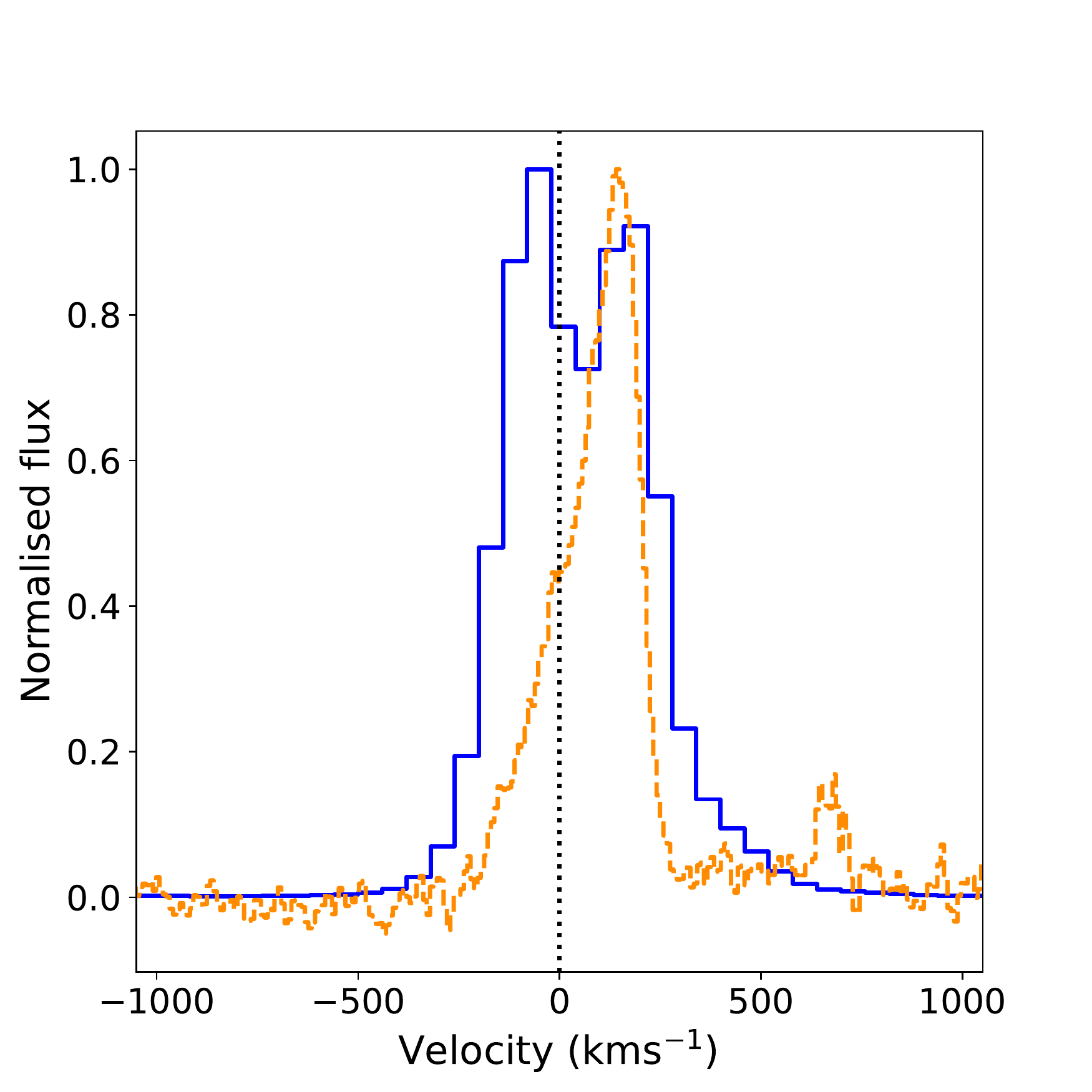}
	\caption{Velocity profiles for the [OIII]$\lambda$5007 (blue solid line) and \mbox{H$_2$1--0S(1)$\mathrm{\lambda}$21218} (orange dashed line) lines in Aperture 3. The blueshifted narrow component seen in the [OIII] line profile (`AP3 Narrow 2'), emitted by the warm ionised gas, is not present in the warm molecular H$_2$ emission. Note that the [OIII]$\lambda$5007 line lies within the VIS Xshooter arm, and has been resampled to a wavelength step of 1\AA/pixel.}
	\label{fig: ap3_oiii_h2}
\end{figure}

The H$_2 \lambda$21218 emission line can be used to probe the warm molecular gas phase, and comparing its line profile to those of forbidden and recombination lines allows for an investigation of the different outflow phases present in each aperture \citep{Tadhunter2014}. Interestingly, the blueshifted narrow component in Aperture 3 seen in the optical line profiles (`AP3 Narrow 2') is absent in the H$_2 \lambda$21218 line profile (Figure \ref{fig: ap3_oiii_h2}), although it is present in the [FeII] and NIR recombination line profiles. This indicates that excited warm molecular gas is not kinematically associated with the blueshifted component in Aperture 3.

\subsubsection{Ionisation and excitation of the near-infrared [FeII], Pa$\mathrm{\beta}$ and H$_2$ lines}
\label{section: nir_excitation}

In order to supplement our investigation of the dominant outflow ionisation mechanism (Section \ref{section: electron_temperatures}), we used line ratios of the [FeII]$\lambda$12570, [FeII]$\lambda$16400 and H$_2 \lambda$21218 emission lines to NIR recombination lines \citep{Rodriguez-Ardila2005, Riffel2013a, Colina2015, Riffel2021}. We made use of the diagnostic plot of [FeII]$\lambda$12570/Pa$\mathrm{\beta}$ vs H$_2\lambda$21218/Br$\mathrm{\gamma}$ for this purpose, with the limits presented by \citet{Riffel2013a} and \citet{Riffel2021} that separate star-forming galaxies (SF), AGN and high line ratio (HLR) objects. The HLR region of the [FeII]$\lambda$12570/Pa$\mathrm{\beta}$ vs H$_2\lambda$21218/Br$\mathrm{\gamma}$ diagnostic plot contains objects such as Supernova Remnants (SNRs) and Low Ionisation Nuclear-Emission line Regions (LINERs; \citealt{Heckman1980}). It is notable that \citet{Riffel2021} found correlations between these emission line ratios and the emission line width metric $W_{80}$ (the line width containing 80 per cent of the total line flux). This indicates that shocks may be the dominant ionisation/excitation mechanism in the HLR region in some objects.

\begin{figure}
	\includegraphics[width=\linewidth]{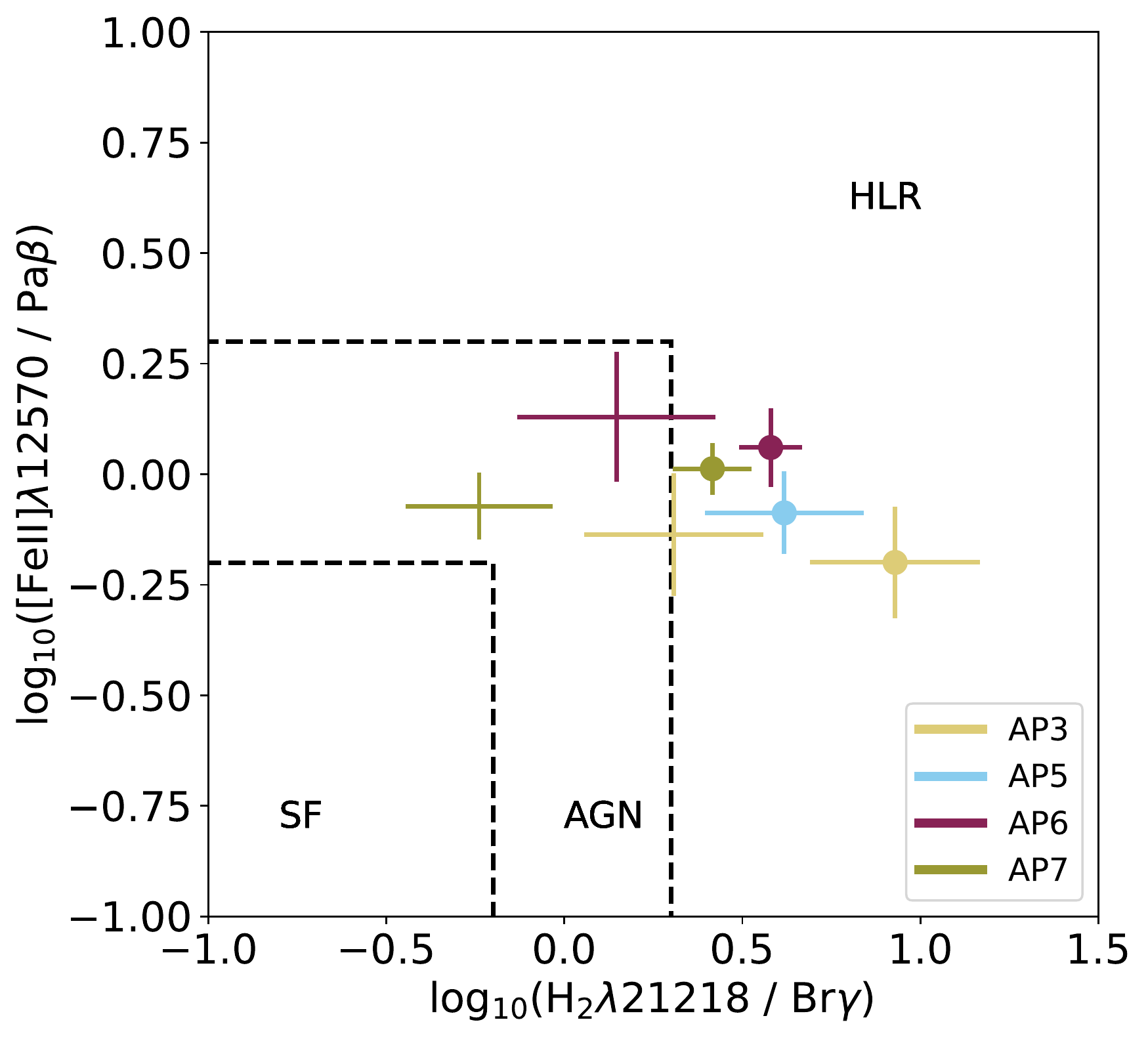}
	\caption{Values for the [FeII]$\lambda$12570/Pa$\mathrm{\beta}$ and H$_2 \lambda$21218/Br$\mathrm{\gamma}$ ratios \mbox{(crosses: narrow components; circles: broad components)}, with regions defined by \citet{Riffel2021}: `SF' denotes observed line ratios which are associated with excitation from star formation, `AGN' denotes line ratios which are associated with the central regions of AGN and `HLR' denotes values higher than those observed from AGN excitation alone. It has been proposed that line ratios in the HLR region indicate contribution from shock excitation, either from supernovae (SNe) or jet-ISM interactions. Here, we find that kinematic components associated with outflows fall within the HLR, with the quiescent components falling within the AGN region (with the exception of Aperture 3). The color and marker scheme is the same as in Figure \ref{fig: tr_grid}.}
	\label{fig: feii_pab}
\end{figure}

We plot the line ratios of the different kinematic components in each aperture on this diagnostic diagram in Figure \ref{fig: feii_pab}. The line ratios are also listed in Table \ref{tab: nir_ratios}. It is striking that the broad components of the NIR emission line profiles have higher H$_2\lambda$21218/Br$\mathrm{\gamma}$ ratios relative to the narrow components, with the broad components falling in the HLR region and the narrow components falling within the AGN region. Considering that the H$_2\lambda$21218/Br$\mathrm{\gamma}$ ratio probes the warm molecular gas, this indicates that the narrow components of this phase are predominately AGN-excited, while the broad components have composite shock-AGN excitation. In contrast, there is no clear difference in the [FeII]$\lambda$12570/Pa$\mathrm{\beta}$ ratios of the broad and narrow components, indicating that the quiescent and outflowing warm ionised gas is predominantly AGN-photoionised (consistent with our results in Section \ref{section: electron_temperatures}).

\begin{table*}
\begin{tabular}{llll}
\hline
Component    & H$_2\lambda$21218/Br$\mathrm{\gamma}$ & [FeII]$\lambda$12570/Pa$\mathrm{\beta}$ & [FeII]$\lambda$16400/Br$\mathrm{\gamma}$ \\ \hline
AP3 Narrow 1 & 2.0$\pm$1.2    & 0.7$\pm$0.2                   & 4.5$\pm$2.5                  \\
AP3 Broad    & 8.5$\pm$4.7    & 0.6$\pm$0.2                   & 3.1$\pm$1.8                  \\ \hline
AP5 Total    & 4.2$\pm$2.1    & 0.8$\pm$0.2                   & 5.8$\pm$3.2                  \\ \hline
AP6 Narrow   & 1.4$\pm$0.9    & 1.4$\pm$0.5                   & 3.6$\pm$2.2                  \\
AP6 Broad    & 3.8$\pm$0.8    & 1.2$\pm$0.2                   & 6.5$\pm$1.6                  \\ \hline
AP7 Narrow   & 0.6$\pm$0.3    & 0.9$\pm$0.2                   & 5.3$\pm$1.3                  \\ 
AP7 Broad    & 2.6$\pm$0.7    & 1.0$\pm$0.1                   & 4.1$\pm$1.2                  \\ \hline
\end{tabular}
\caption{Ratios of NIR H$_2$, [FeII] and recombination lines for the components that we identify as representing outflowing gas. These values are shown on a diagnostic plot in Figure \ref{fig: feii_pab}, which is used to investigate the excitation mechanisms of the outflowing gas.}
\label{tab: nir_ratios}
\end{table*}

\section{Discussion}
\label{section: discussion}

The results presented in Section \ref{section: results} provide evidence for outflowing gas at the NW and SE radio lobes of IC\;5063 as a result of jet-ISM interactions, supporting the findings of previous studies (e.g. \citealt{Morganti2007, Tadhunter2014, Morganti2015}). In this section, we discuss in detail the relationship between the different gas phases at the NW lobe, investigate which is the dominant outflow phase in terms of mass and kinetic power through comparison to previous observations, and propose a physical relation between the different gas phases and components along the radio axis. Finally, we discuss the implications of our findings for the transauroral line technique of deriving electron densities.

\subsection{Placing the warm ionised outflows at the NW lobe in a multi-phase context}
\subsubsection{Kinematics of the different gas phases}

The [OIII] kinematics that we observe along the radio axis in IC\;5063 are consistent with a combination of regular gravitational disk rotation (traced by the narrow components, with the exception of the `Narrow 2' component in Aperture 3) and outflows of velocity 300\;\textless\;$v_\mathrm{out}$\;\textless\;700\;kms$^{-1}$ (traced by the broad components). As noted in previous work, the close association between highly disturbed emission-line kinematics and the radio structure provides strong evidence that the warm gas outflows are driven by the expanding radio source (e.g. \citealt{Morganti2007}, \citealt{Tadhunter2014}).

\begin{figure}
	\includegraphics[width=1\linewidth]{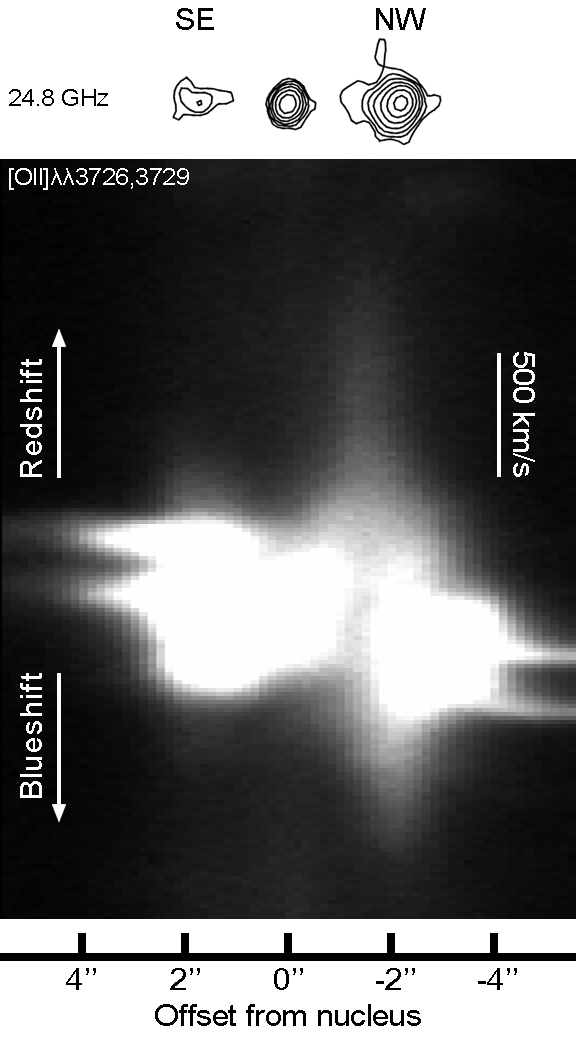}
	\caption{2-dimensional position-velocity profile of the [OII]$\lambda\lambda$3726,3729 doublet. Here, the spatial direction is horizontal and the velocity (spectral) direction is vertical. Corresponding 24.8\;GHz continuum imaging by \citet{Morganti2007} is shown above the profile, indicating the shape and extent of the radio structure. The velocity scale bar represents a shift of $\Delta$V=500\;km\,s$^{-1}$. Very broad velocity wings can be seen at several locations, including at the centroids of the SE and NW lobes, with the most extended being a $\sim$1500\;km\,s$^{-1}$ red wing between the nucleus and the centroid of the NW lobe.}
	\label{fig: o2_pv_diagram}
\end{figure}

Kinematically, our results for the warm-ionised phase differ from what has previously been found for the cold molecular gas. Here, we find extreme warm ionised outflow kinematics at several positions along the radio axis, including the regions between the nucleus and the centroids of the radio lobes, whereas previous observations for the cold molecular phase --- as traced by CO emission lines --- find more extreme kinematics at the outer limit of NW lobe than elsewhere \citep{Morganti2015, Oosterloo2017}. In our 2-dimensional Xshooter spectra for the warm ionised gas (Figure \ref{fig: o2_pv_diagram}), we do indeed see high velocity blue wings extending to $\sim$1000\;km\,s$^{-1}$ at the NW lobe, consistent with what is seen for the cold molecular phase. However, unlike the cold molecular phase, we also find red wings extending to $\sim$800\;km\,s$^{-1}$ in the SE lobe for the warm ionised emission. Most strikingly, we find the most extended velocity wings of the warm ionised phase, redshifted up to a maximum velocity of $\sim$1500km\,s$^{-1}$ at zero intensity, at a location \textit{between} the nucleus and the centroid of the NW lobe. The spatial centroid of this wing between \mbox{1000\;km\,s$^{-1}$\textless $v$ \textless1500\;km\,s$^{-1}$}, as measured by Gaussian fits to the continuum-subtracted spatial flux profile (as in Section \ref{section: spatial_optical}) of the [NII]$\lambda$6583 line, lies 1.41\;arcsec north west of the continuum centroid, or $\sim$0.6\;arcsec (0.14\;kpc) closer to the nucleus than the centroid of the NW lobe.

\begin{figure}
	\includegraphics[width=1\linewidth]{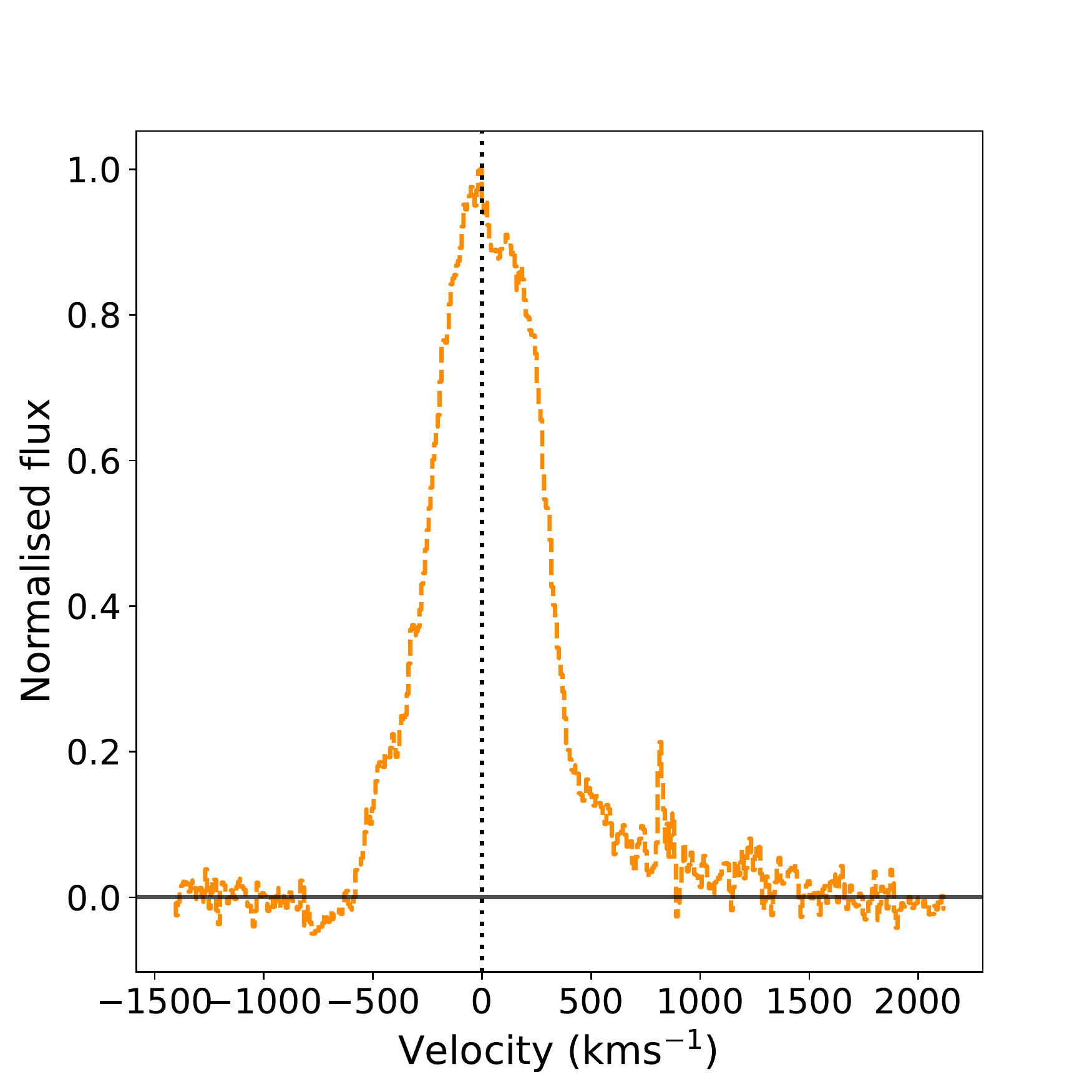}
	\caption{The velocity profile of the H$_2\lambda$21218 (dashed orange) line in the combined apertures 5 and 6, corresponding to the location between the nucleus and the outer extent of the NW radio lobe (Figure \ref{fig: apertures}). The dotted black vertical line represents a velocity of 0\;km\,s$^{-1}$, while the solid black horizontal line shows the continuum level. A red velocity wing extending to \textgreater 1000\;km\,s$^{-1}$ can be seen, consistent with the warm ionised phase (Figure \ref{fig: o2_pv_diagram}).}
	\label{fig: ap56_h2_velocity_profile}
\end{figure}

The kinematics of the warm molecular phase, as traced with the H$_2\lambda$21218 emission, follow the warm ionised kinematics. A blue velocity wing is seen to extend to $\sim$1000\;km\,s$^{-1}$ at the centroid of the NW lobe, and a red wing extends to \textgreater1000\;km\,s$^{-1}$ (Figure \ref{fig: ap56_h2_velocity_profile}) between this position and the nucleus. This feature may extend to higher velocities, potentially up to $\sim$1500\;km\,s$^{-1}$ (as seen in the warm ionised gas), but blending with the continuum makes it difficult to determine the true velocity extent. These observed H$_2\lambda$21218 kinematics are also consistent with previous observations of the warm molecular phase by \citet{Tadhunter2014}.

The kinematics observed in the warmer gas phases can be explained by an expanding hemispherical bow shock or bubble, the tip of which coincides with --- or extends slightly beyond --- the centroid of the NW lobe. Considering that this structure would be seen side-on, the highest velocity widths are expected closer to the nucleus than the centroid of the lobe due to projection effects (i.e. the gas moving along the observer's line of sight). This is consistent with the highest observed velocities being between the nucleus and the outer limit of the NW lobe. Likewise, lower projected velocities would be found at the centroid of the lobe (closer to the tip of the bow shock), as the outflowing gas there would be moving in a direction that is close to the plane of the sky. Our observations are also consistent with the kinematics seen in hydrodynamic modelling of IC\;5063 for a jet of power P$_\mathrm{jet}$=10$^{37-38}$\;W propagating through the disk, as presented by \citet{Mukherjee2018}, where red- and blue-shifted line wings are seen at all locations where the radio source is interacting with the ISM in the galaxy's disk.

The difference in observed kinematics between the warm and cold molecular gas may be explained if the different molecular phases constitute a post-shock cooling sequence, as has been previously proposed \citep{VillarMartin1999, Zubovas2014, Tadhunter2014}. In this scenario, the warm H$_2$ emission represents a transition phase, emitted as gas cools behind the shock through T$\sim$2000\;K, and has a constant mass and luminosity, assuming material passes through the shock at a constant rate. As the gas cools further it enters the cold molecular phase, which emits the CO emission lines. This represents the end-state of the cooling sequence, and its mass therefore accumulates as a function of time. Assuming that the conditions for cold molecular gas formation are met, and the molecular gas is not destroyed by AGN radiation, the ratio of cold molecular gas (traced by CO emission) to warm molecular gas (traced by H$_2$ emission) will thus increase as a function of time. In this scenario, if the gas seen in projection in between the nucleus and the maximum extent of the NW radio lobe has only just started to enter the shock, it is plausible that not enough gas has so far accumulated in the cold molecular phase to be detectable via its CO emission. In contrast, gas may have been entering the shock at the tip of the hemispherical bow shock (the maximum extent of the NW lobe) for longer, allowing more time for CO-emitting cold molecular gas to accumulate. This would explain why we see disturbed kinematics at the centroid of the NW radio lobe for all phases, but only the warmer phases show more extreme kinematics in between this location and the nucleus.

\subsubsection{Energetics of the multi-phase outflows}
\label{section: discussion-multi-phase-energetics}

In Table \ref{tab: all_phases}, we present masses, mass outflow rates and coupling efficiencies for the the different outflow phases in IC\;5063, with values for the colder phases from previous studies recalculated to ensure consistency (see Appendix \ref{section: energetics_recalculation}); all coupling efficiencies ($\epsilon_\mathrm{f}$) have been calculated assuming a bolometric luminosity of L$_\mathrm{bol}$=7.6$\times 10^{37}$\;W \citep{Nicastro2003, Morganti2007}. The mass outflow rate that we find at the NW lobe is higher than that for the warm ionised gas found by \citet{Morganti2007} ($\dot{M}_\mathrm{ion} \sim 0.08 $M$_\odot$yr$^{-1}$). This difference is likely due to the different outflow radii used: \citet{Morganti2007} take the distance of a given aperture from the nucleus ($R$) in Equation \ref{eq: mout_rate}, whereas we instead use the aperture width ($\Delta R$), which is smaller and thus produces higher outflow rates.

However, our derived mass outflow rates for the warm ionised gas are still significantly lower than those of the neutral atomic ($\sim$35\;$\dot{M}_{\odot}$\,yr$^{-1}$) and cold molecular ($\sim$0.8\;$\dot{M}_{\odot}$\,yr$^{-1}$) phases. In addition, the coupling efficiency of the warm ionised phase (\mbox{$\epsilon_\mathrm{f}$=(2.7$\pm$1.7)$\times10^{-3}$ per cent}) is much lower than that of the neutral molecular phase ($\epsilon_\mathrm{f} \approx$ 0.2\;per\;cent), but similar to that of the cold molecular phase ($\epsilon_\mathrm{f} \approx$ $3.1\times10^{-3}$\;per\;cent, although this is likely to be a lower limit: Appendix \ref{section: energetics_recalculation}). Overall, the mass and kinetic power budgets of the outflows are dominated neutral phase, with the warm ionised and cold molecular gas making relatively minor contributions.

\begin{table*}
\begin{tabular}{llllll}
\hline
Phase          & Location        & $M$ ($M_\odot$)                                                   & $\dot{M}_\mathrm{out}$ ($M_\odot$yr$^{-1}$) & $\epsilon_\mathrm{f}$ (per cent)   & Reference                             \\ \hline
Warm ionised   & NW Lobe         & ---                                                                & 0.08                                        & ---                                 & \citet{Morganti2007} \\
               & Galaxy-wide\;$^a$ & 1.5$^{+3.0}_{-0.9} \times 10^6$                                    & ---                                          & ---                                 & \citet{Venturi2021}  \\
               & NW Lobe\;$^b$     & (9.9$\pm0.1) \times 10^4$                                       & (1.8$\pm$0.6) $\times$ $10^{-1}$          & (2.7$\pm$1.7) $\times$ $10^{-3}$ & This work                             \\
               & SE Lobe\;$^c$     & (2.2$\pm0.3) \times 10^4$                                       & (2.6$\pm$0.4) $\times$ $10^{-2}$          & (1.2$\pm$0.2) $\times$ $10^{-4}$ & This work                             \\ \hline
Neutral HI     & NW Lobe         & ---                                              & 3.5$\times10^1$                                          & 1.8$\times10^{-1}$                                  & \citet{Morganti2005}\;$^d$ \\ \hline
Cold molecular & NW Lobe         & 1.3$\times$ $10^6$ & 7.9$\times10^{-1}$                                     & 3.1$\times10^{-3}$                          & \citet{Oosterloo2017}\;$^{d}$ \\ \hline
               &                 &                                                                   &                                             &                                    &                                      
\end{tabular} \\
$^a$For the gas with [OIII] W70 (85th minus 15th velocity percentile) \mbox{\textgreater\;300\;km\,s$^{-1}$} across all outflow regions. \\
$^b$For the NW lobe, the outflow mass is the sum of the masses in apertures 5 and 6, whereas the mass outflow rate and coupling efficiency represent the average values for apertures 5 and 6 (see Table \ref{tab: mout_fkin}). \\
$^c$Taken from the values for the broad component in Aperture 3. \\
$^d$Values have been recalculated using consistent methodology with the values presented in \citet{Morganti2005} and \citet{Oosterloo2017}, and are likely to be lower limits --- see discussion in Appendix \ref{section: energetics_recalculation}. \\
\caption{Masses, mass outflow rates and coupling efficiencies (for the nuclear bolometric case; see Section \ref{section: mout_fkin}) for the different outflow phases in IC\;5063, as reported in this work and calculated using the results from previous observational studies (Appendix \ref{section: energetics_recalculation}).}
\label{tab: all_phases}
\end{table*}

\subsection{Outflow acceleration and ionisation/excitation mechanisms in IC\;5063}
\subsubsection{Ionisation and excitation of the warm gas}

If the outflows in IC\;5063 are both accelerated and ionised by shocks induced by jet-ISM interactions, then it might expected that the broad kinematic components would have higher electron temperatures than the quiescent narrow components \citep{Fosbury1978, VillarMartin1999}. However, we do not find clear evidence for higher electron temperatures associated with outflowing components (Table \ref{tab: temps}). Only Aperture 3 shows a significant (\mbox{\textgreater3$\sigma$}) difference in temperature between kinematic components, with the broad component having a \textit{lower} (instead of a higher) electron temperature. Furthermore, the electron temperatures that we find for all components --- probed with the [OIII](5007/4363) ratio --- are consistent with pure AGN photoionisation (Figure \ref{fig: heii_hbeta}).

Further potential evidence regarding the natures of the ionisation/excitation mechanisms is provided by the [FeII]$\lambda$12570/Pa$\mathrm{\beta}$ vs H$_2 \lambda$21218/Br$\mathrm{\gamma}$ diagnostic diagram (Figure \ref{fig: feii_pab}). The two ratios used in this diagram each probe a different gas phase: warm ionised and warm molecular, respectively. It is interesting that the measured values of the [FeII]$\lambda$12570 /Pa$\mathrm{\beta}$ ratio are similar for the broad and narrow components, consistent with the [OIII](5007/4363) vs HeII4686/H$\mathrm{\beta}$ diagram (Figure \ref{fig: heii_hbeta}), and indicating that both the quiescent and outflowing warm ionised gas is predominantly AGN-photoionised (in agreement with the results of a previous investigation of the warm ionised outflows in IC\;5063 by \citealt{Morganti2007}). However, in the H$_2 \lambda$21218/Br$\mathrm{\gamma}$ ratios, probing the excitation of the warm molecular phase, we find a clear difference between broad and narrow components. Along this ratio axis, the quiescent gas falls within the region of the diagnostic diagram where AGN ionisation is thought to be dominant (with perhaps some contribution from shocks; see \citealt{Riffel2021}), whereas the outflowing gas falls within the HLR region, within which AGN excitation alone cannot account for the high line ratios.

\subsubsection{The physical relation between the different gas phases}

Taken together, our investigation into the ionisation and excitation of the UVB+VIS and NIR lines in IC\;5063 indicates that the quiescent and outflowing warm-ionised gas, along with the quiescent warm-molecular gas, are predominantly ionised/excited by the central AGN, while the outflowing warm molecular gas has a significant contribution from shock excitation. This rules out the idea that we are observing a post-shock cooling sequence alone: in such a scenario, all the warm ionised and warm molecular gas in the outflow would be ionised and accelerated close to the shock front, and the cold molecular gas would represent the post-shock gas further from the shock front (and closer to the central AGN) that has cooled through the sequence.

\begin{figure*}
    \centering
    \includegraphics[width=1\linewidth]{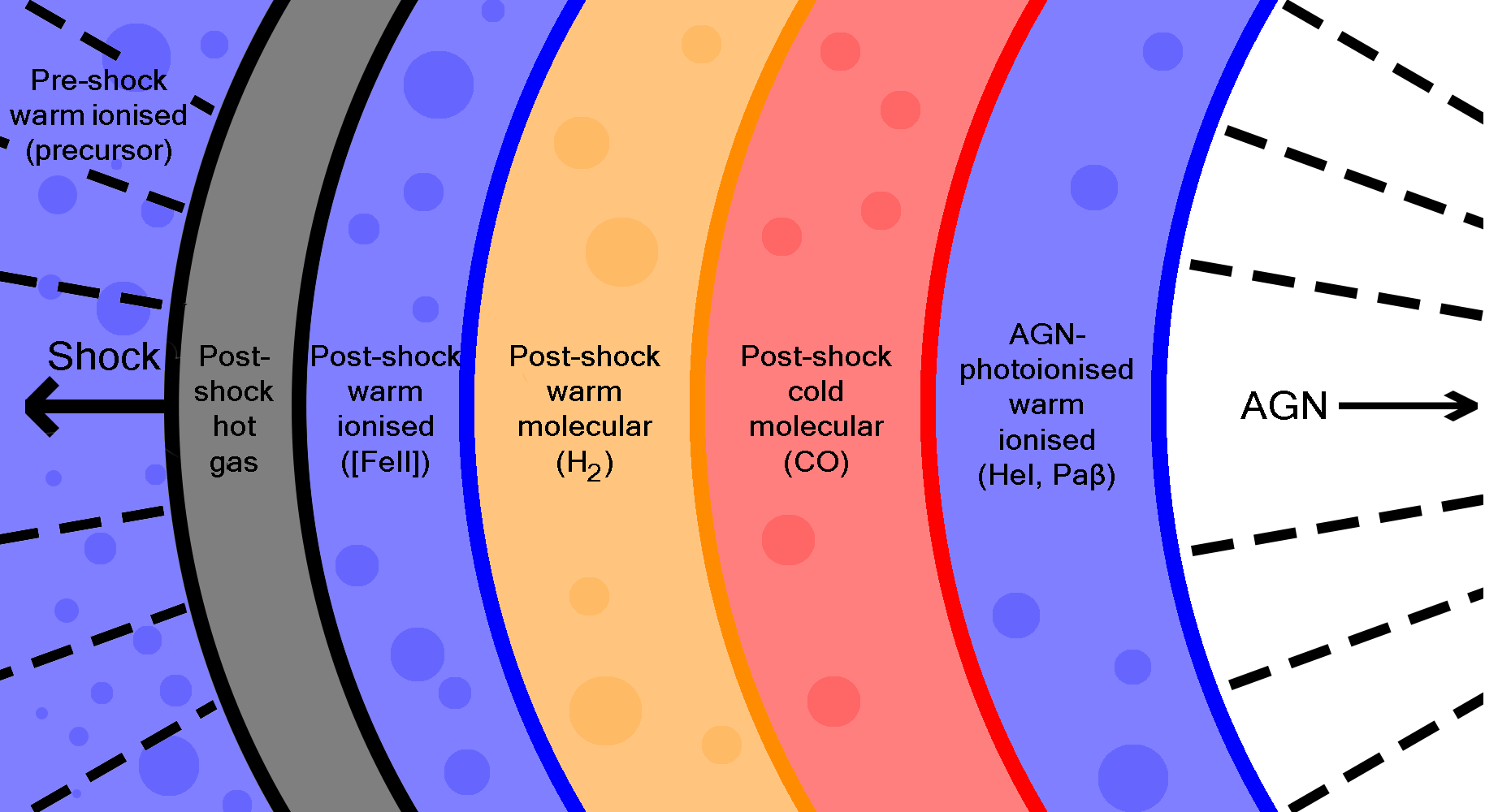}
    \caption{A schematic showing a possible stratification in the emission regions that we propose to explain both the measured line ratios (Figures \ref{fig: heii_hbeta} and \ref{fig: feii_pab}) and the spatial distributions of the various emission lines (Figure \ref{fig: spatial_nir} and Table \ref{tab: spatial_nir}) in IC\;5063. The different phases (labelled) are shown as different coloured regions, and photoionising radiation is shown as black dashed lines. Note that this could represent the structure in individual clouds, or an ensemble of clouds. In this schematic, no attempt has been made to reflect the true relative sizes of the emission regions.}
    \label{fig: cooling_sequence_schematic}
\end{figure*}

Rather, a potential explanation for our results is that the cooled post-shock gas closest to the AGN is photoionised by the AGN, resulting in an additional warm ionised component which has post-shock kinematics and densities while being predominantly AGN-photoionised. In this scenario, presented in Figure \ref{fig: cooling_sequence_schematic}, the post-shock cooled gas that is furthest from the shock front (and partly photoionised by the AGN) would shield the post-shock gas that is closer to the shock front. If this AGN-photoionised component has a higher luminosity than the immediate post-shock warm ionised gas, then it would contribute much more strongly to the emission line ratios, leading to values consistent with AGN-photoionisation being observed. However, the warm molecular phase --- which is only found cooling behind the shock --- would show line ratios consistent with shock excitation, as we find in our measured ratios.

\begin{figure}
    \centering
    \includegraphics[width=1\linewidth]{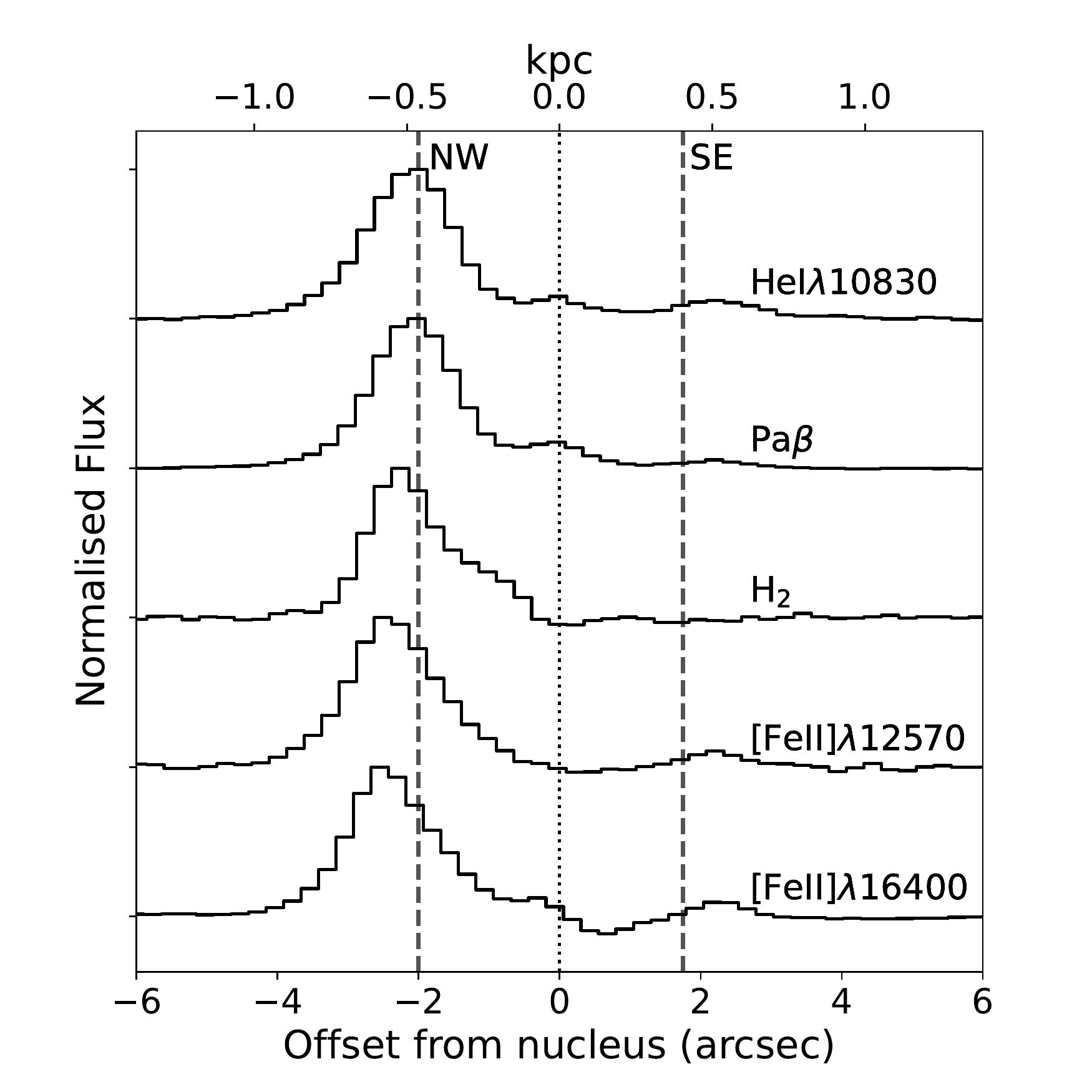}
    \caption{Spatial flux distributions of the \mbox{$-$600\;km\,s$^{-1}$\;\textless\;$v$ \;\textless\;$-$400\;km\,s$^{-1}$} wings of the emission line profiles for [FeII]$\lambda$16400, [FeII]$\lambda$12570, H$_\mathrm{2}\lambda$21218, Pa$\mathrm{\beta}$ and HeI$\lambda$10830, produced using the same methodology as described in Section \ref{section: spatial_optical}. It can be tentatively seen that the [FeII] lines (most likely tracing shocked warm ionised emission) lie the furthest from the nucleus --- at the expected location of the shocks in the NW lobe --- with the warm molecular H$_\mathrm{2}\lambda$21218 emission lying further inward, and the Pa$\mathrm{\beta}$ and HeI$\lambda$10830 lying closest to the nucleus. We present the centroids of Gaussians fitted to these profiles in Table \ref{tab: spatial_nir}.}
    \label{fig: spatial_nir}
\end{figure}

To investigate this situation further, we created spatial flux profiles of the integrated \mbox{$-$600\;km\,s$^{-1}$\;\textless\;$v$ \;\textless\;$-$400\;km\,s$^{-1}$} wings (using the same method as in Section \ref{section: spatial_optical}) of several emission lines present in our NIR apertures: [FeII]$\lambda$16400, [FeII]$\lambda$12570, H$_\mathrm{2}\lambda$21218, Pa$\mathrm{\beta}$ and HeI$\lambda$10830. Here, the [FeII], HeI and Pa$\mathrm{\beta}$ lines trace the warm ionised phase (however the [FeII] lines are expected to be particularly strong in regions associated with shocks: \citealt{Dors2012} and \citealt{Riffel2013a}), while H$_\mathrm{2}\lambda$21218 traces the warm molecular phase. The resulting plot is presented as Figure \ref{fig: spatial_nir}, which shows tentative evidence that the [FeII] emission peaks in flux the furthest away from the nucleus, the peak of H$_\mathrm{2}\lambda$21218 emission lies inward of the [FeII] peak, and the HeI and Pa$\mathrm{\beta}$ emission peaks the closest to the nucleus. This is supported by the centroid positions obtained by fitting Gaussian profiles to the spatial profiles (see Table \ref{tab: spatial_nir}).

Given that the NW lobe is considered to be the location of the strongest shocks, this evidence for stratification in the emission regions supports the geometry presented in Figure \ref{fig: cooling_sequence_schematic}: the [FeII] emission traces the warm ionised gas near the shock front, the H$_\mathrm{2}\lambda$21218 traces post-shock gas that has cooled to the warm molecular phase, and the HeI and Pa$\mathrm{\beta}$ emission traces the inner-most (AGN-photoionised) warm ionised component. However, we note that our measured [FeII]/Pa$\mathrm{\beta}$ ratios (Figure \ref{fig: feii_pab}; Table \ref{tab: nir_ratios}) may not be consistent with this scenario: while the measured H$_\mathrm{2}\lambda21218$/Br$\mathrm{\gamma}$ ratios appear to be enhanced relative to what is expected from AGN excitation, the [FeII]/Pa$\mathrm{\beta}$ ratios show no such relative enhancement. Further near-infrared observations of IC\;5063's NW lobe, taken with an IFU at a higher spatial resolution, are thus needed to investigate this situation and verify the cooling sequence scenario proposed here.

\begin{table}
\centering
\begin{tabular}{l|l|l}
\hline
Emission Line           & \parbox{1cm}{Centroid \\ (pixels)} & \parbox{1cm}{Centroid \\ (arcsec)} \\ \hline
HeI$\lambda$10830 & 8.5$\pm$0.2 & 2.11$\pm$0.05 \\
Pa$\beta$ & 8.6$\pm$0.1 & 2.13$\pm$0.02 \\
H$_\mathrm{2}\lambda$21218 & 9.5$\pm$0.1 & 2.36$\pm$0.02 \\
{[FeII]$\lambda$12570} & 10.1$\pm$0.2 & 2.50$\pm$0.05 \\
{[FeII]$\lambda$16400} & 10.2$\pm$0.2 & 2.53$\pm$0.05 \\                                                                                              
\end{tabular}
\caption{Centroids of Gaussian fits to spatial slices between \mbox{$-$600\;km\,s$^{-1}$\;\textless\;$v$\;\textless\;$-$400\;km\,s$^{-1}$} of lines in our NIR arm data (Figure \ref{fig: spatial_nir}), measured relative to the spatial centroid position of the local continuum. The lines trace distinct phases of gas, which we interpret as a cooling sequence (Figure \ref{fig: cooling_sequence_schematic}).}
\label{tab: spatial_nir}
\end{table}

\subsubsection{Comparison to theoretical predictions of the dominant ionisation mechanism}
Recent relativistic hydrodynamic simulations by \citet{Meenakshi2022} model the relative contribution of AGN photoionisation and jet-induced shock collisional-ionisation for a 5.71$\times10^9$\;M$_\odot$ gas disk of radius 2\;kpc, similar to the gas conditions and properties of IC\;5063, and find that shock-ionisation dominates the ionisation of the warm ionised gas over AGN photoionisation. This is inconsistent with our results for IC\;5063, where we find AGN photoionisation to be dominant. A potential reason for this discrepancy could be that the jet power in IC\;5063 is in reality an order of magnitude lower (P$_\mathrm{jet}$=10$^{37-38}$\;W; \citealt{Mukherjee2018}) than assumed by \citet{Meenakshi2022} (P$_\mathrm{jet}$=10$^{38}$\;W). Moreover, the resolution of the simulations may not be sufficiently high to accurately track the density structure and cooling of the post-shock gas. Further simulation work that is more precisely tailored to the situation in IC\;5063 would allow this to be investigated in more detail, as would higher resolution simulations of AGN and shock ionisation for single clouds, which would allow a more accurate representation of the cooling of the warm gas.

\subsubsection{Compression effects of the jet-induced shocks}
As the outflows along the radio-axis in IC\;5063 are likely to have been accelerated by the jet-induced shocks, some evidence for shock-compression would be expected \citep{Sutherland2017}. Using the TR ratios (Section \ref{section: tr_lines}), we find that the narrow components at all positions have relatively low electron densities in the range \mbox{2.10\;\textless\;log$_{10}$(n$_e$[TR]\;cm$^{-3}$)\;\textless\;2.75}, while outflow components have significantly higher electron densities of \mbox{3.1\;\textless\;log$_{10}$(n$_e$[TR]\;cm$^{-3}$)\;\textless\;3.45}. This indicates that the outflows have higher electron densities than the quiescent gas by approximately a factor of three or four, possibly as a result of compression effects of the jet-induced shocks. This is similar to the findings of other AGN-driven outflow studies, which find that broader components typically have higher densities \citep{Holt2011, Rose2018, Santoro2018, Santoro2020}, and is in line with what is expected for the immediate post-shock gas due to the shock-jump conditions (a factor of four; \citealt{Sutherland2017}). However, the density contrast is much less than the factor of $\sim$100 which may be expected if the broad components originate from gas that has cooled in pressure equilibrium behind a fast shock \citep{Sutherland2017}.

\subsubsection{The nature of the blueshifted narrow [OIII] component seen in Aperture 3}
Along with previous observations of the warm ionised phase of IC\;5063 \citet{Morganti2007}, we identify a narrow kinematic component in the SE radio lobe that is blueshifted relative to the quiescent gas disk. Unlike the broad kinematic components associated with outflows, this component is narrow (\mbox{FWHM$_\mathrm{w}$\;\textless\;200\;km\,s$^{-1}$}) and has a lower electron density (log$_{10}n_\mathrm{e}=2.55 \pm$0.03). Bi-conical outflows have been proposed to explain line splitting in active galaxies \citep{Walker1968, Cecil1990}, but if this is the case, it is unclear why one direction of the outflow is broad and dense (AP3 Broad) and the other is narrow and has a lower density (AP3 Narrow 2). Alternatively, since IC\;5063 is thought to be a post-merger galaxy \citep{Morganti1998}, it would not be surprising to detect infalling and relatively undisturbed low-density clumps of gas. Interestingly, we do not detect H$_2$ emission in Aperture 3 that is kinematically associated with this component, implying that it either does not contain molecular gas or that the warm molecular gas is not being excited. This is consistent with VLT/ISAAC observations presented by \citet{Tadhunter2014}, which shows a blueshifted narrow component in the Br$\mathrm{\gamma}$ profile that is not present in the H$_2$1--0S(1) profile at the south-eastern lobe. Furthermore, this component is not seen in cold molecular PV diagrams (e.g. CO(1--0), CO(2--1), CO(3--2), HCO(4--3); \citealt{Morganti2015, Oosterloo2017}), implying a lack of cold molecular gas. Therefore, we favour the infalling gas explanation and propose that the situation in Aperture 3 is as follows: the broad component represents shock-accelerated outflowing gas, the red-most narrow component represents quiescent gas in IC\;5063's disk, and the blue-most narrow component represents infalling gas (that lacks molecular gas emission) as a result of the recent merger.

\subsection{The use of transauroral line diagnostics in AGN-driven outflows}

This study marks the first time that the transauroral line technique first presented by \citet{Holt2011} has been used to estimate electron density values for spatially-resolved outflow regions in an active galaxy. As previously discussed, a concern with this technique is that the TR lines might be emitted by clouds at different spatial positions to the other important diagnostic lines. The work presented here shows that the transauroral lines have the same line profiles for the same kinematic components and are emitted at the same spatial positions as the high-ionisation optical forbidden and recombination lines, as shown in \mbox{Figures \ref{fig: o3_models_all_lines} and \ref{fig: spatial_optical}}, suggesting that they are emitted by the same cloud complexes. While we cannot conclusively show that the transauroral lines are emitted by the same clouds \textit{within} the apertures (see discussions in \citealt{Sun2017} and \citealt{Rose2018}), this does alleviate concerns regarding their use in spatially unresolved studies made of many moderate-to-high redshift AGN.

In addition, we have used the transauroral line technique alongside the traditional [OII] and [SII] line ratios and [ArIV] line ratio to show that the TR lines give electron densities that are slightly higher than (if not comparable to) those given by the traditional line ratios, but lower than those given by the [ArIV] ratio. \citet{Wang2004} found that the [ArIV] technique provides much higher electron densities in observations of ionised nebulae than the traditional ratios; the findings presented here support this and place the transauroral line technique in-between these two methods in terms of derived densities. 

Further detailed studies making use of the transauroral line ratios alongside other methods of electron density estimation are needed. However, this work highlights the necessity of careful consideration of the technique used to derive electron densities: the traditional [OII] and [SII] lines may underestimate electron density whilst the [ArIV] ratio may overestimate. This, in part, may be due to the [ArIV] ratio probing higher-ionisation gas than the traditional and transauroral [OII] and [SII] techniques. Since the [OIII] lines represent higher-ionisation gas than the [OII] and [SII] lines, then our findings indicate that the density of the [OIII] emitting gas may be higher than probed by the [OII] and [SII] lines. It is unclear how this relates to the density of the H$\mathrm{\beta}$-emitting gas, although it could mean that true mass outflow rates (thus kinetic powers and coupling efficiencies) are even lower than we have estimated, and much lower than typically derived using densities estimated with the traditional [SII] and [OII] ratios.

The high (\mbox{\textgreater10$^3$\;cm$^{-3}$}) electron densities found for the outflowing components using the TR lines are above those typically assumed by past studies for warm outflows in active galaxies (\mbox{10$^{2}$--$10^3$\;cm$^{-3}$}: e.g. \citealt{Liu2013, Harrison2014, Fiore2017}), and above the previously measured maximum density of $n_e\sim10^3$\;cm$^{-3}$ for the radio axis of IC\;5063, as derived from MUSE observations by \citet{Mingozzi2019} (see also \citealt{Venturi2021}). Conversely, our findings support those by other studies that report high electron densities for warm outflows, including those that also use TR lines (e.g. \citealt{Holt2011, Rose2018, Santoro2018, Spence2018, Baron2019b, Davies2020}). This highlights the importance of properly estimating values of electron density and indicates that values determined solely using traditional techniques may be too low, and resulting kinetic powers too high.

Overall, our results show the validity of using the transauroral [SII] and [OII] lines to probe the electron densities and reddenings of AGN-driven outflows. They have the advantages of higher critical densities than the traditional [SII] and [OII] ratios, not suffering from the same blending issues, and being more prominent in galaxy spectra than the [ArIV] doublet.

\section{Conclusions}
\label{section: conclusions}

Using wide wavelength-coverage, high spectral and spatial resolution Xshooter observations of the nearby active galaxy IC\;5063, we have, for the first time, derived electron densities of spatially-resolved AGN-driven outflows using the transauroral [SII] and [OII] lines. The wealth of previous observations of the different gas phases of these outflows have allowed us to place our findings in a multi-phase context. In addition, we have investigated the ionisation and excitation mechanisms of the outflows in an object which shows clear jet-ISM interactions. Our detailed study has found the following.
\begin{itemize}
    \item The transauroral (TR) lines are emitted in the same spatial positions and with the same line profiles as other high-ionisation optical lines which are commonly used as outflow diagnostics, indicating that they are emitted by the same cloud complexes. This alleviates concerns regarding the use of TR lines in spatially-unresolved studies of moderate-to-high redshift AGN.
    \item In the case of IC\;5063, we find tentative evidence that electron densities determined using the transauroral lines are higher than those determined from the commonly used traditional [SII] and [OII] line ratios, while the [ArIV](4711/4740) ratio gives higher densities. Given the implication for derived coupling efficiencies if electron density is underestimated, we highlight that the TR lines should play an important role in future studies of AGN-driven outflows.
    \item The outflowing warm-ionised gas in IC\;5063 is higher-density than the quiescent gas, potentially as a result of compression effects of jet-ISM induced shocks. However, the outflow densities may be below those expected if the post-shock gas is cooling in pressure equilibrium.
    \item The kinetic powers for the warm ionised phase are much below those required by galaxy evolution models. However, when compared to previous observations of cooler phases, it appears that this phase constitutes only a small fraction of the total outflowing mass.
    \item The dominant ionisation mechanism of the warm ionised outflows and quiescent gas is AGN photoionisation, while the warm molecular outflows appear to be excited by composite AGN-shock excitation. A possible scenario is that the post-shock gas closest to the AGN is maintained in an ionised state by the AGN, and forms the ionised part of an ISM component that shields the immediate post-shock gas further out, allowing it to cool to colder phases.
\end{itemize}
 
\section*{Acknowledgements}

LRH and CNT acknowledge support from STFC. Based on observations collected at the European Organisation for Astronomical Research in the Southern Hemisphere under ESO programme 0101.B0409(A). This work makes use of the Starlink software (Currie et al. 2014), which is currently supported by the East Asian Observatory. This research has made use of the NASA/IPAC Infrared Science Archive, which is funded by the National Aeronautics and Space Administration and operated by the California Institute of Technology. The STARLIGHT project is supported by the Brazilian agencies CNPq, CAPES and FAPESP and by the France-Brazil CAPES/Cofecub program. LRH thanks Rebecca J. Houghton for her helpful comments in preparing this manuscript and Alex J. Brown for his assistance in preparing plots for increased visual accessibility. LRH and CNT thank Moun Meenakshi and Dipanjan Mukherjee for their useful discussion.

\section*{Data Availability}
The data used in this report is available from the ESO Science Archive Facility (\url{http://archive.eso.org/cms.html}) with Run/Program ID 0101.B-0409(A).
 



\bibliographystyle{mnras}
\bibliography{xshoo_IC5063} 

\begin{thebibliography}{}
\makeatletter
\relax
\def\mn@urlcharsother{\let\do\@makeother \do\$\do\&\do\#\do\^\do\_\do\%\do\~}
\def\mn@doi{\begingroup\mn@urlcharsother \@ifnextchar [ {\mn@doi@}
  {\mn@doi@[]}}
\def\mn@doi@[#1]#2{\def\@tempa{#1}\ifx\@tempa\@empty \href
  {http://dx.doi.org/#2} {doi:#2}\else \href {http://dx.doi.org/#2} {#1}\fi
  \endgroup}
\def\mn@eprint#1#2{\mn@eprint@#1:#2::\@nil}
\def\mn@eprint@arXiv#1{\href {http://arxiv.org/abs/#1} {{\tt arXiv:#1}}}
\def\mn@eprint@dblp#1{\href {http://dblp.uni-trier.de/rec/bibtex/#1.xml}
  {dblp:#1}}
\def\mn@eprint@#1:#2:#3:#4\@nil{\def\@tempa {#1}\def\@tempb {#2}\def\@tempc
  {#3}\ifx \@tempc \@empty \let \@tempc \@tempb \let \@tempb \@tempa \fi \ifx
  \@tempb \@empty \def\@tempb {arXiv}\fi \@ifundefined
  {mn@eprint@\@tempb}{\@tempb:\@tempc}{\expandafter \expandafter \csname
  mn@eprint@\@tempb\endcsname \expandafter{\@tempc}}}

\bibitem[\protect\citeauthoryear{{Alatalo} et~al.,}{{Alatalo}
  et~al.}{2011}]{Alatalo2011}
{Alatalo} K.,  et~al., 2011, \mn@doi [ApJ] {10.1088/0004-637X/735/2/88}, \href
  {https://ui.adsabs.harvard.edu/abs/2011ApJ...735...88A} {735, 88}

\bibitem[\protect\citeauthoryear{{Allen}, {Groves}, {Dopita}, {Sutherland}  \&
  {Kewley}}{{Allen} et~al.}{2008}]{Allen2008}
{Allen} M.~G.,  {Groves} B.~A.,  {Dopita} M.~A.,  {Sutherland} R.~S.,
  {Kewley} L.~J.,  2008, \mn@doi [\apjs] {10.1086/589652}, \href
  {https://ui.adsabs.harvard.edu/abs/2008ApJS..178...20A} {178, 20}

\bibitem[\protect\citeauthoryear{{Astropy Collaboration} et~al.,}{{Astropy
  Collaboration} et~al.}{2013}]{AstropyCollaboration2013}
{Astropy Collaboration} et~al., 2013, \mn@doi [A\&A]
  {10.1051/0004-6361/201322068}, \href
  {https://ui.adsabs.harvard.edu/abs/2013A&A...558A..33A} {558, A33}

\bibitem[\protect\citeauthoryear{{Astropy Collaboration} et~al.,}{{Astropy
  Collaboration} et~al.}{2018}]{AstropyCollaboration2018}
{Astropy Collaboration} et~al., 2018, \mn@doi [ApJ] {10.3847/1538-3881/aabc4f},
  \href {https://ui.adsabs.harvard.edu/abs/2018AJ....156..123A} {156, 123}

\bibitem[\protect\citeauthoryear{{Baldwin}, {Phillips}  \&
  {Terlevich}}{{Baldwin} et~al.}{1981}]{Baldwin1981}
{Baldwin} J.~A.,  {Phillips} M.~M.,   {Terlevich} R.,  1981, \mn@doi [\pasp]
  {10.1086/130766}, \href
  {https://ui.adsabs.harvard.edu/abs/1981PASP...93....5B} {93, 5}

\bibitem[\protect\citeauthoryear{{Baron} \& {Netzer}}{{Baron} \&
  {Netzer}}{2019}]{Baron2019b}
{Baron} D.,  {Netzer} H.,  2019, \mn@doi [MNRAS] {10.1093/mnras/stz1070}, \href
  {https://ui.adsabs.harvard.edu/abs/2019MNRAS.486.4290B} {486, 4290}

\bibitem[\protect\citeauthoryear{{Binette}, {Wilson}  \&
  {Storchi-Bergmann}}{{Binette} et~al.}{1996}]{Binette1996}
{Binette} L.,  {Wilson} A.~S.,   {Storchi-Bergmann} T.,  1996, \aap, \href
  {https://ui.adsabs.harvard.edu/abs/1996A&A...312..365B} {312, 365}

\bibitem[\protect\citeauthoryear{{Bruzual} \& {Charlot}}{{Bruzual} \&
  {Charlot}}{2003}]{Bruzual2003}
{Bruzual} G.,  {Charlot} S.,  2003, \mn@doi [MNRAS]
  {10.1046/j.1365-8711.2003.06897.x}, \href
  {https://ui.adsabs.harvard.edu/abs/2003MNRAS.344.1000B} {344, 1000}

\bibitem[\protect\citeauthoryear{{Cardelli}, {Clayton}  \& {Mathis}}{{Cardelli}
  et~al.}{1989}]{Cardelli1989}
{Cardelli} J.~A.,  {Clayton} G.~C.,   {Mathis} J.~S.,  1989, \mn@doi [ApJ]
  {10.1086/167900}, \href
  {https://ui.adsabs.harvard.edu/abs/1989ApJ...345..245C} {345, 245}

\bibitem[\protect\citeauthoryear{{Carnall}}{{Carnall}}{2017}]{Carnall2017}
{Carnall} A.~C.,  2017, arXiv e-prints, \href
  {https://ui.adsabs.harvard.edu/abs/2017arXiv170505165C} {p. arXiv:1705.05165}

\bibitem[\protect\citeauthoryear{{Cecil}, {Bland}  \& {Tully}}{{Cecil}
  et~al.}{1990}]{Cecil1990}
{Cecil} G.,  {Bland} J.,   {Tully} R.~B.,  1990, \mn@doi [ApJ]
  {10.1086/168742}, \href
  {https://ui.adsabs.harvard.edu/abs/1990ApJ...355...70C} {355, 70}

\bibitem[\protect\citeauthoryear{{Cicone} et~al.,}{{Cicone}
  et~al.}{2014}]{Cicone2014}
{Cicone} C.,  et~al., 2014, \mn@doi [A\&A] {10.1051/0004-6361/201322464}, \href
  {https://ui.adsabs.harvard.edu/abs/2014A&A...562A..21C} {562, A21}

\bibitem[\protect\citeauthoryear{{Cicone}, {Brusa}, {Ramos Almeida}, {Cresci},
  {Husemann}  \& {Mainieri}}{{Cicone} et~al.}{2018}]{Cicone2018}
{Cicone} C.,  {Brusa} M.,  {Ramos Almeida} C.,  {Cresci} G.,  {Husemann} B.,
  {Mainieri} V.,  2018, \mn@doi [Nat. Astron.] {10.1038/s41550-018-0406-3},
  \href {https://ui.adsabs.harvard.edu/abs/2018NatAs...2..176C} {2, 176}

\bibitem[\protect\citeauthoryear{{Cid Fernandes}, {Mateus}, {Sodr{\'e}},
  {Stasi{\'n}ska}  \& {Gomes}}{{Cid Fernandes} et~al.}{2005}]{CidFernandes2005}
{Cid Fernandes} R.,  {Mateus} A.,  {Sodr{\'e}} L.,  {Stasi{\'n}ska} G.,
  {Gomes} J.~M.,  2005, \mn@doi [MNRAS] {10.1111/j.1365-2966.2005.08752.x},
  \href {https://ui.adsabs.harvard.edu/abs/2005MNRAS.358..363C} {358, 363}

\bibitem[\protect\citeauthoryear{{Colina} et~al.,}{{Colina}
  et~al.}{2015}]{Colina2015}
{Colina} L.,  et~al., 2015, \mn@doi [\aap] {10.1051/0004-6361/201425567}, \href
  {https://ui.adsabs.harvard.edu/abs/2015A&A...578A..48C} {578, A48}

\bibitem[\protect\citeauthoryear{{Concas}, {Popesso}, {Brusa}, {Mainieri},
  {Erfanianfar}  \& {Morselli}}{{Concas} et~al.}{2017}]{Concas2017}
{Concas} A.,  {Popesso} P.,  {Brusa} M.,  {Mainieri} V.,  {Erfanianfar} G.,
  {Morselli} L.,  2017, \mn@doi [\aap] {10.1051/0004-6361/201629519}, \href
  {https://ui.adsabs.harvard.edu/abs/2017A&A...606A..36C} {606, A36}

\bibitem[\protect\citeauthoryear{{Congiu} et~al.,}{{Congiu}
  et~al.}{2017}]{Congiu2017}
{Congiu} E.,  et~al., 2017, \mn@doi [\mnras] {10.1093/mnras/stx1628}, \href
  {https://ui.adsabs.harvard.edu/abs/2017MNRAS.471..562C} {471, 562}

\bibitem[\protect\citeauthoryear{{Currie}, {Berry}, {Jenness}, {Gibb}, {Bell}
  \& {Draper}}{{Currie} et~al.}{2014}]{Currie2014}
{Currie} M.~J.,  {Berry} D.~S.,  {Jenness} T.,  {Gibb} A.~G.,  {Bell} G.~S.,
  {Draper} P.~W.,  2014, in {Manset} N.,  {Forshay} P.,  eds,  Astronomical
  Society of the Pacific Conference Series Vol. 485, Astronomical Data Analysis
  Software and Systems XXIII. p.~391

\bibitem[\protect\citeauthoryear{{Danziger}, {Goss}  \&
  {Wellington}}{{Danziger} et~al.}{1981}]{Danziger1981}
{Danziger} I.~J.,  {Goss} W.~M.,   {Wellington} K.~J.,  1981, \mn@doi [\mnras]
  {10.1093/mnras/196.4.845}, \href
  {https://ui.adsabs.harvard.edu/abs/1981MNRAS.196..845D} {196, 845}

\bibitem[\protect\citeauthoryear{{Dasyra}, {Combes}, {Oosterloo}, {Oonk},
  {Morganti}, {Salom{\'e}}  \& {Vlahakis}}{{Dasyra} et~al.}{2016}]{Dasyra2016}
{Dasyra} K.~M.,  {Combes} F.,  {Oosterloo} T.,  {Oonk} J.~B.~R.,  {Morganti}
  R.,  {Salom{\'e}} P.,   {Vlahakis} N.,  2016, \mn@doi [A\&A]
  {10.1051/0004-6361/201629689}, \href
  {https://ui.adsabs.harvard.edu/abs/2016A&A...595L...7D} {595, L7}

\bibitem[\protect\citeauthoryear{{Davies} et~al.,}{{Davies}
  et~al.}{2020}]{Davies2020}
{Davies} R.,  et~al., 2020, \mn@doi [MNRAS] {10.1093/mnras/staa2413}, \href
  {https://ui.adsabs.harvard.edu/abs/2020MNRAS.498.4150D} {498, 4150}

\bibitem[\protect\citeauthoryear{{Di Matteo}, {Springel}  \& {Hernquist}}{{Di
  Matteo} et~al.}{2005}]{DiMatteo2005}
{Di Matteo} T.,  {Springel} V.,   {Hernquist} L.,  2005, \mn@doi [Nat]
  {10.1038/nature03335}, \href
  {https://ui.adsabs.harvard.edu/abs/2005Natur.433..604D} {433, 604}

\bibitem[\protect\citeauthoryear{{Dors}, {Riffel}, {Cardaci}, {H{\"a}gele},
  {Krabbe}, {P{\'e}rez-Montero}  \& {Rodrigues}}{{Dors}
  et~al.}{2012}]{Dors2012}
{Dors} Oli~L. J.,  {Riffel} R.~A.,  {Cardaci} M.~V.,  {H{\"a}gele} G.~F.,
  {Krabbe} {\'A}.~C.,  {P{\'e}rez-Montero} E.,   {Rodrigues} I.,  2012, \mn@doi
  [\mnras] {10.1111/j.1365-2966.2012.20600.x}, \href
  {https://ui.adsabs.harvard.edu/abs/2012MNRAS.422..252D} {422, 252}

\bibitem[\protect\citeauthoryear{{Earl} et~al.,}{{Earl}
  et~al.}{2021}]{Earl2021}
{Earl} N.,  et~al., 2021, {astropy/specutils: v1.2},
  \mn@doi{10.5281/zenodo.4603801}

\bibitem[\protect\citeauthoryear{{Fabian}}{{Fabian}}{1999}]{Fabian1999}
{Fabian} A.~C.,  1999, \mn@doi [MNRAS] {10.1046/j.1365-8711.1999.03017.x},
  \href {https://ui.adsabs.harvard.edu/abs/1999MNRAS.308L..39F} {308, L39}

\bibitem[\protect\citeauthoryear{{Ferland} et~al.,}{{Ferland}
  et~al.}{2017}]{Ferland2017}
{Ferland} G.~J.,  et~al., 2017, RMXAA, \href
  {https://ui.adsabs.harvard.edu/abs/2017RMxAA..53..385F} {53, 385}

\bibitem[\protect\citeauthoryear{{Fiore} et~al.,}{{Fiore}
  et~al.}{2017}]{Fiore2017}
{Fiore} F.,  et~al., 2017, \mn@doi [A\&A] {10.1051/0004-6361/201629478}, \href
  {https://ui.adsabs.harvard.edu/abs/2017A&A...601A.143F} {601, A143}

\bibitem[\protect\citeauthoryear{{Fosbury}, {Mebold}, {Goss}  \&
  {Dopita}}{{Fosbury} et~al.}{1978}]{Fosbury1978}
{Fosbury} R.~A.~E.,  {Mebold} U.,  {Goss} W.~M.,   {Dopita} M.~A.,  1978,
  \mn@doi [\mnras] {10.1093/mnras/183.4.549}, \href
  {https://ui.adsabs.harvard.edu/abs/1978MNRAS.183..549F} {183, 549}

\bibitem[\protect\citeauthoryear{{Freudling}, {Romaniello}, {Bramich},
  {Ballester}, {Forchi}, {Garc{\'{\i}}a-Dabl{\'o}}, {Moehler}  \&
  {Neeser}}{{Freudling} et~al.}{2013}]{Freudling2013}
{Freudling} W.,  {Romaniello} M.,  {Bramich} D.~M.,  {Ballester} P.,  {Forchi}
  V.,  {Garc{\'{\i}}a-Dabl{\'o}} C.~E.,  {Moehler} S.,   {Neeser} M.~J.,  2013,
  \mn@doi [A\&A] {10.1051/0004-6361/201322494}, \href
  {http://adsabs.harvard.edu/abs/2013A%26A...559A..96F} {559, A96}

\bibitem[\protect\citeauthoryear{{Gaibler}, {Khochfar}, {Krause}  \&
  {Silk}}{{Gaibler} et~al.}{2012}]{Gaibler2012}
{Gaibler} V.,  {Khochfar} S.,  {Krause} M.,   {Silk} J.,  2012, \mn@doi
  [\mnras] {10.1111/j.1365-2966.2012.21479.x}, \href
  {https://ui.adsabs.harvard.edu/abs/2012MNRAS.425..438G} {425, 438}

\bibitem[\protect\citeauthoryear{Harris et~al.,}{Harris
  et~al.}{2020}]{Harris2020}
Harris C.~R.,  et~al., 2020, \mn@doi [Nat.] {10.1038/s41586-020-2649-2}, 585,
  357

\bibitem[\protect\citeauthoryear{{Harrison}, {Alexander}, {Mullaney}  \&
  {Swinbank}}{{Harrison} et~al.}{2014}]{Harrison2014}
{Harrison} C.~M.,  {Alexander} D.~M.,  {Mullaney} J.~R.,   {Swinbank} A.~M.,
  2014, \mn@doi [MNRAS] {10.1093/mnras/stu515}, \href
  {https://ui.adsabs.harvard.edu/abs/2014MNRAS.441.3306H} {441, 3306}

\bibitem[\protect\citeauthoryear{{Harrison}, {Costa}, {Tadhunter},
  {Fl{\"u}tsch}, {Kakkad}, {Perna}  \& {Vietri}}{{Harrison}
  et~al.}{2018}]{Harrison2018}
{Harrison} C.~M.,  {Costa} T.,  {Tadhunter} C.~N.,  {Fl{\"u}tsch} A.,  {Kakkad}
  D.,  {Perna} M.,   {Vietri} G.,  2018, \mn@doi [Nat. Astron]
  {10.1038/s41550-018-0403-6}, \href
  {https://ui.adsabs.harvard.edu/abs/2018NatAs...2..198H} {2, 198}

\bibitem[\protect\citeauthoryear{{Heckman}}{{Heckman}}{1980}]{Heckman1980}
{Heckman} T.~M.,  1980, \aap, \href
  {https://ui.adsabs.harvard.edu/abs/1980A&A....87..152H} {87, 152}

\bibitem[\protect\citeauthoryear{{Heckman}}{{Heckman}}{2002}]{Heckman2002}
{Heckman} T.~M.,  2002, in {Mulchaey} J.~S.,  {Stocke} J.~T.,  eds,
  Astronomical Society of the Pacific Conference Series Vol. 254, Extragalactic
  Gas at Low Redshift. p.~292 (\mn@eprint {arXiv} {astro-ph/0107438})

\bibitem[\protect\citeauthoryear{{Holt}, {Tadhunter}  \& {Morganti}}{{Holt}
  et~al.}{2003}]{Holt2003}
{Holt} J.,  {Tadhunter} C.~N.,   {Morganti} R.,  2003, \mn@doi [MNRAS]
  {10.1046/j.1365-8711.2003.06532.x}, \href
  {https://ui.adsabs.harvard.edu/abs/2003MNRAS.342..227H} {342, 227}

\bibitem[\protect\citeauthoryear{{Holt}, {Tadhunter}, {Morganti}  \&
  {Emonts}}{{Holt} et~al.}{2011}]{Holt2011}
{Holt} J.,  {Tadhunter} C.~N.,  {Morganti} R.,   {Emonts} B.~H.~C.,  2011,
  \mn@doi [MNRAS] {10.1111/j.1365-2966.2010.17535.x}, \href
  {https://ui.adsabs.harvard.edu/abs/2011MNRAS.410.1527H} {410, 1527}

\bibitem[\protect\citeauthoryear{{Hopkins} \& {Elvis}}{{Hopkins} \&
  {Elvis}}{2010}]{Hopkins2010}
{Hopkins} P.~F.,  {Elvis} M.,  2010, \mn@doi [MNRAS]
  {10.1111/j.1365-2966.2009.15643.x}, \href
  {https://ui.adsabs.harvard.edu/abs/2010MNRAS.401....7H} {401, 7}

\bibitem[\protect\citeauthoryear{{Inglis}, {Brindle}, {Hough}, {Young}, {Axon},
  {Bailey}  \& {Ward}}{{Inglis} et~al.}{1993}]{Inglis1993}
{Inglis} M.~D.,  {Brindle} C.,  {Hough} J.~H.,  {Young} S.,  {Axon} D.~J.,
  {Bailey} J.~A.,   {Ward} W.~J.,  1993, \mn@doi [\mnras]
  {10.1093/mnras/263.4.895}, \href
  {https://ui.adsabs.harvard.edu/abs/1993MNRAS.263..895I} {263, 895}

\bibitem[\protect\citeauthoryear{{Kausch} et~al.,}{{Kausch}
  et~al.}{2015}]{Kausch2015}
{Kausch} W.,  et~al., 2015, \mn@doi [A\&A] {10.1051/0004-6361/201423909}, \href
  {https://ui.adsabs.harvard.edu/abs/2015A&A...576A..78K} {576, A78}

\bibitem[\protect\citeauthoryear{{Larkin}, {Armus}, {Knop}, {Soifer}  \&
  {Matthews}}{{Larkin} et~al.}{1998}]{Larkin1998}
{Larkin} J.~E.,  {Armus} L.,  {Knop} R.~A.,  {Soifer} B.~T.,   {Matthews} K.,
  1998, \mn@doi [\apjs] {10.1086/313063}, \href
  {https://ui.adsabs.harvard.edu/abs/1998ApJS..114...59L} {114, 59}

\bibitem[\protect\citeauthoryear{{Le Borgne} et~al.,}{{Le Borgne}
  et~al.}{2003}]{LeBorgne2003}
{Le Borgne} J.~F.,  et~al., 2003, \mn@doi [\aap] {10.1051/0004-6361:20030243},
  \href {https://ui.adsabs.harvard.edu/abs/2003A&A...402..433L} {402, 433}

\bibitem[\protect\citeauthoryear{{Liu}, {Zakamska}, {Greene}, {Nesvadba}  \&
  {Liu}}{{Liu} et~al.}{2013}]{Liu2013}
{Liu} G.,  {Zakamska} N.~L.,  {Greene} J.~E.,  {Nesvadba} N. P.~H.,   {Liu} X.,
   2013, \mn@doi [\mnras] {10.1093/mnras/stt1755}, \href
  {https://ui.adsabs.harvard.edu/abs/2013MNRAS.436.2576L} {436, 2576}

\bibitem[\protect\citeauthoryear{{Maddox}, {Hess}, {Obreschkow}, {Jarvis}  \&
  {Blyth}}{{Maddox} et~al.}{2015}]{Maddox2015}
{Maddox} N.,  {Hess} K.~M.,  {Obreschkow} D.,  {Jarvis} M.~J.,   {Blyth} S.~L.,
   2015, \mn@doi [\mnras] {10.1093/mnras/stu2532}, \href
  {https://ui.adsabs.harvard.edu/abs/2015MNRAS.447.1610M} {447, 1610}

\bibitem[\protect\citeauthoryear{{Maksym} et~al.,}{{Maksym}
  et~al.}{2021}]{Maksym2020}
{Maksym} W.~P.,  et~al., 2021, \mn@doi [\apj] {10.3847/1538-4357/ac0976}, \href
  {https://ui.adsabs.harvard.edu/abs/2021ApJ...917...85M} {917, 85}

\bibitem[\protect\citeauthoryear{{Mateus}, {Sodr{\'e}}, {Cid Fernandes},
  {Stasi{\'n}ska}, {Schoenell}  \& {Gomes}}{{Mateus} et~al.}{2006}]{Mateus2006}
{Mateus} A.,  {Sodr{\'e}} L.,  {Cid Fernandes} R.,  {Stasi{\'n}ska} G.,
  {Schoenell} W.,   {Gomes} J.~M.,  2006, \mn@doi [MNRAS]
  {10.1111/j.1365-2966.2006.10565.x}, \href
  {https://ui.adsabs.harvard.edu/abs/2006MNRAS.370..721M} {370, 721}

\bibitem[\protect\citeauthoryear{{Meenakshi}, {Mukherjee}, {Wagner},
  {Nesvadba}, {Morganti}, {Janssen}  \& {Bicknell}}{{Meenakshi}
  et~al.}{2022}]{Meenakshi2022}
{Meenakshi} M.,  {Mukherjee} D.,  {Wagner} A.~Y.,  {Nesvadba} N. P.~H.,
  {Morganti} R.,  {Janssen} R. M.~J.,   {Bicknell} G.~V.,  2022, \mn@doi
  [\mnras] {10.1093/mnras/stac167}, \href
  {https://ui.adsabs.harvard.edu/abs/2022MNRAS.511.1622M} {511, 1622}

\bibitem[\protect\citeauthoryear{{Mingozzi} et~al.,}{{Mingozzi}
  et~al.}{2019}]{Mingozzi2019}
{Mingozzi} M.,  et~al., 2019, \mn@doi [\aap] {10.1051/0004-6361/201834372},
  \href {https://ui.adsabs.harvard.edu/abs/2019A&A...622A.146M} {622, A146}

\bibitem[\protect\citeauthoryear{Montgomery}{Montgomery}{2012}]{Montgomery2012}
Montgomery D.~C.,  2012, Introduction to linear regression analysis, fifth
  edition. edn.
Wiley series in probability and statistics, John Wiley $\&$ Sons Ltd, Hoboken,
  New Jersey

\bibitem[\protect\citeauthoryear{{Morganti}, {Oosterloo}  \&
  {Tsvetanov}}{{Morganti} et~al.}{1998}]{Morganti1998}
{Morganti} R.,  {Oosterloo} T.,   {Tsvetanov} Z.,  1998, \mn@doi [ApJ]
  {10.1086/300236}, \href
  {https://ui.adsabs.harvard.edu/abs/1998AJ....115..915M} {115, 915}

\bibitem[\protect\citeauthoryear{{Morganti}, {Tadhunter}  \&
  {Oosterloo}}{{Morganti} et~al.}{2005}]{Morganti2005}
{Morganti} R.,  {Tadhunter} C.~N.,   {Oosterloo} T.~A.,  2005, \mn@doi [A\&A]
  {10.1051/0004-6361:200500197}, \href
  {https://ui.adsabs.harvard.edu/abs/2005A&A...444L...9M} {444, L9}

\bibitem[\protect\citeauthoryear{{Morganti}, {Holt}, {Saripalli}, {Oosterloo}
  \& {Tadhunter}}{{Morganti} et~al.}{2007}]{Morganti2007}
{Morganti} R.,  {Holt} J.,  {Saripalli} L.,  {Oosterloo} T.~A.,   {Tadhunter}
  C.~N.,  2007, \mn@doi [A\&A] {10.1051/0004-6361:20077888}, \href
  {https://ui.adsabs.harvard.edu/abs/2007A&A...476..735M} {476, 735}

\bibitem[\protect\citeauthoryear{{Morganti}, {Frieswijk}, {Oonk}, {Oosterloo}
  \& {Tadhunter}}{{Morganti} et~al.}{2013}]{Morganti2013}
{Morganti} R.,  {Frieswijk} W.,  {Oonk} R.~J.~B.,  {Oosterloo} T.,
  {Tadhunter} C.,  2013, \mn@doi [A\&A] {10.1051/0004-6361/201220734}, \href
  {https://ui.adsabs.harvard.edu/abs/2013A&A...552L...4M} {552, L4}

\bibitem[\protect\citeauthoryear{{Morganti}, {Oosterloo}, {Oonk}, {Frieswijk}
  \& {Tadhunter}}{{Morganti} et~al.}{2015}]{Morganti2015}
{Morganti} R.,  {Oosterloo} T.,  {Oonk} J.~B.~R.,  {Frieswijk} W.,
  {Tadhunter} C.,  2015, \mn@doi [A\&A] {10.1051/0004-6361/201525860}, \href
  {https://ui.adsabs.harvard.edu/abs/2015A&A...580A...1M} {580, A1}

\bibitem[\protect\citeauthoryear{{Mukherjee}, {Wagner}, {Bicknell}, {Morganti},
  {Oosterloo}, {Nesvadba}  \& {Sutherland}}{{Mukherjee}
  et~al.}{2018}]{Mukherjee2018}
{Mukherjee} D.,  {Wagner} A.~Y.,  {Bicknell} G.~V.,  {Morganti} R.,
  {Oosterloo} T.,  {Nesvadba} N.,   {Sutherland} R.~S.,  2018, \mn@doi [MNRAS]
  {10.1093/mnras/sty067}, \href
  {https://ui.adsabs.harvard.edu/abs/2018MNRAS.476...80M} {476, 80}

\bibitem[\protect\citeauthoryear{{Mullaney}, {Alexander}, {Fine}, {Goulding},
  {Harrison}  \& {Hickox}}{{Mullaney} et~al.}{2013}]{Mullaney2013}
{Mullaney} J.~R.,  {Alexander} D.~M.,  {Fine} S.,  {Goulding} A.~D.,
  {Harrison} C.~M.,   {Hickox} R.~C.,  2013, \mn@doi [MNRAS]
  {10.1093/mnras/stt751}, \href
  {https://ui.adsabs.harvard.edu/abs/2013MNRAS.433..622M} {433, 622}

\bibitem[\protect\citeauthoryear{{Nesvadba}, {Lehnert}, {Eisenhauer},
  {Gilbert}, {Tecza}  \& {Abuter}}{{Nesvadba} et~al.}{2006}]{Nesvadba2006}
{Nesvadba} N.~P.~H.,  {Lehnert} M.~D.,  {Eisenhauer} F.,  {Gilbert} A.,
  {Tecza} M.,   {Abuter} R.,  2006, \mn@doi [ApJ] {10.1086/507266}, \href
  {https://ui.adsabs.harvard.edu/abs/2006ApJ...650..693N} {650, 693}

\bibitem[\protect\citeauthoryear{{Nicastro}, {Martocchia}  \&
  {Matt}}{{Nicastro} et~al.}{2003}]{Nicastro2003}
{Nicastro} F.,  {Martocchia} A.,   {Matt} G.,  2003, \mn@doi [ApJ]
  {10.1086/375715}, \href
  {https://ui.adsabs.harvard.edu/abs/2003ApJ...589L..13N} {589, L13}

\bibitem[\protect\citeauthoryear{{Oosterloo}, {Morganti}, {Tzioumis},
  {Reynolds}, {King}, {McCulloch}  \& {Tsvetanov}}{{Oosterloo}
  et~al.}{2000}]{Oosterloo2000}
{Oosterloo} T.~A.,  {Morganti} R.,  {Tzioumis} A.,  {Reynolds} J.,  {King} E.,
  {McCulloch} P.,   {Tsvetanov} Z.,  2000, \mn@doi [ApJ] {10.1086/301358},
  \href {https://ui.adsabs.harvard.edu/abs/2000AJ....119.2085O} {119, 2085}

\bibitem[\protect\citeauthoryear{{Oosterloo}, {Raymond Oonk}, {Morganti},
  {Combes}, {Dasyra}, {Salom{\'e}}, {Vlahakis}  \& {Tadhunter}}{{Oosterloo}
  et~al.}{2017}]{Oosterloo2017}
{Oosterloo} T.,  {Raymond Oonk} J.~B.,  {Morganti} R.,  {Combes} F.,  {Dasyra}
  K.,  {Salom{\'e}} P.,  {Vlahakis} N.,   {Tadhunter} C.,  2017, \mn@doi [A\&A]
  {10.1051/0004-6361/201731781}, \href
  {https://ui.adsabs.harvard.edu/abs/2017A&A...608A..38O} {608, A38}

\bibitem[\protect\citeauthoryear{{Osterbrock} \& {Ferland}}{{Osterbrock} \&
  {Ferland}}{2006}]{Osterbrock2006}
{Osterbrock} D.~E.,  {Ferland} G.~J.,  2006, {Astrophysics of gaseous nebulae
  and active galactic nuclei}

\bibitem[\protect\citeauthoryear{{Riffel}, {Rodr{\'\i}guez-Ardila}, {Aleman},
  {Brotherton}, {Pastoriza}, {Bonatto}  \& {Dors}}{{Riffel}
  et~al.}{2013}]{Riffel2013a}
{Riffel} R.,  {Rodr{\'\i}guez-Ardila} A.,  {Aleman} I.,  {Brotherton} M.~S.,
  {Pastoriza} M.~G.,  {Bonatto} C.,   {Dors} O.~L.,  2013, \mn@doi [\mnras]
  {10.1093/mnras/stt026}, \href
  {https://ui.adsabs.harvard.edu/abs/2013MNRAS.430.2002R} {430, 2002}

\bibitem[\protect\citeauthoryear{{Riffel}, {Bianchin}, {Riffel},
  {Storchi-Bergmann}, {Sch{\"o}nell}, {Dahmer-Hahn}, {Dametto}  \&
  {Diniz}}{{Riffel} et~al.}{2021}]{Riffel2021}
{Riffel} R.~A.,  {Bianchin} M.,  {Riffel} R.,  {Storchi-Bergmann} T.,
  {Sch{\"o}nell} A.~J.,  {Dahmer-Hahn} L.~G.,  {Dametto} N.~Z.,   {Diniz}
  M.~R.,  2021, \mn@doi [\mnras] {10.1093/mnras/stab788}, \href
  {https://ui.adsabs.harvard.edu/abs/2021MNRAS.503.5161R} {503, 5161}

\bibitem[\protect\citeauthoryear{{Rodr{\'\i}guez-Ardila}, {Riffel}  \&
  {Pastoriza}}{{Rodr{\'\i}guez-Ardila} et~al.}{2005}]{Rodriguez-Ardila2005}
{Rodr{\'\i}guez-Ardila} A.,  {Riffel} R.,   {Pastoriza} M.~G.,  2005, \mn@doi
  [\mnras] {10.1111/j.1365-2966.2005.09638.x}, \href
  {https://ui.adsabs.harvard.edu/abs/2005MNRAS.364.1041R} {364, 1041}

\bibitem[\protect\citeauthoryear{{Rodr{\'\i}guez Zaur{\'\i}n}, {Tadhunter},
  {Rose}  \& {Holt}}{{Rodr{\'\i}guez Zaur{\'\i}n} et~al.}{2013}]{Zaurin2013}
{Rodr{\'\i}guez Zaur{\'\i}n} J.,  {Tadhunter} C.~N.,  {Rose} M.,   {Holt} J.,
  2013, \mn@doi [MNRAS] {10.1093/mnras/stt423}, \href
  {https://ui.adsabs.harvard.edu/abs/2013MNRAS.432..138R} {432, 138}

\bibitem[\protect\citeauthoryear{{Rose}, {Tadhunter}, {Ramos Almeida},
  {Rodr{\'\i}guez Zaur{\'\i}n}, {Santoro}  \& {Spence}}{{Rose}
  et~al.}{2018}]{Rose2018}
{Rose} M.,  {Tadhunter} C.,  {Ramos Almeida} C.,  {Rodr{\'\i}guez Zaur{\'\i}n}
  J.,  {Santoro} F.,   {Spence} R.,  2018, \mn@doi [MNRAS]
  {10.1093/mnras/stx2590}, \href
  {https://ui.adsabs.harvard.edu/abs/2018MNRAS.474..128R} {474, 128}

\bibitem[\protect\citeauthoryear{{Rupke}, {Veilleux}  \& {Sanders}}{{Rupke}
  et~al.}{2002}]{Rupke2002}
{Rupke} D.~S.,  {Veilleux} S.,   {Sanders} D.~B.,  2002, \mn@doi [\apj]
  {10.1086/339789}, \href
  {https://ui.adsabs.harvard.edu/abs/2002ApJ...570..588R} {570, 588}

\bibitem[\protect\citeauthoryear{{Rupke}, {Veilleux}  \& {Sanders}}{{Rupke}
  et~al.}{2005}]{Rupke2005}
{Rupke} D.~S.,  {Veilleux} S.,   {Sanders} D.~B.,  2005, \mn@doi [ApJ]
  {10.1086/444451}, \href
  {https://ui.adsabs.harvard.edu/abs/2005ApJ...632..751R} {632, 751}

\bibitem[\protect\citeauthoryear{{Santoro}, {Rose}, {Morganti}, {Tadhunter},
  {Oosterloo}  \& {Holt}}{{Santoro} et~al.}{2018}]{Santoro2018}
{Santoro} F.,  {Rose} M.,  {Morganti} R.,  {Tadhunter} C.,  {Oosterloo} T.~A.,
   {Holt} J.,  2018, \mn@doi [A\&A] {10.1051/0004-6361/201833248}, \href
  {https://ui.adsabs.harvard.edu/abs/2018A&A...617A.139S} {617, A139}

\bibitem[\protect\citeauthoryear{{Santoro}, {Tadhunter}, {Baron}, {Morganti}
  \& {Holt}}{{Santoro} et~al.}{2020}]{Santoro2020}
{Santoro} F.,  {Tadhunter} C.,  {Baron} D.,  {Morganti} R.,   {Holt} J.,  2020,
  \mn@doi [A\&A] {10.1051/0004-6361/202039077}, \href
  {https://ui.adsabs.harvard.edu/abs/2020A&A...644A..54S} {644, A54}

\bibitem[\protect\citeauthoryear{{Schlafly} \& {Finkbeiner}}{{Schlafly} \&
  {Finkbeiner}}{2011}]{Schlafly2011}
{Schlafly} E.~F.,  {Finkbeiner} D.~P.,  2011, \mn@doi [ApJ]
  {10.1088/0004-637X/737/2/103}, \href
  {https://ui.adsabs.harvard.edu/abs/2011ApJ...737..103S} {737, 103}

\bibitem[\protect\citeauthoryear{{Schlegel}, {Finkbeiner}  \&
  {Davis}}{{Schlegel} et~al.}{1998}]{Schlegel1998}
{Schlegel} D.~J.,  {Finkbeiner} D.~P.,   {Davis} M.,  1998, \mn@doi [ApJ]
  {10.1086/305772}, \href
  {https://ui.adsabs.harvard.edu/abs/1998ApJ...500..525S} {500, 525}

\bibitem[\protect\citeauthoryear{{Sharp} \& {Bland-Hawthorn}}{{Sharp} \&
  {Bland-Hawthorn}}{2010}]{Sharp2010}
{Sharp} R.~G.,  {Bland-Hawthorn} J.,  2010, \mn@doi [\apj]
  {10.1088/0004-637X/711/2/818}, \href
  {https://ui.adsabs.harvard.edu/abs/2010ApJ...711..818S} {711, 818}

\bibitem[\protect\citeauthoryear{{Shaw} \& {Dufour}}{{Shaw} \&
  {Dufour}}{1995}]{Shaw1995}
{Shaw} R.~A.,  {Dufour} R.~J.,  1995, \mn@doi [PASP] {10.1086/133637}, \href
  {https://ui.adsabs.harvard.edu/abs/1995PASP..107..896S} {107, 896}

\bibitem[\protect\citeauthoryear{{Silk} \& {Rees}}{{Silk} \&
  {Rees}}{1998}]{Silk1998}
{Silk} J.,  {Rees} M.~J.,  1998, A\&A, \href
  {https://ui.adsabs.harvard.edu/abs/1998A&A...331L...1S} {331, L1}

\bibitem[\protect\citeauthoryear{{Smette} et~al.,}{{Smette}
  et~al.}{2015}]{Smette2015}
{Smette} A.,  et~al., 2015, \mn@doi [A\&A] {10.1051/0004-6361/201423932}, \href
  {https://ui.adsabs.harvard.edu/abs/2015A&A...576A..77S} {576, A77}

\bibitem[\protect\citeauthoryear{{Spence}, {Tadhunter}, {Rose}  \&
  {Rodr{\'\i}guez Zaur{\'\i}n}}{{Spence} et~al.}{2018}]{Spence2018}
{Spence} R.~A.~W.,  {Tadhunter} C.~N.,  {Rose} M.,   {Rodr{\'\i}guez
  Zaur{\'\i}n} J.,  2018, \mn@doi [MNRAS] {10.1093/mnras/sty1046}, \href
  {https://ui.adsabs.harvard.edu/abs/2018MNRAS.478.2438S} {478, 2438}

\bibitem[\protect\citeauthoryear{{Springel}, {Di Matteo}  \&
  {Hernquist}}{{Springel} et~al.}{2005}]{Springel2005}
{Springel} V.,  {Di Matteo} T.,   {Hernquist} L.,  2005, \mn@doi [MNRAS]
  {10.1111/j.1365-2966.2005.09238.x}, \href
  {https://ui.adsabs.harvard.edu/abs/2005MNRAS.361..776S} {361, 776}

\bibitem[\protect\citeauthoryear{{Sun}, {Greene}  \& {Zakamska}}{{Sun}
  et~al.}{2017}]{Sun2017}
{Sun} A.-L.,  {Greene} J.~E.,   {Zakamska} N.~L.,  2017, \mn@doi [ApJ]
  {10.3847/1538-4357/835/2/222}, \href
  {https://ui.adsabs.harvard.edu/abs/2017ApJ...835..222S} {835, 222}

\bibitem[\protect\citeauthoryear{{Sutherland} \& {Dopita}}{{Sutherland} \&
  {Dopita}}{2017}]{Sutherland2017}
{Sutherland} R.~S.,  {Dopita} M.~A.,  2017, \mn@doi [\apjs]
  {10.3847/1538-4365/aa6541}, \href
  {https://ui.adsabs.harvard.edu/abs/2017ApJS..229...34S} {229, 34}

\bibitem[\protect\citeauthoryear{{Tadhunter}, {Morganti}, {Rose}, {Oonk}  \&
  {Oosterloo}}{{Tadhunter} et~al.}{2014}]{Tadhunter2014}
{Tadhunter} C.,  {Morganti} R.,  {Rose} M.,  {Oonk} J.~B.~R.,   {Oosterloo} T.,
   2014, \mn@doi [Nat.] {10.1038/nature13520}, \href
  {https://ui.adsabs.harvard.edu/abs/2014Natur.511..440T} {511, 440}

\bibitem[\protect\citeauthoryear{{Tadhunter}, {Holden}, {Ramos Almeida}  \&
  {Batcheldor}}{{Tadhunter} et~al.}{2019}]{Tadhunter2019}
{Tadhunter} C.,  {Holden} L.,  {Ramos Almeida} C.,   {Batcheldor} D.,  2019,
  \mn@doi [MNRAS] {10.1093/mnras/stz1755}, \href
  {https://ui.adsabs.harvard.edu/abs/2019MNRAS.488.1813T} {488, 1813}

\bibitem[\protect\citeauthoryear{{Tazaki}, {Ueda}, {Terashima}  \&
  {Mushotzky}}{{Tazaki} et~al.}{2011}]{Tazaki2011}
{Tazaki} F.,  {Ueda} Y.,  {Terashima} Y.,   {Mushotzky} R.~F.,  2011, \mn@doi
  [\apj] {10.1088/0004-637X/738/1/70}, \href
  {https://ui.adsabs.harvard.edu/abs/2011ApJ...738...70T} {738, 70}

\bibitem[\protect\citeauthoryear{{Tody}}{{Tody}}{1986}]{Tody1986}
{Tody} D.,  1986, in {Crawford} D.~L.,  ed.,  Society of Photo-Optical
  Instrumentation Engineers (SPIE) Conference Series Vol. 627, Instrumentation
  in astronomy VI. p.~733, \mn@doi{10.1117/12.968154}

\bibitem[\protect\citeauthoryear{{Tody}}{{Tody}}{1993}]{Tody1993}
{Tody} D.,  1993, in {Hanisch} R.~J.,  {Brissenden} R.~J.~V.,   {Barnes} J.,
  eds,  Astronomical Society of the Pacific Conference Series Vol. 52,
  Astronomical Data Analysis Software and Systems II. p.~173

\bibitem[\protect\citeauthoryear{{Travascio}, {Fabbiano}, {Paggi}, {Elvis},
  {Maksym}, {Morganti}, {Oosterloo}  \& {Fiore}}{{Travascio}
  et~al.}{2021}]{Travascio2021}
{Travascio} A.,  {Fabbiano} G.,  {Paggi} A.,  {Elvis} M.,  {Maksym} W.~P.,
  {Morganti} R.,  {Oosterloo} T.,   {Fiore} F.,  2021, \mn@doi [\apj]
  {10.3847/1538-4357/ac18c7}, \href
  {https://ui.adsabs.harvard.edu/abs/2021ApJ...921..129T} {921, 129}

\bibitem[\protect\citeauthoryear{{Tremonti} et~al.,}{{Tremonti}
  et~al.}{2004}]{Tremonti2004}
{Tremonti} C.~A.,  et~al., 2004, \mn@doi [\apj] {10.1086/423264}, \href
  {https://ui.adsabs.harvard.edu/abs/2004ApJ...613..898T} {613, 898}

\bibitem[\protect\citeauthoryear{{Venturi} et~al.,}{{Venturi}
  et~al.}{2021}]{Venturi2021}
{Venturi} G.,  et~al., 2021, \mn@doi [\aap] {10.1051/0004-6361/202039869},
  \href {https://ui.adsabs.harvard.edu/abs/2021A&A...648A..17V} {648, A17}

\bibitem[\protect\citeauthoryear{{Vignali}, {Comastri}, {Cappi}  \&
  {Palumbo}}{{Vignali} et~al.}{1997}]{Vignali1997}
{Vignali} C.,  {Comastri} A.,  {Cappi} M.,   {Palumbo} G.~G.~C.,  1997,
  \memsai, \href {https://ui.adsabs.harvard.edu/abs/1997MmSAI..68..139V} {68,
  139}

\bibitem[\protect\citeauthoryear{{Villar-Mart{\'\i}n}, {Tadhunter}, {Morganti},
  {Axon}  \& {Koekemoer}}{{Villar-Mart{\'\i}n} et~al.}{1999}]{VillarMartin1999}
{Villar-Mart{\'\i}n} M.,  {Tadhunter} C.,  {Morganti} R.,  {Axon} D.,
  {Koekemoer} A.,  1999, \mn@doi [MNRAS] {10.1046/j.1365-8711.1999.02603.x},
  \href {https://ui.adsabs.harvard.edu/abs/1999MNRAS.307...24V} {307, 24}

\bibitem[\protect\citeauthoryear{{Wagner} \& {Bicknell}}{{Wagner} \&
  {Bicknell}}{2011}]{Wagner2011}
{Wagner} A.~Y.,  {Bicknell} G.~V.,  2011, \mn@doi [ApJ]
  {10.1088/0004-637X/728/1/29}, \href
  {https://ui.adsabs.harvard.edu/abs/2011ApJ...728...29W} {728, 29}

\bibitem[\protect\citeauthoryear{{Walker}}{{Walker}}{1968}]{Walker1968}
{Walker} M.~F.,  1968, \mn@doi [ApJ] {10.1086/149420}, \href
  {https://ui.adsabs.harvard.edu/abs/1968ApJ...151...71W} {151, 71}

\bibitem[\protect\citeauthoryear{{Wang}, {Liu}, {Zhang}  \& {Barlow}}{{Wang}
  et~al.}{2004}]{Wang2004}
{Wang} W.,  {Liu} X.~W.,  {Zhang} Y.,   {Barlow} M.~J.,  2004, \mn@doi [A\&A]
  {10.1051/0004-6361:20041470}, \href
  {https://ui.adsabs.harvard.edu/abs/2004A&A...427..873W} {427, 873}

\bibitem[\protect\citeauthoryear{{Zubovas} \& {King}}{{Zubovas} \&
  {King}}{2014}]{Zubovas2014}
{Zubovas} K.,  {King} A.~R.,  2014, \mn@doi [MNRAS] {10.1093/mnras/stt2472},
  \href {https://ui.adsabs.harvard.edu/abs/2014MNRAS.439..400Z} {439, 400}

\bibitem[\protect\citeauthoryear{pandas~development team}{pandas~development
  team}{2020}]{reback2020pandas}
pandas~development team T.,  2020, pandas-dev/pandas: Pandas,
  \mn@doi{10.5281/zenodo.3509134}, \url
  {https://doi.org/10.5281/zenodo.3509134}

\makeatother
\end{thebibliography}



\appendix

\section{STARLIGHT stellar continuum modelling}
\label{section: starlight_appendix}

\begin{figure}
    \includegraphics[width=\linewidth]{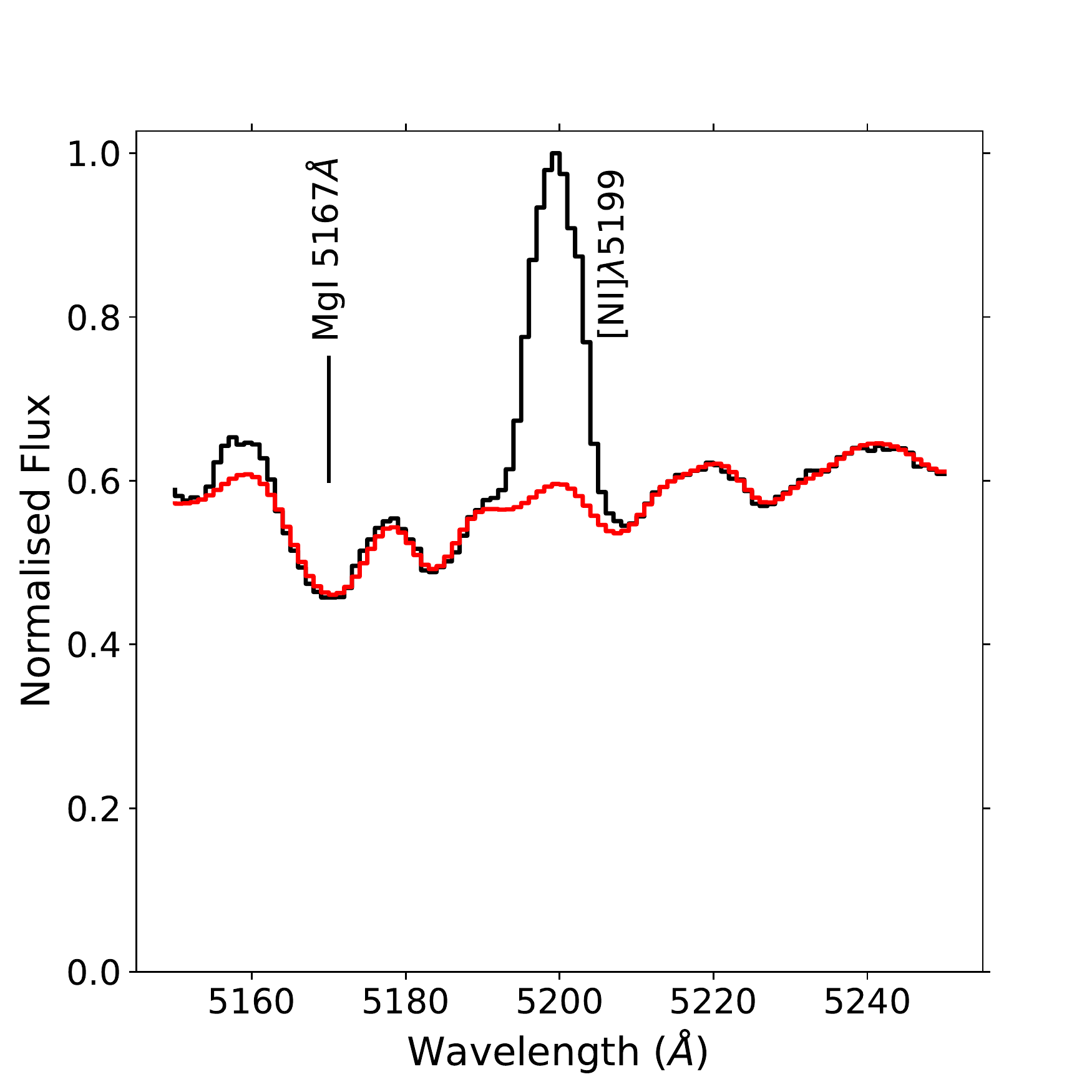}
    \caption{\textsc{STARLIGHT} fit (red solid line) to the stellar continua in the spectral region of the MgI 5167\;{\AA} absorption feature and the [NI]$\lambda$5199 emission line present in Aperture 3.}
    \label{fig: starlight_mgi_ni}
\end{figure}

\begin{figure}
    \includegraphics[width=\linewidth]{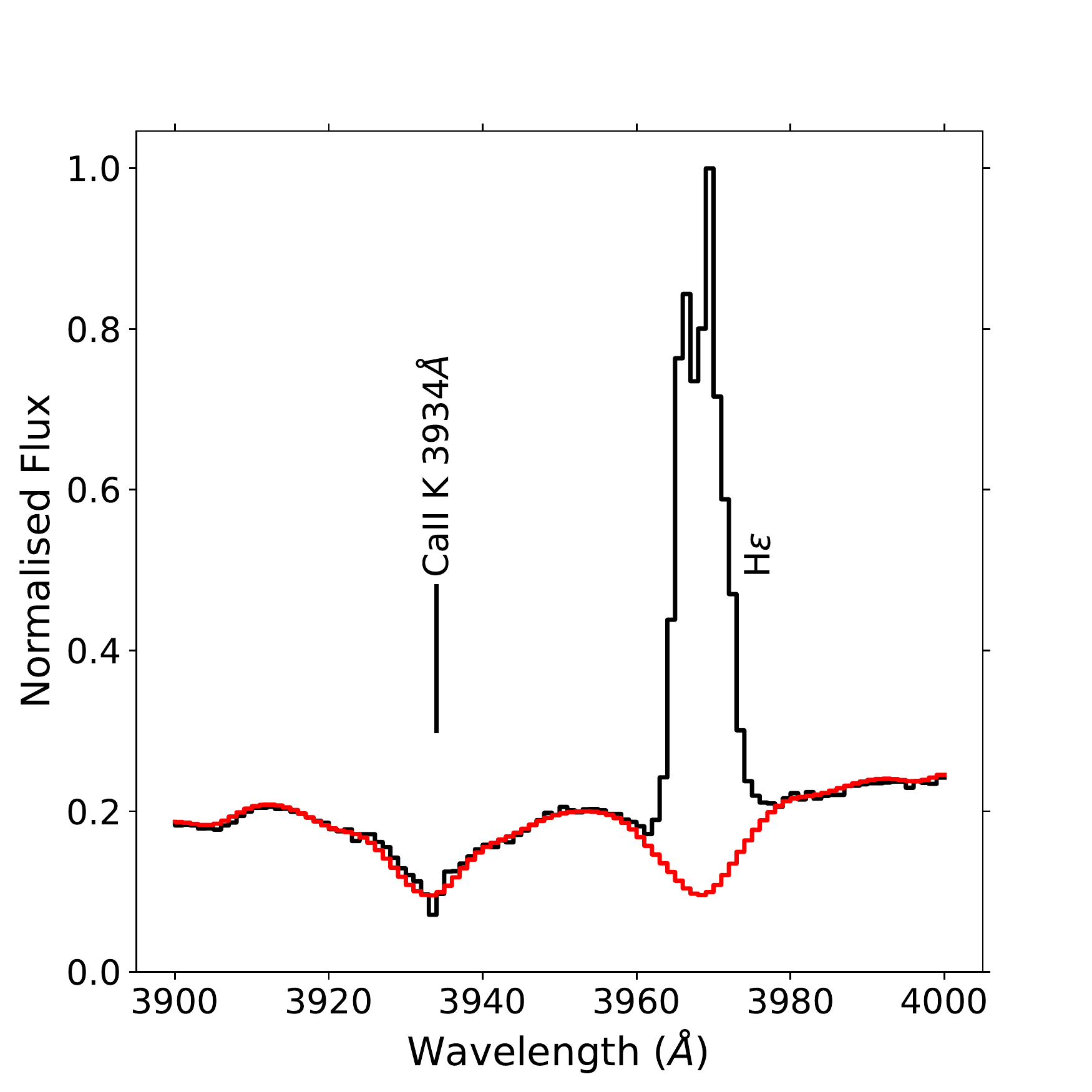}
    \caption{\textsc{STARLIGHT} fit to the stellar continua in the spectral region of the CaII K absorption feature at 3934\;{\AA}, and the H$\mathrm{\varepsilon}$ recombination line in Aperture 3.}
    \label{fig: starlight_caii_hk_heta}
\end{figure}

\begin{figure}
    \includegraphics[width=\linewidth]{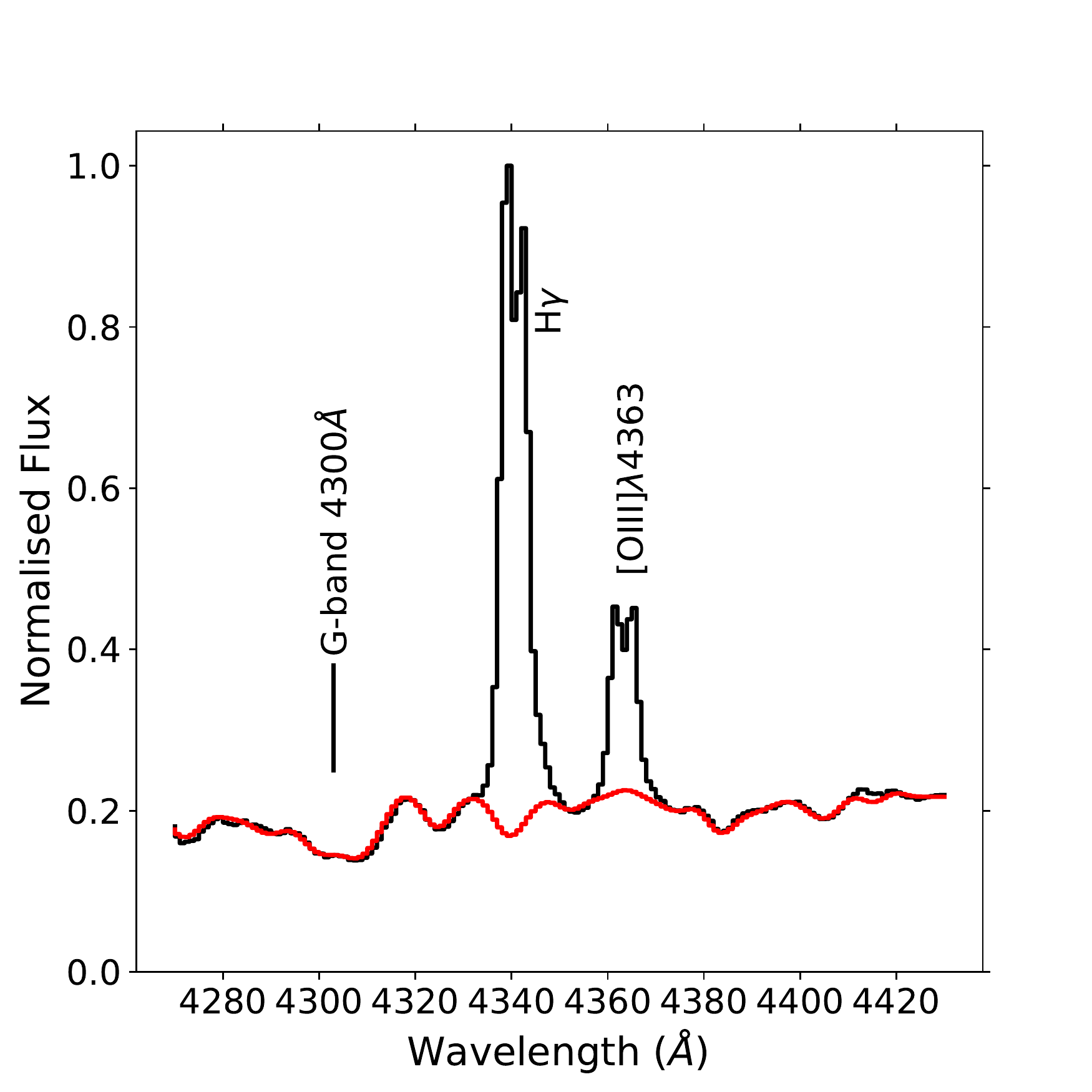}
    \caption{\textsc{STARLIGHT} fit to the stellar continua in the spectral region of the H$\mathrm{\gamma}$ and [OIII]$\lambda$4363 emission lines Aperture 3. The fit to the G-band stellar absorption feature at $\sim$4300\;{\AA} can also be seen. Complex stellar continuum structure, which may significantly affect the derived line fluxes, can be seen underneath the lines.}
    \label{fig: starlight_hgamma_oiii}
\end{figure}

\begin{figure}
    \includegraphics[width=\linewidth]{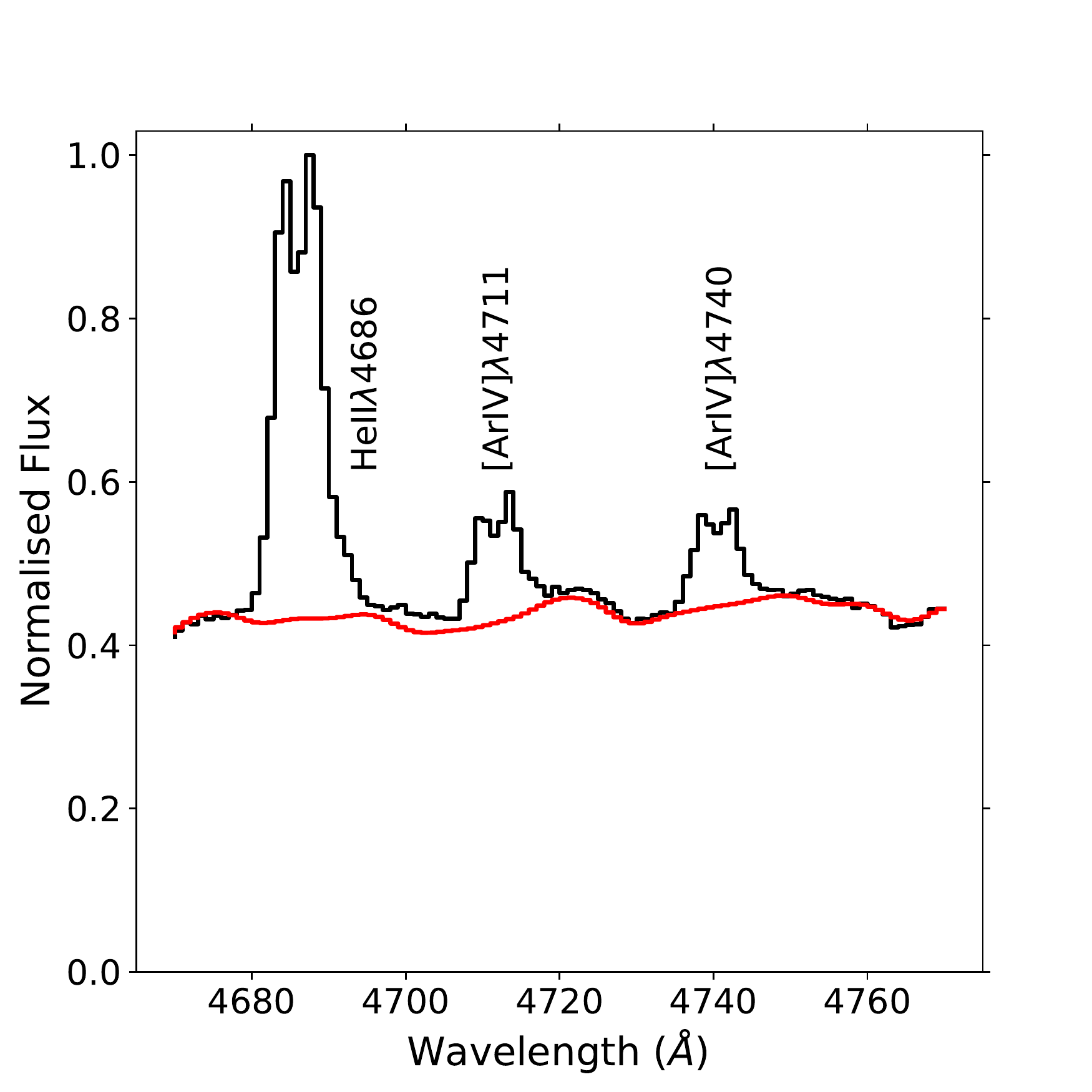}
    \caption{\textsc{STARLIGHT} fit to the stellar continua in the spectral region of the HeII$\lambda$4686 and [ArIV]$\lambda\lambda$4711,4740 emission lines in Aperture 3. The stellar continuum contributes significantly to the profiles of the [ArIV] lines seen here, thus proper modelling and subtraction was necessary to ensure accurate measurements of line fluxes.}
    \label{fig: starlight_heii_ariv}
\end{figure}

In this section, we discuss results from the \textsc{STARLIGHT} stellar continua modelling described in Section \ref{section: starlight}. Accurate subtraction of the stellar continua present in our Xshooter apertures was essential for accurate measurement of line fluxes: failing to do so may have had significant effects on derived line fluxes and ratios due to stellar absorption and emission underneath key diagnostic lines. Therefore, as an example, we show here the \textsc{STARLIGHT} fits in the spectral regions of the MgI absorption feature at 5167\;{\AA} (Figure \ref{fig: starlight_mgi_ni}) and the CaII K absorption feature at 3934\;{\AA} (Figure \ref{fig: starlight_caii_hk_heta}) for Aperture 3. These lines and features were used to check the adequacy of the \textsc{STARLIGHT} fits in each aperture. Furthermore, we show the stellar continuum fits to the H$\mathrm{\gamma}$ and [OIII]$\lambda$4363 emission lines (Figure \ref{fig: starlight_hgamma_oiii}) and the HeII and [ArIV]$\lambda\lambda4711,4740$ emission lines (Figure \ref{fig: starlight_heii_ariv}) --- these demonstrate the contribution of stellar light to the line profiles of several key emission lines that used are in our analysis, and highlight the necessity of proper continuum subtraction.

\section{AGN photoionisation modelling}
\label{section: photoionisation_modelling}
 
As noted in Section \ref{section: electron_temperatures}, the modelled values of the [OIII](5007/4363) and HeII$\lambda$4686/H$\beta$ ratios will depend on the electron density and metallicity of the gas and the spectral index of the AGN continuum, as well as the ionisation parameter. In order to investigate the extent of these effects, we present different photoionisation models here. Our principal goal was to determine if there exists a reasonable set of parameters which can explain our measured ratios as being entirely due to AGN photoionisation, which would remove the need to include a contribution from shocks.

The $\textsc{Cloudy}$ photoionisation code was used for this purpose. We set up the models in the same way as described in Section \ref{section: tr_lines}, used to produce grids for the transauroral line ratios. Single-slab, plane-parallel, radiation bounded models of dust-free photoionised gas with an electron density of $n_\mathrm{e}$=10$^3$\;cm$^{-3}$ were produced for varying metallicities, spectral indices and ionisation parameters.

\begin{figure}
    \includegraphics[width = \linewidth]{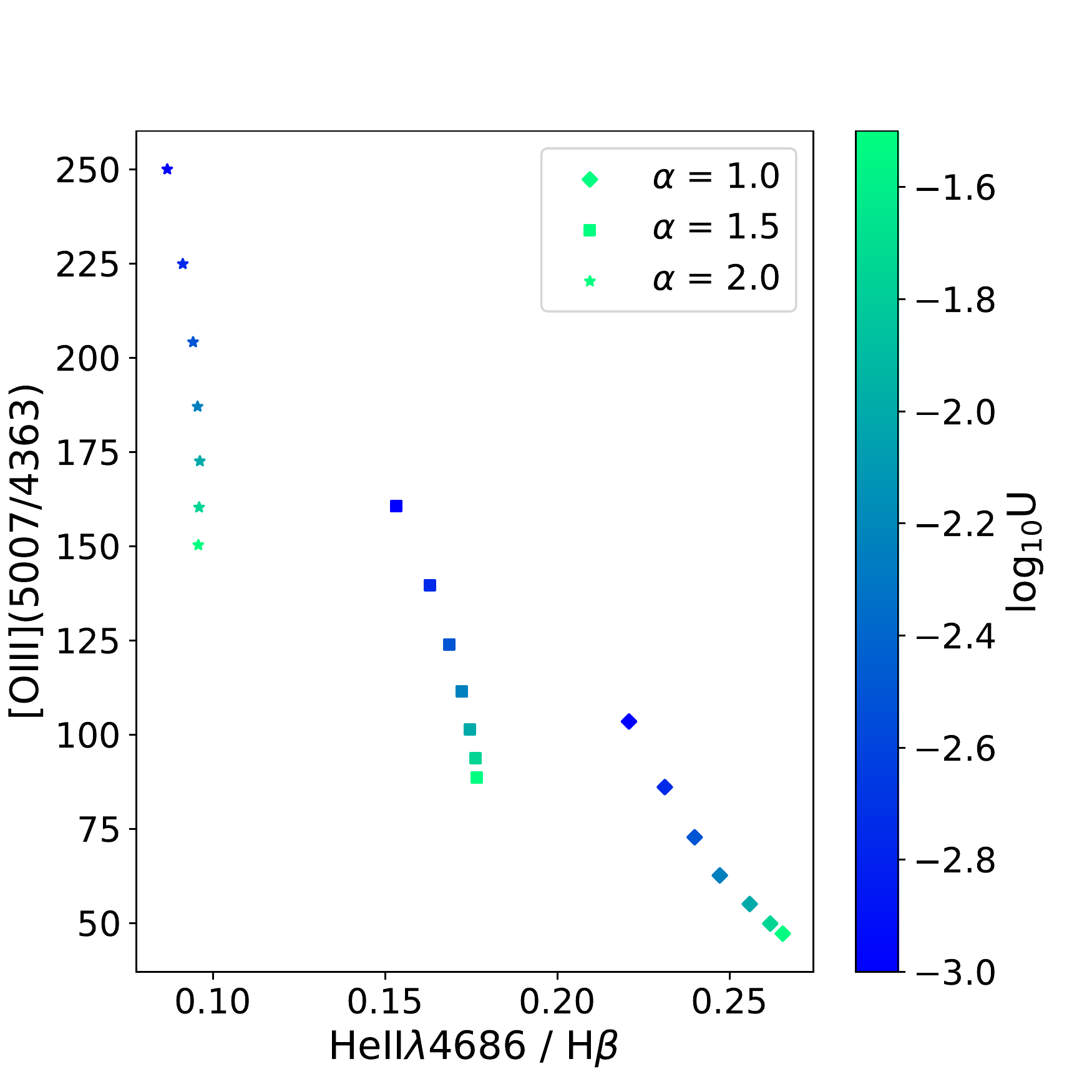}
    \caption{HeII$\lambda$4686/H$\mathrm{\beta}$ vs [OIII]$\lambda$5007/4363 line ratios from photoionisation modelling of a solar-composition gas with plane-parallel geometry, a density of n$_\mathrm{e}$=10$^3$\;cm$^{-3}$ and an ionising source of varying spectral index and ionisation parameter. Spectral indices of 1.0, 1.5 and 2.0 are shown by diamonds, squares and stars respectively, and the variation in ionisation parameter is shown by the colour bar (ranging between \mbox{$-$3.0\;\textless\;log$_{10}U$\;\textless\;$-$1.5} in steps of 0.25). Note that the axis limits are different than those used in Figures \ref{fig: heii_hbeta}, \ref{fig: heii_hbeta_z} and \ref{fig: heii_hbeta_ne}.}
    \label{fig: heii_hbeta_alpha}
\end{figure}

In Figure \ref{fig: heii_hbeta_alpha} we present the modelled ratios for a solar-composition gas of density n$_\mathrm{e}$=10$^3$\;cm$^{-3}$ with three different the spectral indices of the AGN-photoionising continuum ($\alpha$ = 1.0, 1.5 and 2.0, each marked with different symbols) and ionisation parameters ranging between \mbox{$-$3.0\;\textless\;log$_{10}U$ \textless\;$-$1.5}. Varying the ionisation parameter for a given spectral index can change the [OIII](5007/4363) ratio by a factor of two, while the effect on the HeII$\lambda$4686/H$\mathrm{\beta}$ ratio is only a factor of $\sim$1.25 in the most extreme case. Changing the spectral index likewise has a significant effect on the [OIII] ratio: the modelled ratio for $\alpha$=1.5 is a factor of $\sim$1.5 higher than for $\alpha$=1.0. The chosen spectral index also has a non-negligible effect on the HeII$\lambda$4686/H$\mathrm{\beta}$ ratio, with the ratio for $\alpha$=1.5 being a factor of $\sim$1.5 higher than for $\alpha$=1.0.

\begin{figure}
    \includegraphics[width = \linewidth]{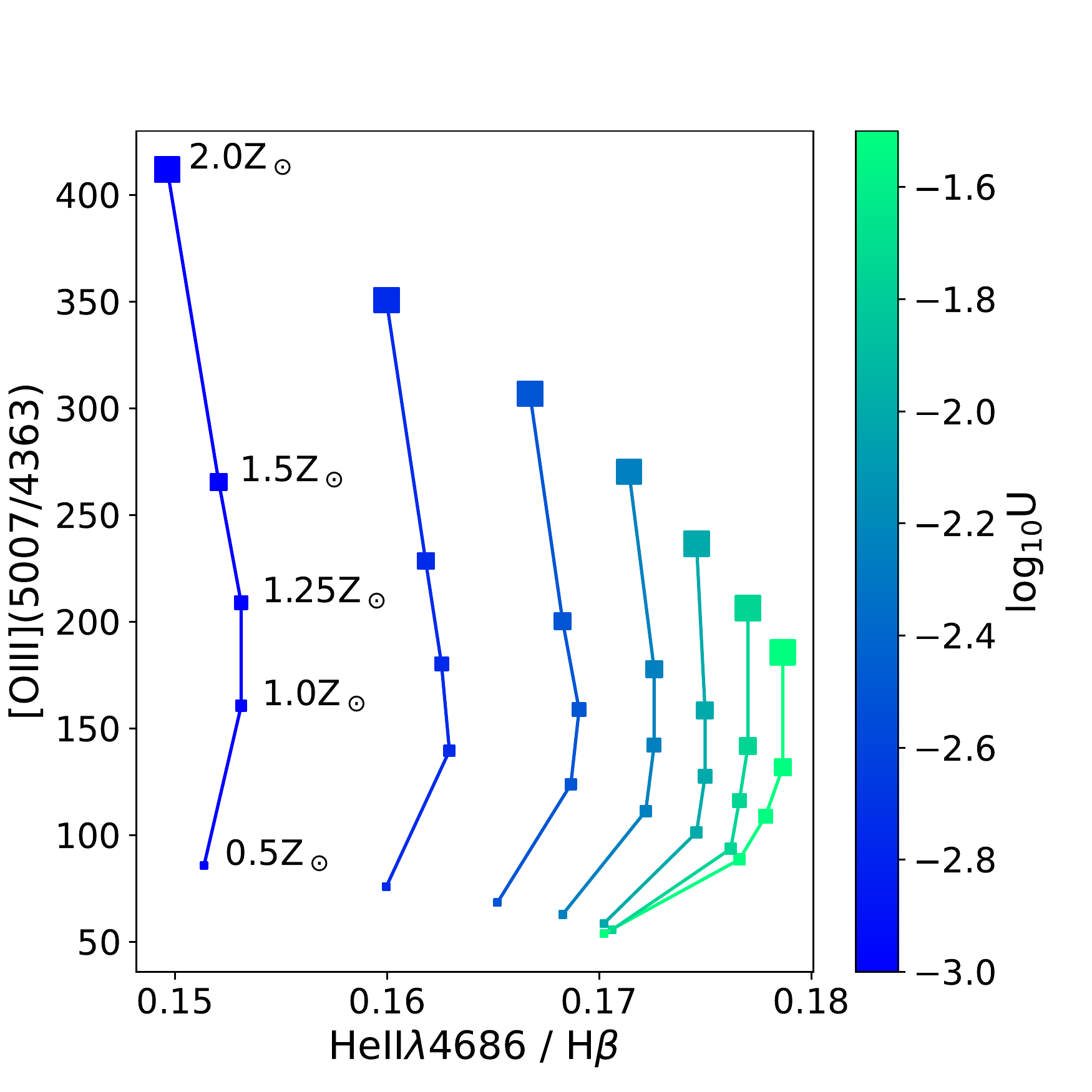}
    \caption{The same photoionisation modelling as shown in Figure \ref{fig: heii_hbeta_alpha}, but with only a single spectral index of $\alpha$=1.5 and varying metallicities as a fraction of solar abundance (1$Z_\odot$). The different fractions of solar abundance, ranging from 0.5--2$Z_\odot$ in steps of 0.5$Z_\odot$, are shown with different marker sizes and are labelled. Points with the same ionisation parameter are joined to show the variation with metallicity. Note that the limits for both axes are different to those shown in Figures \ref{fig: heii_hbeta}, \ref{fig: heii_hbeta_alpha} and \ref{fig: heii_hbeta_ne}.}
    \label{fig: heii_hbeta_z}
\end{figure}

The variation in the modelled ratios for $\alpha$=1.5 with gas metallicities ranging from 0.5--2.0\;$Z_\odot$ is shown in Figure \ref{fig: heii_hbeta_z}, with larger metallicities having larger marker sizes. As with spectral index and ionisation parameter, the chosen gas metallicity also has a significant effect on the modelled [OIII] ratio, however the effect on HeII$\lambda$4686/H$\mathrm{\beta}$ is negligible.

\begin{figure}
    \includegraphics[width = \linewidth]{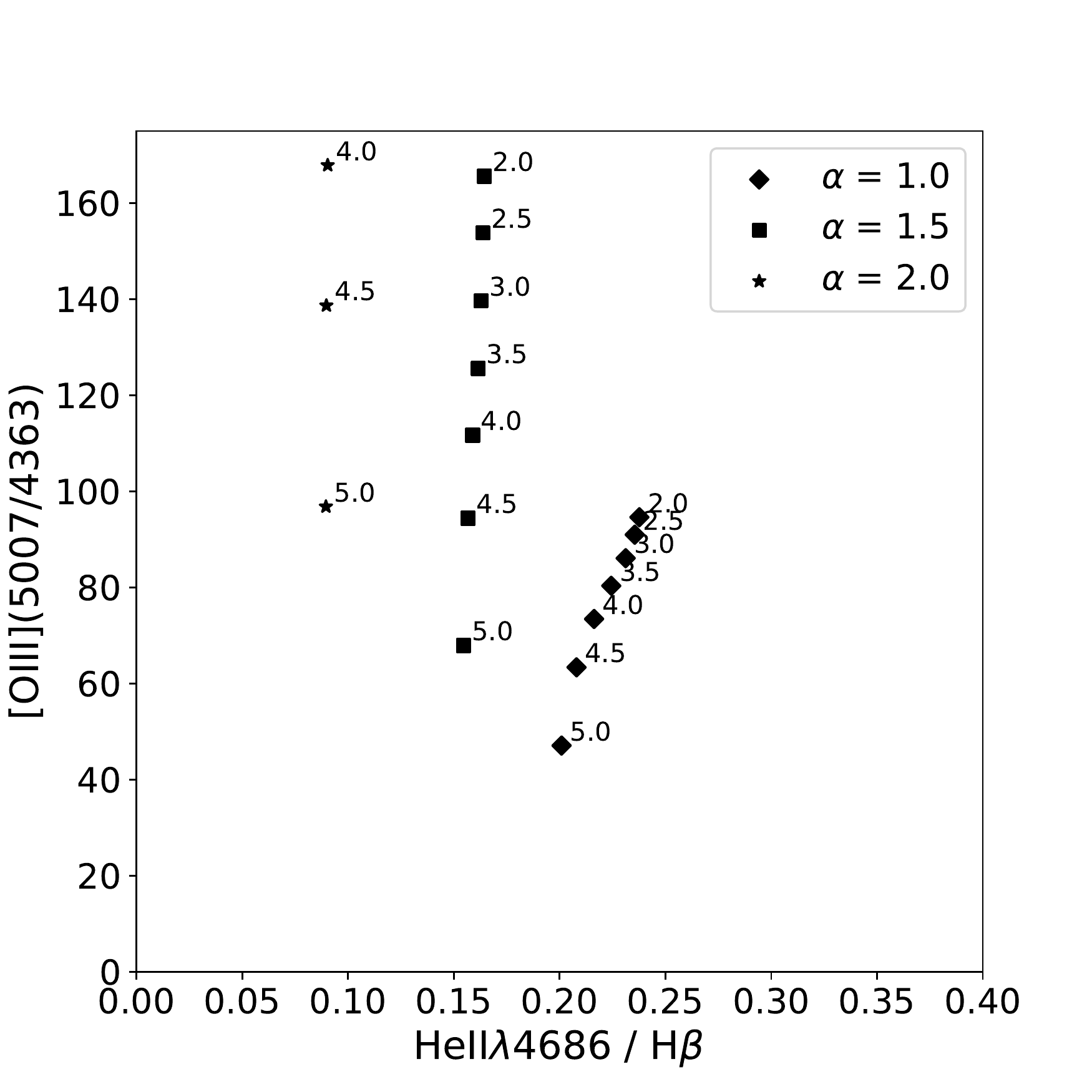}
    \caption{The same photoionisation modelling as shown in Figure \ref{fig: heii_hbeta_alpha}, but for a fixed ionisation parameter of log$U$=$-$2.75 and varied electron densities in the range \mbox{2\;\textless\;log$n_\mathrm{e}$\;\textless\;5} (labelled). Note that the limits for both axes are different to those shown in Figures \ref{fig: heii_hbeta}, \ref{fig: heii_hbeta_alpha} and \ref{fig: heii_hbeta_z}.}
    \label{fig: heii_hbeta_ne}
\end{figure}

Finally, we investigate the effect that electron density has on the modelled ratios by fixing the ionisation parameter to log$U=-2.75$ for a solar-metallicity gas with three spectral indices ($\alpha$=1.0, 1.5 and 2.0), and varying the electron density between \mbox{2\;\textless\;log$n_\mathrm{e}$\;\textless\;5}, as is shown in Figure \ref{fig: heii_hbeta_ne}.

Given the ratios we measure from our apertures of the outflowing regions in IC\;5063 (Figure \ref{fig: heii_hbeta}), we find that for $\alpha$=1.5, only metal-poor gas ($\sim$0.5$Z_\odot$), a higher electron density than we measure (\mbox{$n_\mathrm{e}$ \textgreater 3.5\;cm$^{-3}$}; Section \ref{section: tr_lines}) and a much higher ionisation parameter can explain our observations as being solely due to AGN photoionisation. The gas mass contained in the disk of IC\;5063 is $M_\mathrm{disk}\gtrapprox1\times 10^9$\;M$_\odot$ (from the molecular phase; \citealt{Morganti2015}), corresponding to an approximate stellar mass of $M_*\sim10^9$\;M$_\odot$ \citep{Maddox2015}, which would imply a metallicity of 12+log(O/H)$\approx$8.6 using the relation given by \citet{Tremonti2004}. A metallicity of 0.5\;Z$_\odot$ corresponds to 12+log(O/H)$\approx$8.3, lower than would be expected of IC\;5063 given its gas mass. 

However, the spectral index of $\alpha$=1.5 is an assumption --- it is feasible that some combination of ionisation parameter (which vary between kinematic components), along with a different spectral index, can produce line ratios similar to those found for the warm ionised gas in IC\;5063 without the need for an additional shock component.
 
\section{Recalculated mass outflow rates and energetics for the colder gas phases}
\label{section: energetics_recalculation}

In order to ensure that the mass outflow rates and energetics of the neutral atomic and cold molecular phases are consistent with our results for the warm ionised phase --- and to take into account recent results of the cold molecular gas kinematics \citep{Morganti2015, Oosterloo2017} --- we here recalculate values from past studies of IC\;5063 using a consistent methodology. These recalculated values are presented in Table \ref{tab: all_phases}, and are discussed within the context of our results for the warm ionised outflows in Section \ref{section: discussion-multi-phase-energetics}.

\subsection{Energetics of the neutral atomic phase}

We first calculated the mass outflow rate of the neutral atomic (HI) outflow at IC\;5063's NW lobe using
\begin{equation}
    \dot{M}_\mathrm{out} = 30\cdot\frac{\Omega}{4\pi}\frac{r_\ast}{1\;\mathrm{kpc}}\cdot\frac{N_\mathrm{H}}{10^{21}\;\mathrm{cm}^{-2}}\cdot\frac{v_\mathrm{out}}{300\;\mathrm{kms}^{-1}}M_\odot\mathrm{yr}^{-1},
\label{eq: appendix_hi_mout}
\end{equation}
from \citet{Morganti2005} (following the methodology given in \citealt{Heckman2002} and \citealt{Rupke2002}), where $N_H$ is the column density of the gas and $\Omega$ is the solid angle through which the gas is flowing at a radius $r_\ast$ with a velocity of $v_\mathrm{out}$. Using the values and assumptions given in \citet{Morganti2005} --- namely that \mbox{$\Omega=\pi$}, \mbox{$r_\ast=0.4$\;kpc}, $N_\mathrm{H}=10\times10^{21}$\;cm$^{-2}$, and $v_\mathrm{out}=\mathrm{FWZI}/2=350$\;kms$^{-1}$ --- we determine a mass outflow rate of 35\;M$_\odot$yr$^{-1}$.

We highlight that it is assumed that the gas is outflowing through a solid angle of $\pi$, which is highly uncertain. Moreover, taking $v_\mathrm{out}$ = $\mathrm{FWZI}/2$ may underestimate the true outflow velocity in this case, since much of the blueshifted absorption in the HI 21\;cm profile of the NW lobe in IC\;5063 is concentrated close to the maximum blue shifted velocity (approximately -700\;kms$^{-1}$),  in contrast to the other objects considered in \citet{Morganti2005}. Using $v_\mathrm{out}=700$\;kms$^{-1}$ rather than $v_\mathrm{out}=350$\;kms$^{-1}$ would result in a mass outflow rate that is a factor of two higher, and a coupling efficiency (see below) that is a factor of eight higher.

Next, we calculated the kinetic power of the HI outflow using the relation from \citet{Morganti2015}:
\begin{equation}
    \dot{E}_\mathrm{kin} = \frac{1}{2}\dot{M}_\mathrm{out}\Bigl(v^2_\mathrm{out}+\frac{v^2_\mathrm{turb}}{5.5}\Bigr),
\label{eq: appendix_ekin}
\end{equation}
where $v^2_\mathrm{turb}$ is the velocity component representing the turbulence of the outflowing gas, taken in \citet{Morganti2015} to be the FWHM of the CO(2-1) emission line that is associated with the outflow. Assuming that the CO and HI-emitting gas have similar kinematics, then we take $v_\mathrm{turb}=\mathrm{FWHM}=100$\;kms$^{-1}$ \citep{Morganti2015} along with the determined mass outflow rate calculated here ($\dot{M}_\mathrm{out}=35$\;M$_\odot$yr$^{-1}$) and the outflow velocity taken from \citet{Morganti2005} ($v_\mathrm{out}=$350\;kms$^{-1}$) --- this results in a neutral HI outflow kinetic power of \mbox{E$_\mathrm{kin}=1.4\times10^{35}$\;W}. Comparing this to the estimated nuclear bolometric luminosity of IC\;5063 (7.6$\times10^{37}$\;W: \citealt{Nicastro2003, Morganti2007}) using Equation \ref{eq: fkin}, we calculate a coupling efficiency of $\epsilon_\mathrm{f}=0.18$\;per\;cent for the neutral HI outflow at the NW radio lobe of IC\;5063.

\subsection{Energetics of the cold molecular phase}

From \citet{Oosterloo2017}, we take the estimated mass of cold molecular gas outflowing at the NW lobe to be \mbox{M$_\mathrm{out}=1.3\times10^{6}$\;M$_\odot$} (assuming optically thin gas, $T_\mathrm{ex}=29$\;K and $\alpha_\mathrm{CO} =
0.25$\;K\;kms$^{-1}$pc$^2$); we then calculate the mass outflow rate using a modified version of the relation given by \citet{Oosterloo2017} that is consistent with the method we use for the warm ionised gas \footnote{We do not include the factor of 3 used by \citet{Oosterloo2017}, as this accounts for a spherical outflow geometry, which we do not assume here.}
\begin{equation}
    \dot{M}_\mathrm{out}=\frac{v_\mathrm{out}M_\mathrm{out}}{R},
\end{equation}
where $R$ is the size of the outflow region. Taking $R=0.5$\;kpc, $v_\mathrm{out}=300$\;kms$^{-1}$ (as in \citealt{Oosterloo2017}) and \mbox{$M_\mathrm{out}=1.3\times10^{6}$\;M$_\odot$}, we calculate a mass outflow rate of \mbox{$\dot{M}_\mathrm{out}=0.79$\;M$_\odot$yr$^{-1}$.}

Taking \mbox{$\dot{M}_\mathrm{out}=0.79$\;M$_\odot$yr$^{-1}$}, $v_\mathrm{out}=300$\;kms$^{-1}$, and $v_\mathrm{turb}=\mathrm{FWHM}=100$\;kms$^{-1}$ \citep{Morganti2015}, Equation \ref{eq: appendix_ekin} thus gives a kinetic power of $2.3\times10^{33}$\;W. Therefore, we use Equation \ref{eq: fkin} to calculate the coupling efficiency of the cold molecular phase to be $\epsilon_\mathrm{f}=3.1\times10^{-3}$\;per\;cent.

We note that the true mass of the cold molecular outflow is uncertain, and that the value calculated here is likely a lower limit: the calculated outflow mass would be higher if the gas is assumed to be optically thick instead of optically thin; assuming instead the highest excitation temperature observed by \citet{Oosterloo2017} ($T_\mathrm{ex}\sim$55\;K) would approximately double the estimated mass, and uncertainties regarding separating the outflowing and non-outflowing mass may mean that the true gas mass is a factor of a few higher than calculated here (see \citealt{Oosterloo2017} for a detailed discussion).


\bsp	
\label{lastpage}
\end{document}